\documentclass[prb,amsfonts,amssymb,showpacs,twocolumn,floatfix]{revtex4}
\usepackage{graphics,bm,natbib}
\begin{document}
\widetext
\title{Lattice Distortion and Magnetism of $3d$-$t_{2g}$ Perovskite Oxides}
\author{I. V. Solovyev}
\email[Electronic address: ]{solovyev.igor@nims.go.jp}
\affiliation{
Computational Materials Science Center (CMSC),\\
National Institute for Materials Science (NIMS),\\
1-2-1 Sengen, Tsukuba, Ibaraki 305-0047, Japan
}
\date{\today}

\widetext
\begin{abstract}
Several puzzling aspects of interplay of the
experimental lattice distortion
and the
the magnetic properties of four
narrow $t_{2g}$-band perovskite oxides
(YTiO$_3$, LaTiO$_3$, YVO$_3$, and LaVO$_3$)
are clarified
using results of first-principles electronic structure calculations.
First, we derive parameters of the effective Hubbard-type Hamiltonian
for the isolated $t_{2g}$ bands
using newly developed downfolding method for the kinetic-energy part
and a hybrid approach, based on the combination of
the random-phase approximation
and
the constraint local-density approximation,
for the screened Coulomb interaction part.
Apart form the above-mentioned approximation, the procedure of
constructing the model Hamiltonian is totally parameter-free.
The results are discussed in terms of the Wannier functions localized
around transition-metal sites.
The obtained Hamiltonian was solved using a number of techniques, including
the mean-field Hartree-Fock (HF) approximation, the second-order perturbation theory
for the correlation energy, and a variational superexchange theory, which
takes into account
the multiplet structure of the atomic states.
We argue that
the crystal distortion has a profound effect not only on the
values of the
crystal-field (CF) splitting, but also on the behavior of transfer integrals
and even the screened Coulomb interactions.
Even though the CF splitting
is not particularly large to quench the orbital degrees of freedom,
the crystal distortion imposes a severe constraint on the form of the
possible orbital states,
which favor the formation of the
experimentally observed magnetic structures in
YTiO$_3$, YVO$_3$, and LaVO$_3$
even at the level of mean-field HF approximation.
It is remarkable that for all three compounds,
the main results of all-electron calculations can be successfully reproduced
in our minimal model derived for the isolated $t_{2g}$ bands.
We confirm that
such an agreement is only possible when
the nonsphericity of the Madelung potential is explicitly
included into the model.
Beyond the HF approximation,
the correlations effects
systematically improve the agreement with the
experimental data.
Using the
same type of approximations we could not
reproduce the correct magnetic ground state of LaTiO$_3$.
However, we expect that the situation may change by systematically
improving the level of approximations
for dealing with the correlation effects.
\end{abstract}



\maketitle


\onecolumngrid

\section{\label{sec:Intro}Introduction}

  The transition-metal perovskite oxides $AB$O$_3$ (with $A$$=$ Y, La, or
other trivalent rare-earth ion, and $B$$=$ Ti or V) are regarded as some of the
key materials for understanding the strong coupling among spin, orbital, and lattice
degrees of freedom in correlated electron systems.\cite{reviews}

\begin{figure}[h!]
\begin{center}
\resizebox{7cm}{!}{\includegraphics{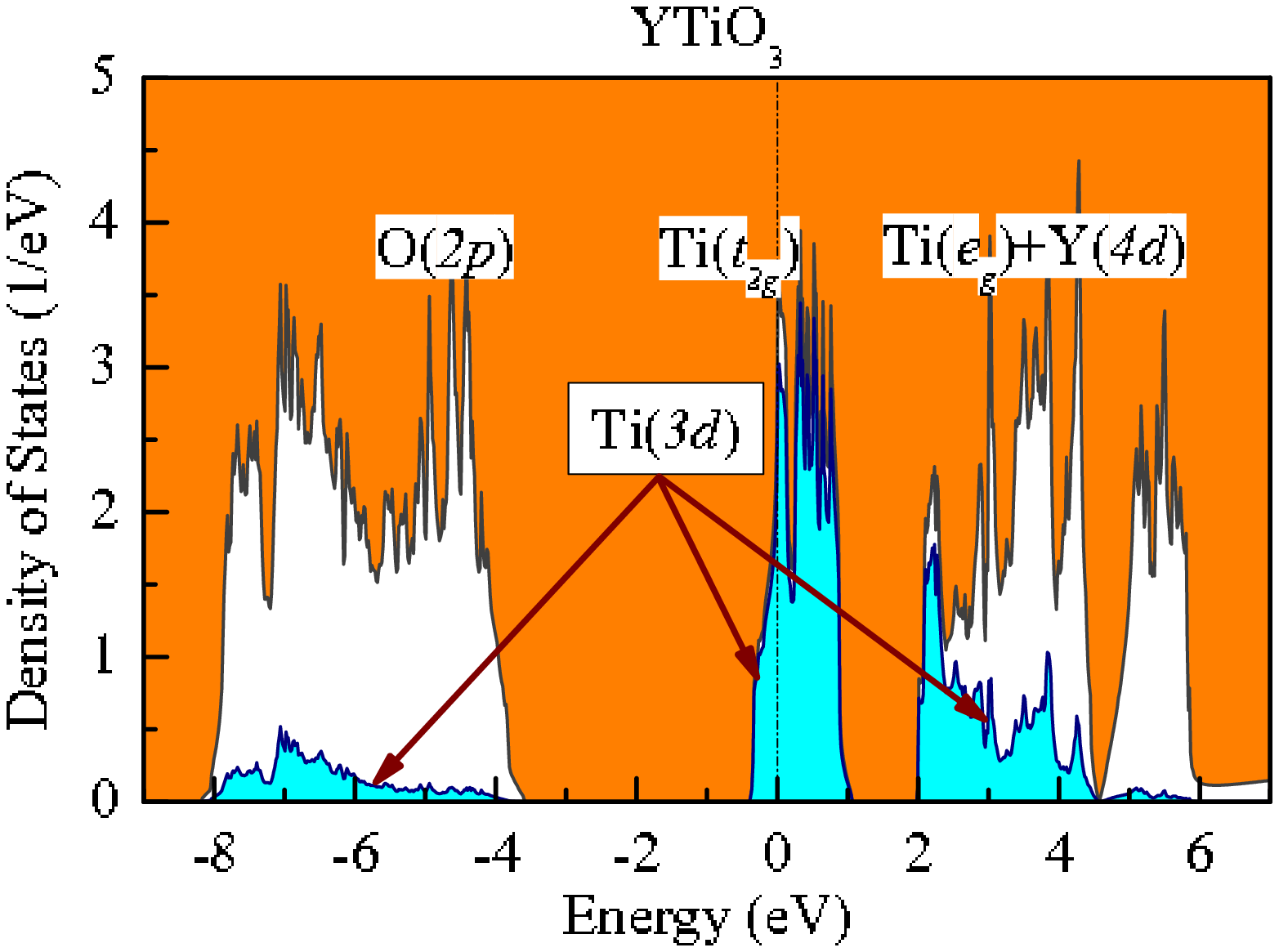}}
\resizebox{7cm}{!}{\includegraphics{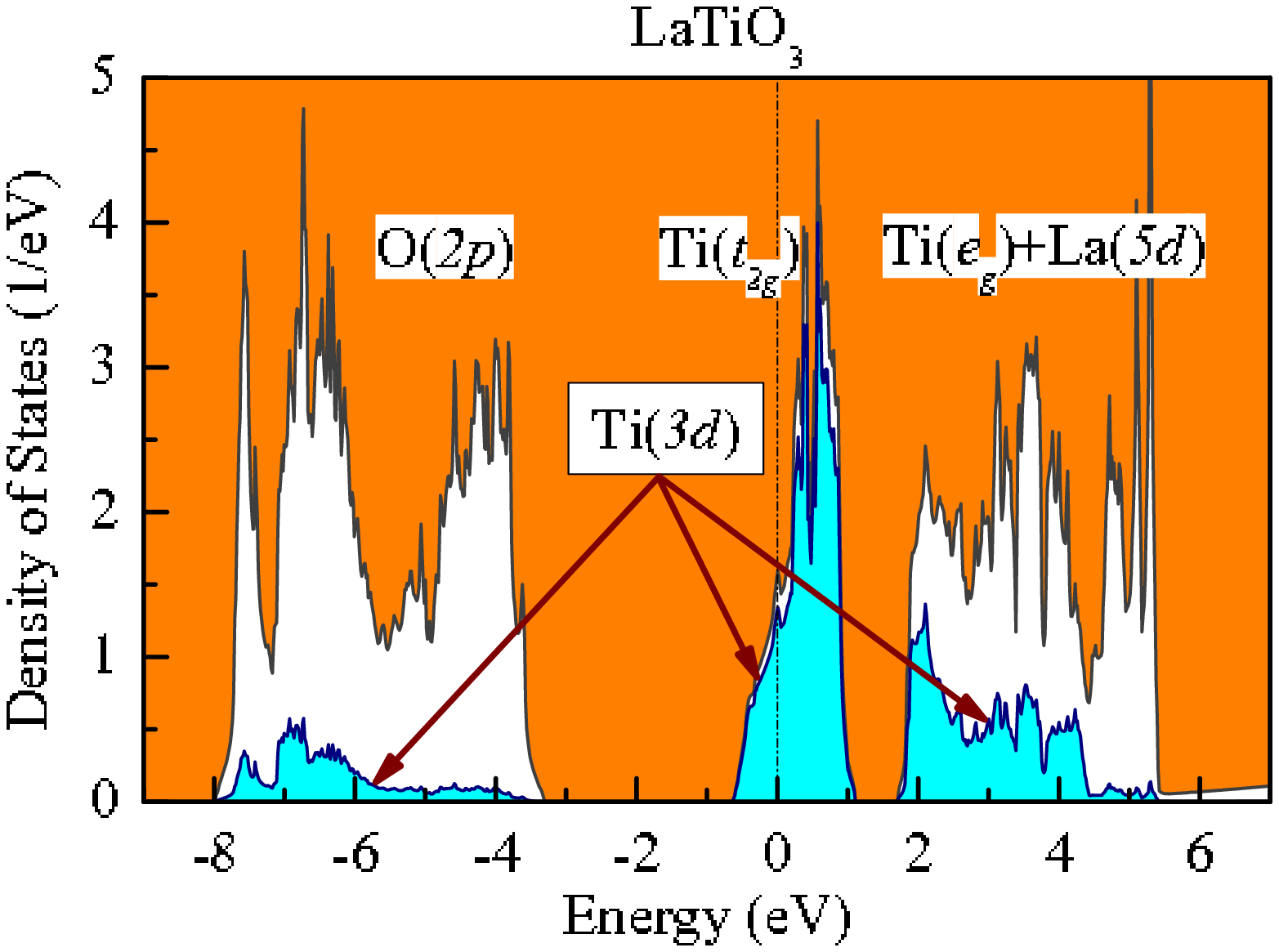}}
\end{center}
\begin{center}
\resizebox{7cm}{!}{\includegraphics{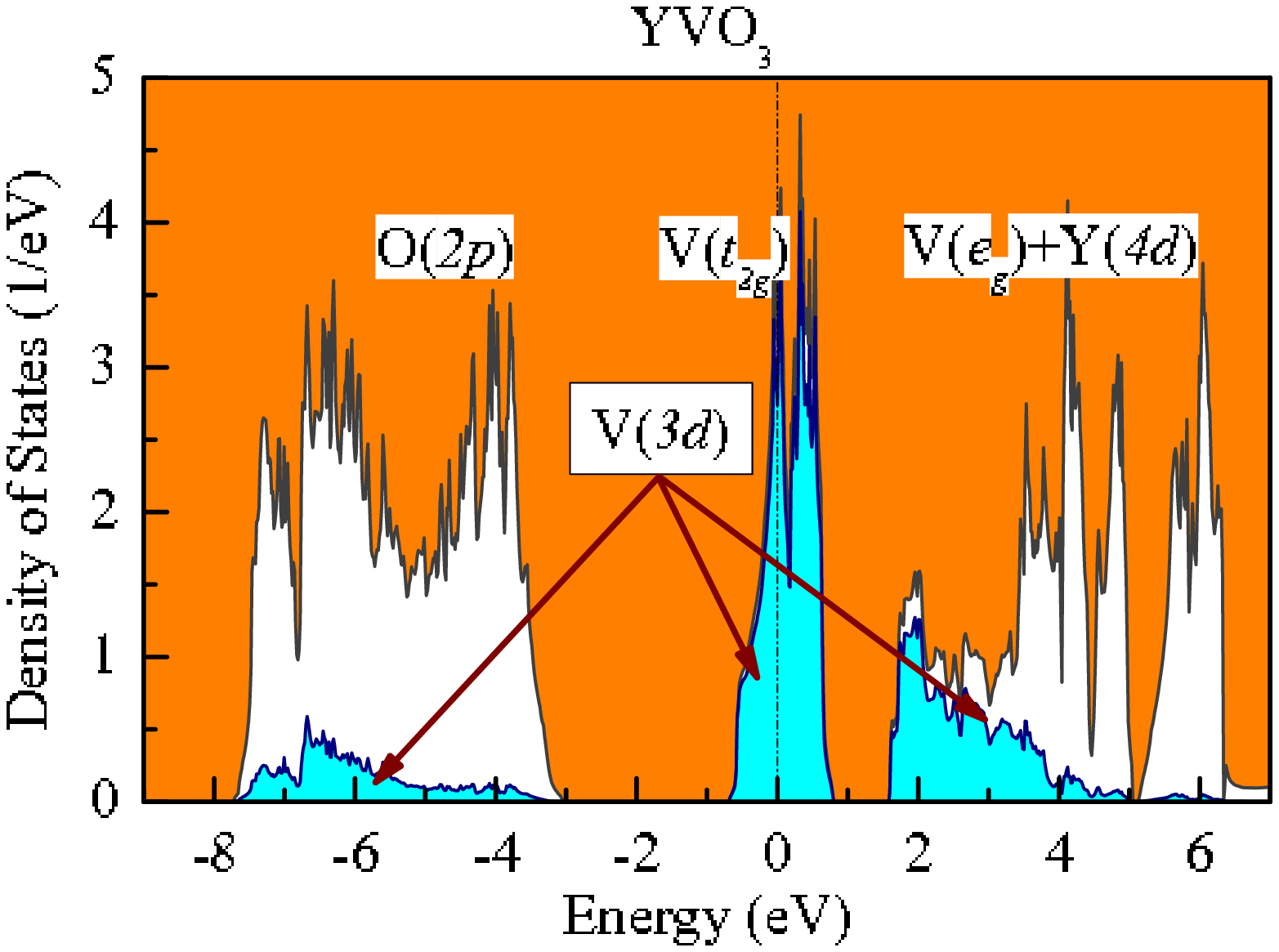}}
\resizebox{7cm}{!}{\includegraphics{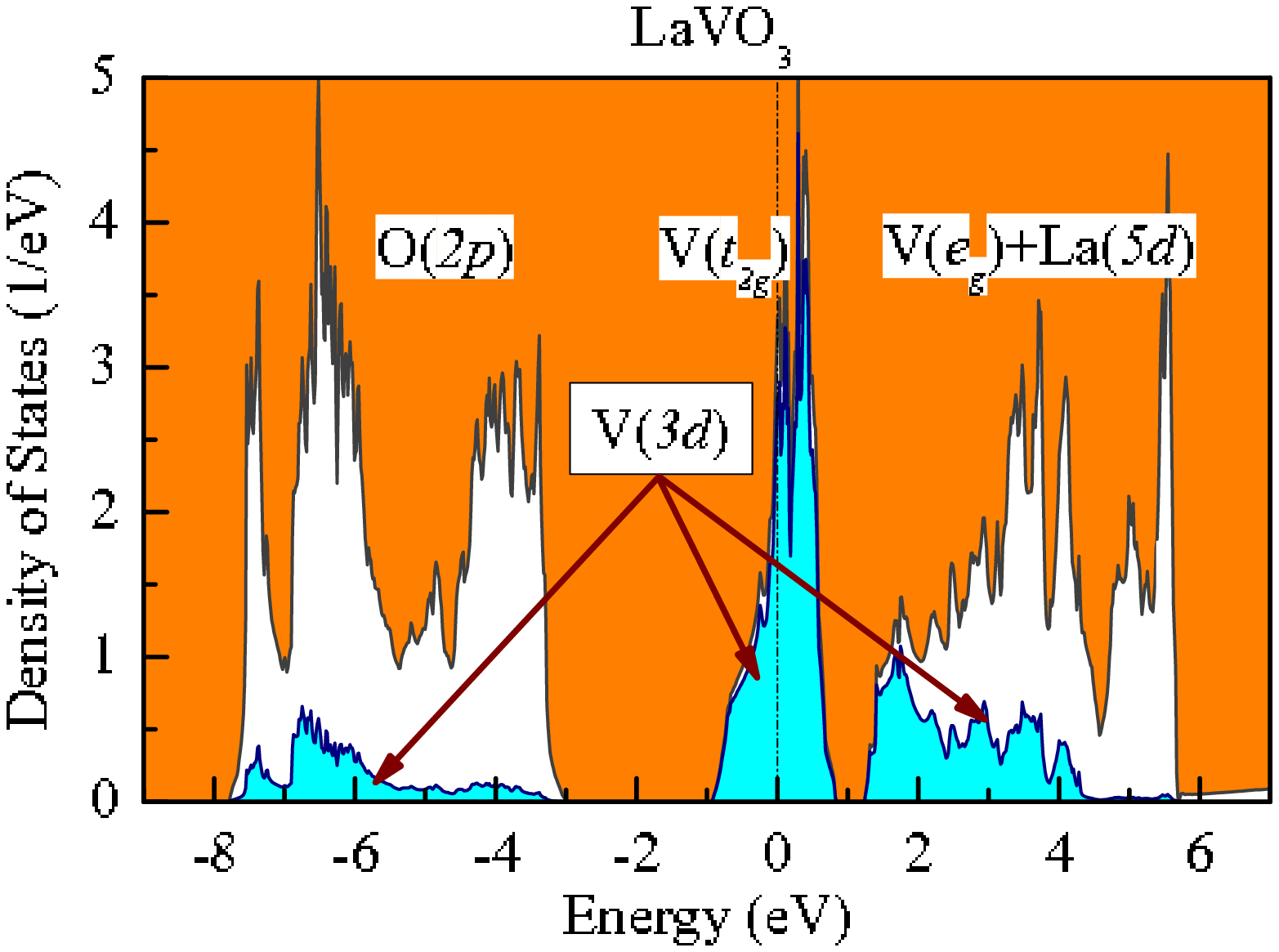}}
\end{center}
\caption{\label{fig.DOSsummary} (Color online) Total and partial densities of states
for YTiO$_3$, LaTiO$_3$, YVO$_3$ (orthorhombic phase),
and LaVO$_3$ in the local-density approximation.
The shaded area shows the contributions of the transition-metal $3d$-states.
Other symbols show positions of the main bands.
The Fermi level is at zero energy.}
\end{figure}

\begin{figure}[h!]
\begin{center}
\resizebox{5cm}{!}{\includegraphics{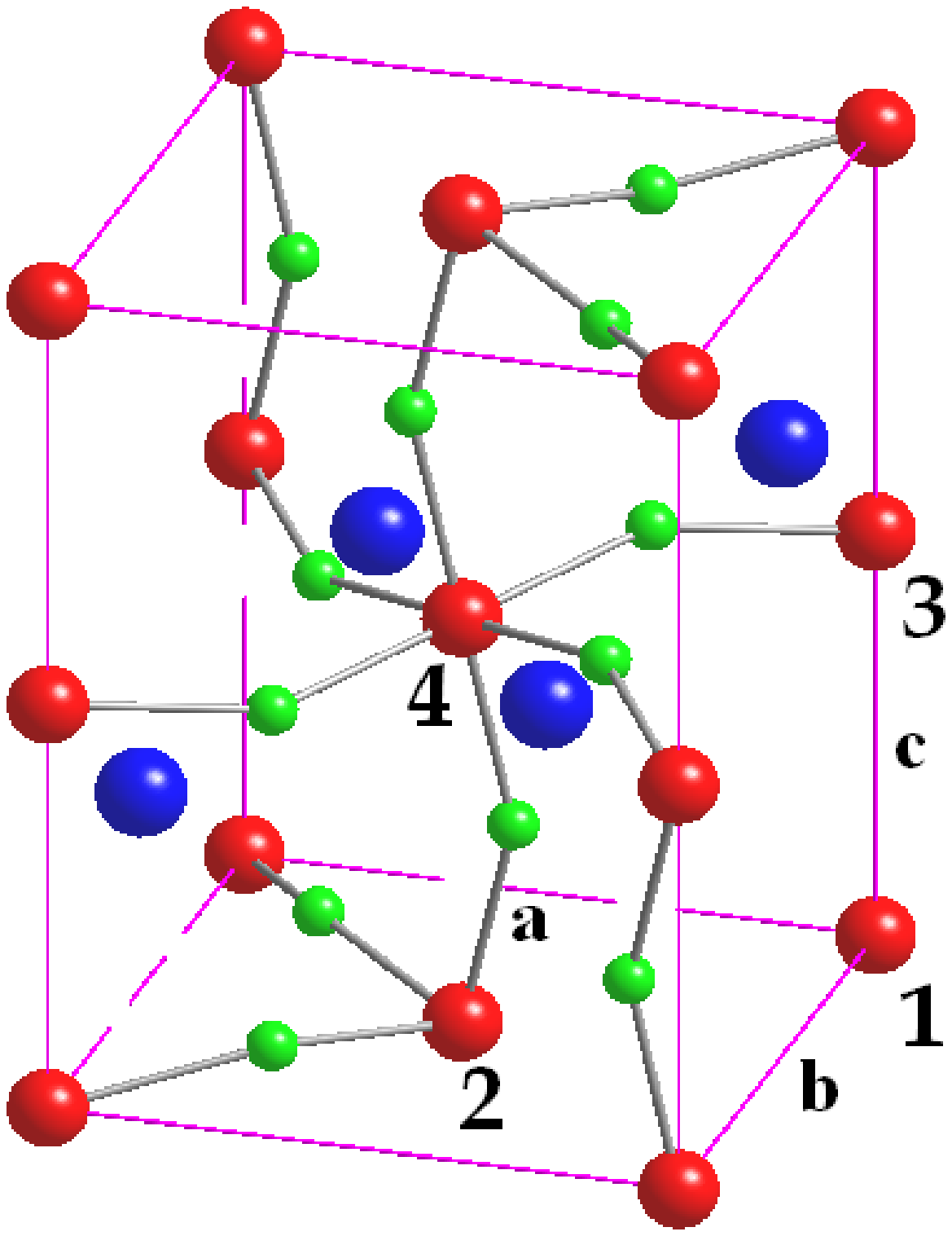}}
\resizebox{5cm}{!}{\includegraphics{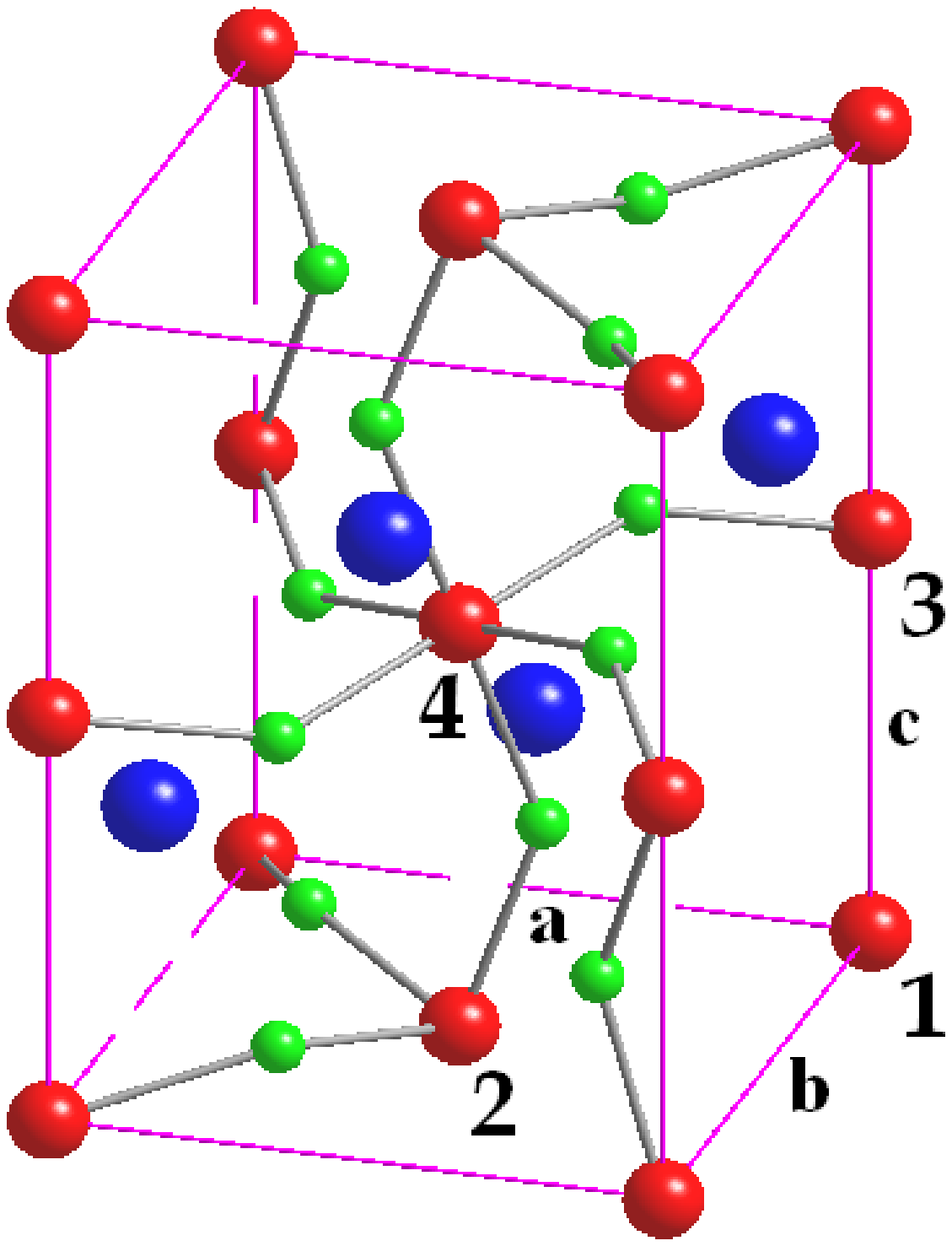}}
\resizebox{5cm}{!}{\includegraphics{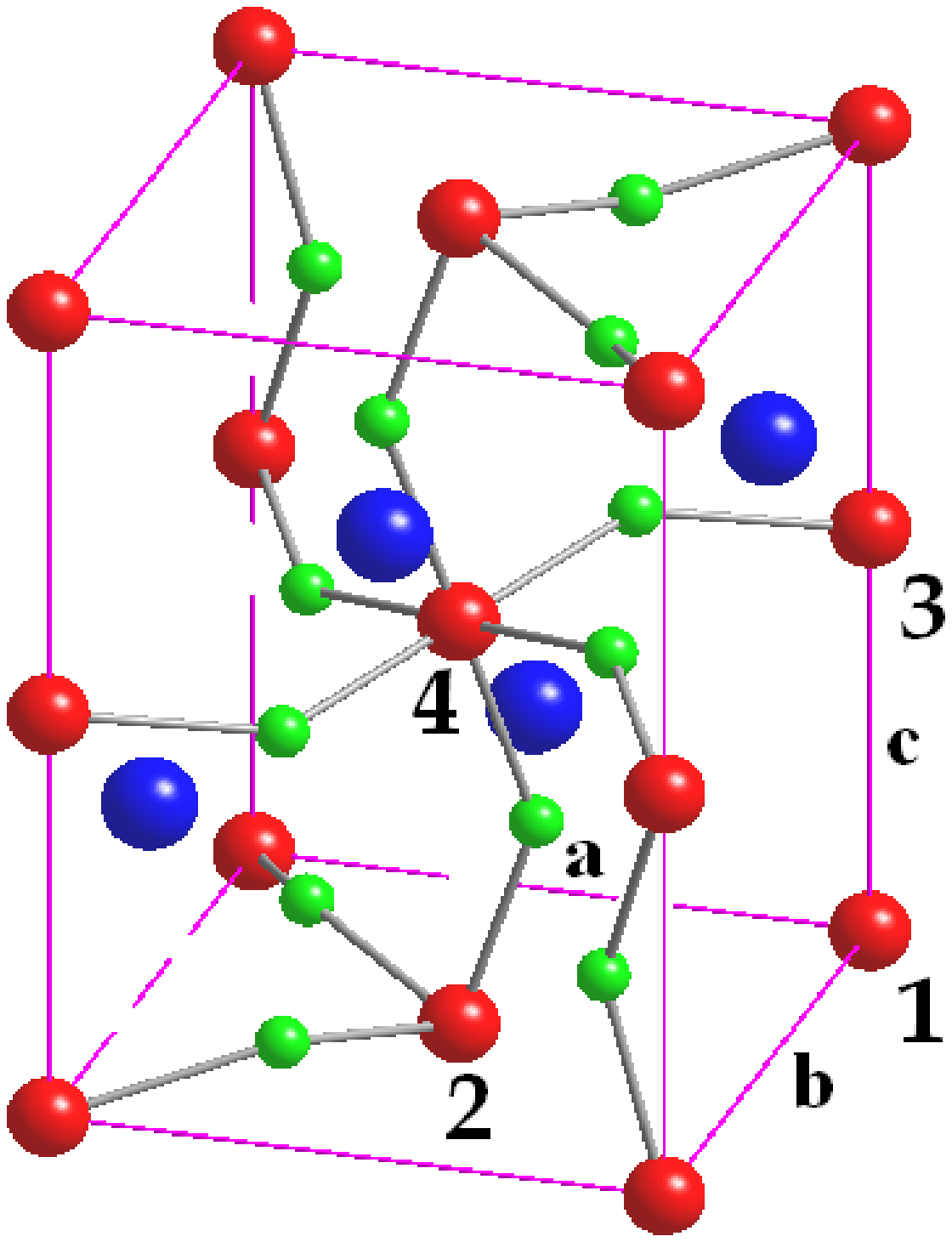}}
\end{center}
\caption{\label{fig.structure} (Color online) Crystal structure
of orthorhombic LaTiO$_3$ (left),
YTiO$_3$ (middle), and YVO$_3$ (right). La and Y sites are indicated
by big blue spheres, Ti and V sites -- by medium red spheres, and oxygen sites --
by small green spheres. The symbols ${\bf a}$, ${\bf b}$, and ${\bf c}$ stand for
the orthorhombic translations.
The symbols $1$-$4$ denote the
transition-metal sites, which forms the
unit cell of the distorted perovskite oxides.}
\end{figure}

  According to the electronic structure calculations in the local-density approximation (LDA),
all these compounds can be classified as ``$t_{2g}$ systems'', as all of them have a common
transition-metal $t_{2g}$-band, located near the Fermi level, which is well separated from the
oxygen-$2p$ band
and a hybrid transition-metal $e_g$ and either
Y($4d$) or La($5d$) band, located correspondingly in the lower- and upper-part
of the spectrum (Fig.~\ref{fig.DOSsummary}).
The number of electrons that are donated by each Ti and V site into
the $t_{2g}$-band is correspondingly one and two. These electrons
are subjected to the strong intraatomic Coulomb repulsion, which is
not properly treated by LDA and requires some considerable
improvement of this approximation, which currently processes in the
direction of merging LDA with various model approaches for the
strongly-correlated systems.\cite{AZA,PRB94,LSDADMFT,Imai}
Nevertheless, LDA continues
play an important role for these systems
as it naturally incorporates into the
model analysis the effects of the lattice distortion, and does it
formally without any adjustable parameters. Although the origin of
the lattice distortion in the $t_{2g}$ perovskite oxides is not
fully understood, is is definitely strong and exhibits an
appreciable material-dependence, which can be seen even visually in
Fig.~\ref{fig.structure}.
The interplay of this lattice distortion with the Coulomb
correlations seems to be the key factor for understanding the
large variation of the magnetic properties among the $t_{2g}$
perovskite oxides. The difference exists not only between Ti- and
V-based compounds, but also within each group of formally isovalent
materials, depending on whether it is composed of the Y or La
atoms. The latter seems to be a clear experimental manifestation
of the distortion effect, which is related with the difference
of the ionic radii of Y and La. All together this leads to the
famous phase diagram of the distorted $t_{2g}$ perovskite oxides,
where each material exhibits quite a distinct magnetic behavior: YTiO$_3$ is a
ferromagnet;\cite{Maclean,Itoh,Akimitsu,Ulrich2002,Iga} LaTiO$_3$ is
a three-dimensional (G-type) antiferromagnet;\cite{Keimer,Cwik}
YVO$_3$ has the low-temperature G-type antiferromagnetic (AFM) phase,
which at around $77$ K transforms into a chain-like (C-type)
antiferromagnetic phase;\cite{Ren,Blake,Tsvetkov,Ulrich2003} and
LaVO$_3$ is the C-type antiferromagnet.\cite{Zubkov,Bordet}

  On the theoretical side, the large variety of these magnetic phases
has been intensively studied
using
model approaches
(Refs.~\onlinecite{MizokawaFujimori,KhaliullinMaekawa,Khaliullin01,Khaliullin02,
MochizukiImada,Schmitz})
as well as the first-principles electronic structure
calculations
(Refs.~\onlinecite{FujitaniAsano,Sawada96,SawadaTerakura,PRB04,FangNagaosa,
Pavarini1,Pavarini2,Streltsov,Okatov}). The problem is still far from being
understood, and remains to be the subject of numerous contradictions and debates.
Surprisingly that at present there is no clear consensus not only between
model and first-principles electronic structure
communities, but also between researchers working in each of these groups.
Presumably, the most striking example is LaTiO$_3$, where in order to explain the
experimentally observed G-type AFM ground state,
two different models, which practically exclude
each other, have been proposed. One is the model of orbital liquid, \textit{which implies
the degeneracy of the
atomic $t_{2g}$ levels in the crystalline environment}.\cite{KhaliullinMaekawa}
Another model is based on the theory of crystal-field (CF) splitting,
\textit{which lifts the orbital degeneracy} and leads to the one particular type
the orbital ordering compatible with the
G-type antiferromagnetism.\cite{MochizukiImada,Schmitz}
The situation in the area of first-principles electronic structure
calculations is controversial as well.
Although majority of the researchers now agree that in order to describe properly
the electronic structure of $t_{2g}$ perovskite oxides, one should go beyond the
conventional LDA and incorporate the effects of intraatomic
Coulomb correlations, this merging is typically done in a semi-empirical way, as it
relies on a certain number of adjustable parameters, postulates,
and the form of the basis functions used for the implementation of
various corrections on the top of LDA.\cite{AZA,PRB94,LSDADMFT}
There are also certain differences regarding both
the definition and the approximations used for the CF splitting
in the electronic structure calculations,
which will be considered in details in Sec.~\ref{sec:kinetic}.
Since the magnetic properties of $t_{2g}$ perovskite oxides are extremely
sensitive to all such details, it is not surprising that there is a
substantial variation in the results of first-principles
calculations, which sometimes yield even qualitatively different conclusions
about the CF splitting and the magnetic structure of the distorted
$t_{2g}$ perovskite oxides.\cite{PRB04,Pavarini2,Streltsov,Okatov}
These discrepancies put forward
a severe demand on the creation of a really parameter-free scheme of
electronic structure calculations for the strongly-correlated systems.

  Therefore, the main motivation of the present work is twofold.\\
(i) In our previous work (Ref.~\onlinecite{condmat05}) we have proposed a method
of construction of the effective Hubbard-type model for the electronic
states near the Fermi level on the basis of first-principles
electronic structure calculations. In the present work we apply this
strategy to the $t_{2g}$ states of the distorted perovskite oxides.
Namely, we will derive the parameters of the Hubbard Hamiltonian
for the $t_{2g}$ bands and solve this Hamiltonian using several
different
techniques, including the Hartree-Fock (HF) approximation, the perturbation theory
for the correlation energy, and the the theory of superexchange interactions
taking into account the effects of the multiplet structure of the atomic states.
Of course, our method is based on a number
of approximation, which
have been introduced in Ref.~\onlinecite{condmat05} and
will be briefly discussed in Sec.~\ref{sec:Model}.
However,
we would like to emphasize from the very beginning that our policy here is
\textit{not to use any adjustable parameters} apart from the approximations
considered in Ref.~\onlinecite{condmat05}.
Thus, we believe that it
poses a severe test for the proposed method, and the
obtained results
should clearly demonstrate that our general strategy, which
can be expressed by the formula
\textit{first-principles electronic structure calculations} $\rightarrow$
\textit{construction of the model Hamiltonian} $\rightarrow$
\textit{solution of the model Hamiltonian},\cite{Imai,condmat05}
is indeed very promising.
For example,
at the HF level,
using relatively simple model Hamiltonian,
which is limited exclusively by the $t_{2g}$ bands,
we will be able to reproduce the main results of all-electron
calculations.\cite{SawadaTerakura,FangNagaosa}
Furthermore,
due to the simplicity of the model Hamiltonian we can easily go beyond the
HF approximation and include the correlation effects.\\
(ii) Why do we need to convert the
results of first-principles electronic structure calculations into a model?
Apart from the purely computational reason, related with the
reduced dimensionality
of the Hilbert space for the solution of the many-electron problem,\cite{Imai}
the story of
distorted $t_{2g}$ perovskite oxides
clearly shows that
the model consideration has yet another advantage, which is typically not sufficiently
appreciated in the computational community.
It is true that the field of first-principles electronic structure calculations is
currently on the rise, and the calculation of the basic properties for many
materials will soon become a matter of routine. However, the methods of
electronic structure calculations are based on some approximations, the limitations
of which should be clearly understood. Furthermore, like the experiment data,
the results of
first-principles electronic structure
calculations will always require some interpretation, which
would transform the world of numbers and trends into a ``parallel world'' of
rationalized
model categories capturing the essence of the
electronic structure calculations.
The understanding of the results of calculations in terms
of these categories
opens a way,
on the one hand,
to the material engineering of compounds
with a desired set of properties, and, on the other way,
to ``engineering'' of the new methods of electronic structure calculations
in the direction of elucidation and
overcoming the existing approximations.
In this work we will illustrate how the results of first-principles
calculations for the distorted perovskite oxides can be
interpreted in terms of a limited number of model parameters, such as
the crystal-field splitting, transfer integrals, and the intraatomic
Coulomb interactions,
which can be regarded as the basic operating blocks
for understanding the properties of these materials as well as
the limitation of approximations existing
in the methods
of electronic structure calculations.
Particularly, we will explicitly show that the
atomic-spheres-approximation (ASA), which was employed in the series of
publications (Refs.~\onlinecite{PRB04,Pavarini1,Pavarini2,Streltsov}),
is not enough as it neglects the nonsphericity of the Madelung potential.
The latter
plays an important role and in many cases predetermines the character of
the magnetic ordering
in the distorted $t_{2g}$ perovskite
oxides.
We will also show that once the parameters of Coulomb interactions
are determined from the first principles, the commonly used
mean-field HF approximation does not necessary guaranty
the right answer for the magnetic properties of $t_{2g}$
perovskite oxides. However, we will argue that this is a normal situation,
and in the majority of cases, a better agreement with the experimental
data can be obtained
by systematically including the
correlation effects beyond the HF approximation.
In this sense, our strategy is completely different from conventional LDA$+$$U$
calculations, where the on-site Coulomb interaction $U$ is typically
treated as an adjustable parameter
(e.g., Refs.~\onlinecite{SawadaTerakura,FangNagaosa,Pavarini2,Streltsov,Okatov}).
By changing $U$,
one can certainly get a better numerical agreement with some experimental data
already at the HF level. However, one should clearly understand that
such an empirical treatment
actually disguises the actual role played by the correlation effects
in the narrow-band compounds.

  The rest of the paper is organized as follows. In Sec.~\ref{sec:structure}
we will briefly remind the main details of the crystal and magnetic structure
of the distorted perovskite oxides. The procedure of constructing the model
Hamiltonian as well as the results of calculations
of the CF splitting, transfer
integrals, and on-site Coulomb interactions
for the isolated $t_{2g}$ band will be briefly explained in Sec.~\ref{sec:Model}.
Particularly, in Sec.~\ref{sec:kinetic} we will discuss highly controversial
situation around the values of
the
CF splitting extracted from electronic structure
calculations,\cite{PRB04,Pavarini1,Pavarini2,Streltsov}
and argue
that the main difference
is caused by two factors:
(i) certain arbitrariness with the choice of the Wannier functions for the
$t_{2g}$ bands of the distorted perovskite oxides;
(ii) additional approximations used for the nonspherical part
of the crystalline potential inside atomic spheres.
The methods of solution of the model Hamiltonian will be described
Sec.~\ref{sec:Method}, and the results of calculations will be
presented in Sec.~\ref{sec:results}. Finally, in Sec.~\ref{sec:summary}
we will summarize the main results of our work.

\section{\label{sec:structure}Crystal and Magnetic Structures}

  The distorted perovskite oxides contain
four formula units in the primitive cell. The transition-metal ($B$)
atoms are located at $(0,0,0)$ (site 1), $({\bf a}/2,{\bf b}/2,0)$ (site 2),
$(0,0,{\bf c}/2)$ (site 3), and $({\bf a}/2,{\bf b}/2,{\bf c}/2)$ (site 4),
in terms of three primitive translations: ${\bf a}$, ${\bf b}$, and ${\bf c}$
(see Fig.~\ref{fig.structure}).
The distortion can be either orthorhombic or monoclinic.

  The space group of the orthorhombic phase is $D^{16}_{2h}$
(in the Sch\"{o}nflies notations or $Pbnm$ in the Hermann-Maguin notations,
No.~62 in the International Tables).
In this case all $B$-sites are equivalent and can be transformed to each other
using symmetry operations of the $D^{16}_{2h}$ group.

  The monoclinic phase has the space group $C^5_{2h}$
($P2_1/a$, No.~14 in the International Tables).\cite{Blake,comment.7}
In this case, there are two nonequivalent pairs of $B$-sites: (1,2) and (3,4).
Each pair is allocated within
the same ${\bf ab}$-plane, so that the atoms can be
transformed to each other using symmetry operations of the $C^5_{2h}$ group.
However, there is no symmetry operation, which connects the atoms
belonging to different ${\bf ab}$-planes.

  We use the
experimental lattice parameters and the atomic positions reported in
Ref.~\onlinecite{Maclean} for YTiO$_3$ (the data corresponds to the
temperature T$=$$293$ K), in
Ref.~\onlinecite{Cwik} for LaTiO$_3$ (T$=$$8$ K), in
Ref.~\onlinecite{Blake} for YVO$_3$ (T$=$$65$ K and $100$ K,
for the orthorhombic
and monoclinic phase, respectively), and
in Ref.~\onlinecite{Bordet} for LaVO$_3$ (T$=$$10$ K).

  There are five possible magnetic structure, which can be obtained
by associating with each transition-metal site either positive ($\uparrow$)
or negative ($\downarrow$) direction of spin, without enlarging
the unit cell. Therefore, each magnetic structure can be denoted by
means of four
vectors associated with the transition-metal
sites (1 2 3 4). They are\\
1. ($\uparrow$$\uparrow$$\uparrow$$\uparrow$), which is called the
ferromagnetic (F) phase;\\
2. ($\uparrow$$\uparrow$$\downarrow$$\downarrow$), the layered (A-type)
antiferromagnetic phase;\\
3. ($\uparrow$$\downarrow$$\uparrow$$\downarrow$), the chainlike (C-type)
antiferromagnetic phase;\\
4. ($\uparrow$$\downarrow$$\downarrow$$\uparrow$), the totally antiferromagnetic
(G-type) phase;\\
5. ($\downarrow$$\uparrow$$\uparrow$$\uparrow$), the spin-flip phase.
In the monoclinic structure, there are two different spin-flip phases:
($\downarrow$$\uparrow$$\uparrow$$\uparrow$) and ($\uparrow$$\uparrow$$\downarrow$$\uparrow$),
which will be denoted as flip-I and flip-II, respectively.\\
Similar classification can be used the orbital ordering.
Typically, two orthogonal orbitals
at the neighboring transition-metal sites
are said to be ordered antiferromagnetically, although such a definition is not
unique.\cite{comment.6}

\section{\label{sec:Model}Model Hamiltonian}

  Our first goal is the construction of the effective
multi-orbital Hubbard model for
the isolated $t_{2g}$ bands:
\begin{equation}
\hat{\cal{H}}= \sum_{{\bf R}{\bf R}'} \sum_{\alpha \beta} h_{{\bf
R}{\bf R}'}^{\alpha \beta}\hat{c}^\dagger_{{\bf
R}\alpha}\hat{c}^{\phantom{\dagger}}_{{\bf R}'\beta} + \frac{1}{2}
\sum_{\bf R}  \sum_{\alpha \beta \gamma \delta} U_{\alpha \beta
\gamma \delta} \hat{c}^\dagger_{{\bf R}\alpha} \hat{c}^\dagger_{{\bf
R}\gamma} \hat{c}^{\phantom{\dagger}}_{{\bf R}\beta}
\hat{c}^{\phantom{\dagger}}_{{\bf R}\delta},
\label{eqn:Hmanybody}
\end{equation}
where $\hat{c}^\dagger_{{\bf R}\alpha}$ ($\hat{c}_{{\bf R}\alpha}$)
creates (annihilates) an electron in the Wannier orbital
$\tilde{W}_{\bf R}^\alpha$ of the site ${\bf R}$, and $\alpha$ is
a joint index, incorporating all remaining (spin and orbital)
degrees of freedom. The matrix $\hat{h}_{{\bf R}{\bf R}'}$$=
$$\| h_{{\bf R}{\bf R}'}^{\alpha \beta} \|$
parameterizes the kinetic energy of electrons, where
the site-diagonal part (${\bf R}$$=$${\bf R}'$) describes
the local level-splitting, caused by the crystal field and (or) the
spin-orbit interaction, and the off-diagonal part
(${\bf R}$$\neq$${\bf R}'$) stands for the transfer integrals.
$U_{\alpha \beta \gamma \delta}
=
\int d{\bf r} \int d{\bf r}' \tilde{W}_{\bf R}^{\alpha \dagger}({\bf r})
\tilde{W}_{\bf R}^\beta({\bf r}) v_{\rm scr}({\bf r}$$-$${\bf r}')
\tilde{W}_{\bf R}^{\gamma \dagger}({\bf r}') \tilde{W}_{\bf R}^\delta({\bf r}')
\equiv \langle \tilde{W}_{\bf R}^\alpha \tilde{W}_{\bf R}^\gamma | v_{\rm scr} |
\tilde{W}_{\bf R}^\beta \tilde{W}_{\bf R}^\delta \rangle$
are
the matrix elements of \textit{screened} Coulomb interaction
$v_{\rm scr}({\bf r}$$-$${\bf r}')$, which are supposed to be diagonal with
respect to the site indices.

  The parameters of the Hubbard Hamiltonian (\ref{eqn:Hmanybody}) can
be derived ``from the first principles'', starting from
the electronic structure in LDA.
This procedure has been already discussed in details in Ref.~\onlinecite{condmat05}.
Here we only remind the main ideas and present the results for the
distorted perovskite compounds.

  All calculations have been performed using linear muffin-tin-orbital (LMTO)
method in the atomic-spheres-approximation.\cite{LMTO}
We have also considered the additional correction to the crystal-field
splitting, coming from the nonsphericity of electron-ion interactions,
beyond conventional ASA.\cite{condmat05}

\subsection{\label{sec:kinetic}Kinetic-Energy Part,
Controversy about the Crystal-Field Splitting}

  The kinetic-energy part of the Hubbard Hamiltonian was constructed using the
downfolding method.\cite{condmat05,PRB04} It yields certain set
of parameters $\{ h_{{\bf R}{\bf R}'}^{\alpha \beta} \}$.
The Wannier functions $\{ \tilde{W}_{\bf R}^\alpha \}$ for
the the isolated $t_{2g}$ bands can be formally reconstructed from $\{ h_{{\bf R}{\bf R}'}^{\alpha \beta} \}$
using the definition
$ h_{{\bf R}{\bf R}'}^{\alpha \beta} $$=$$
\langle \tilde{W}_{\bf R}^\alpha| \hat{H}^{\rm LDA} |
\tilde{W}_{\bf R}^\beta \rangle$, where $\hat{H}^{\rm LDA}$
is the LDA Hamiltonian in ASA.\cite{comment.2}

\begin{figure}[h!]
\begin{center}
\resizebox{5cm}{!}{\includegraphics{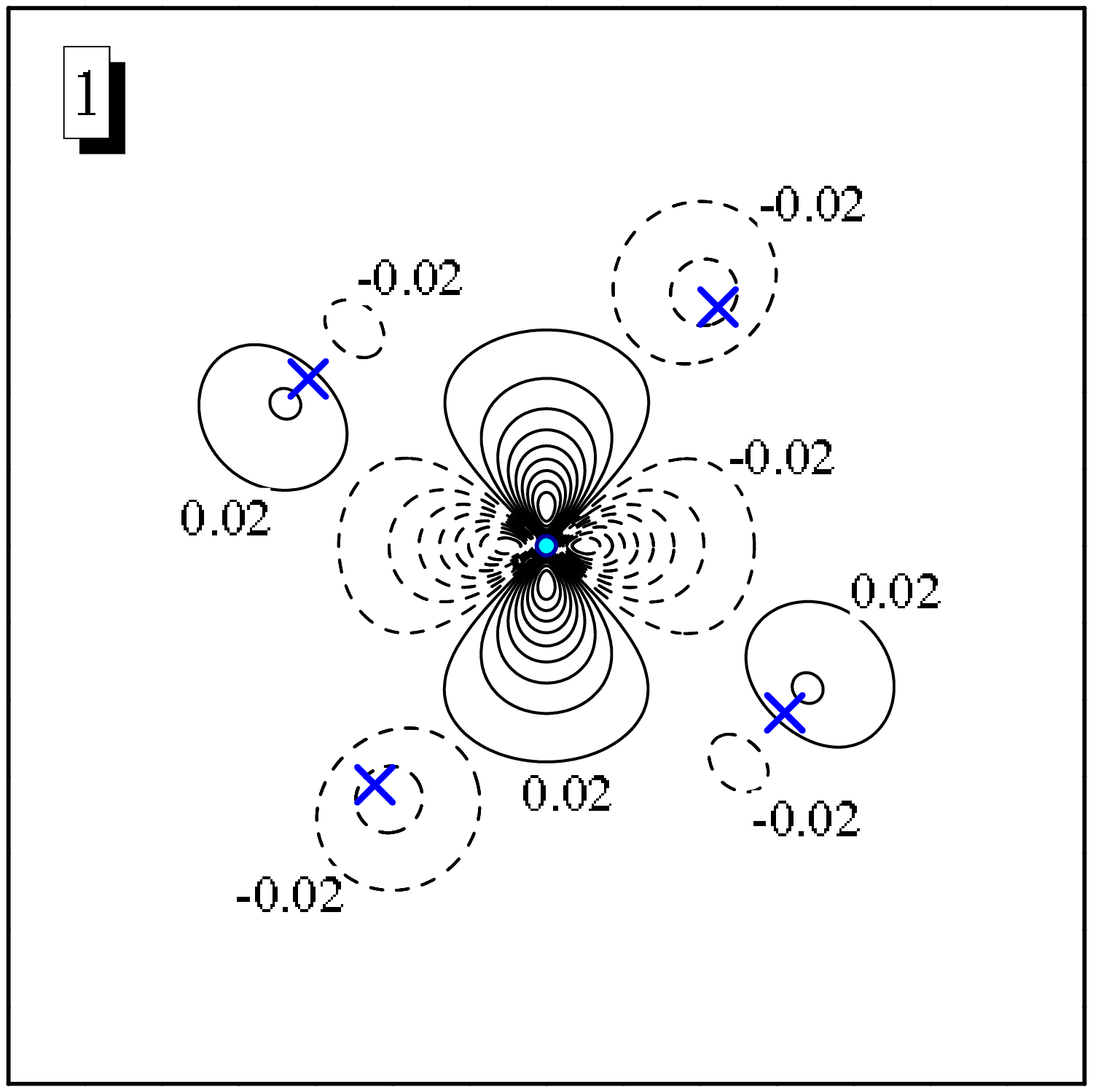}}
\resizebox{5cm}{!}{\includegraphics{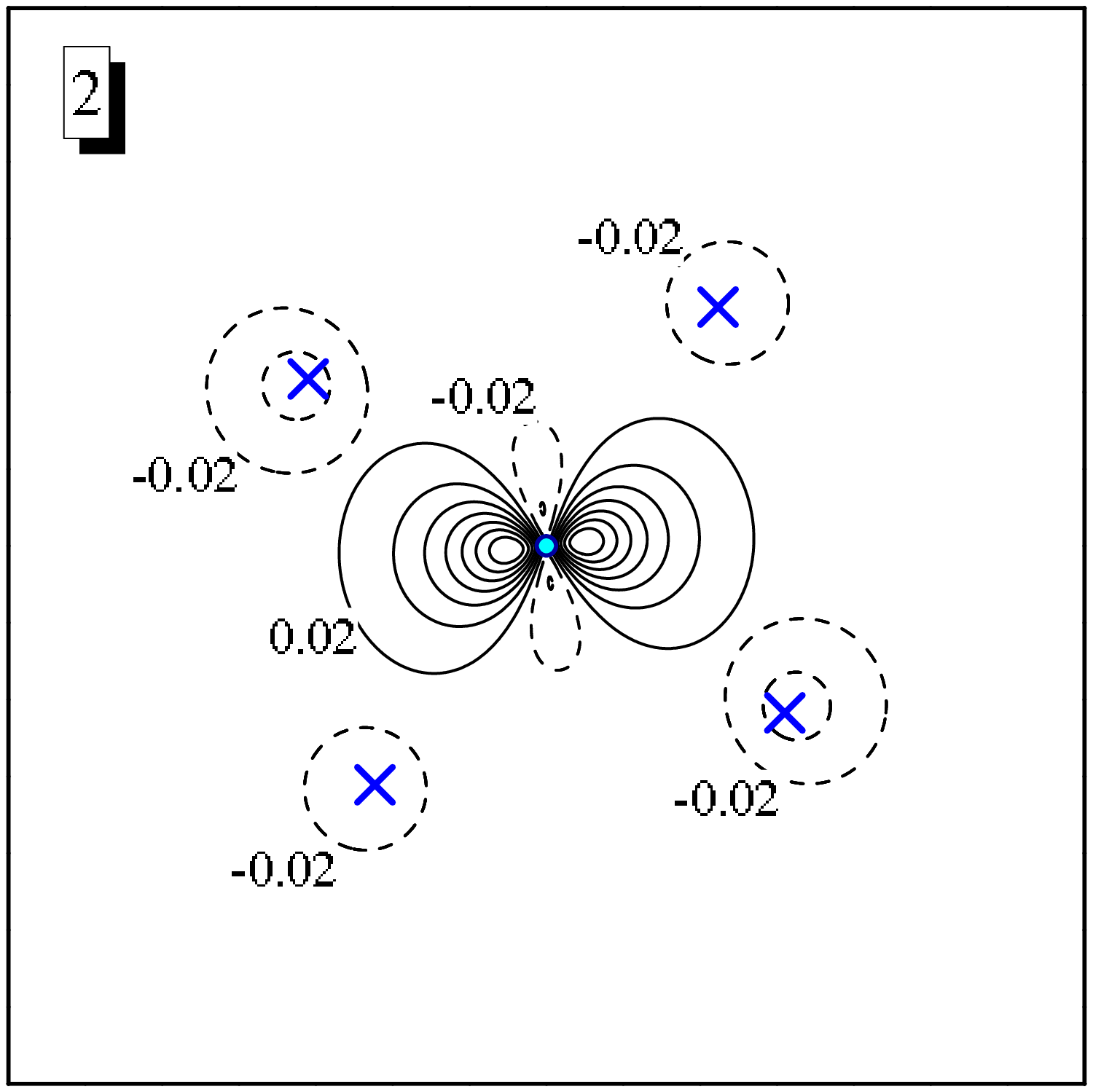}}
\resizebox{5cm}{!}{\includegraphics{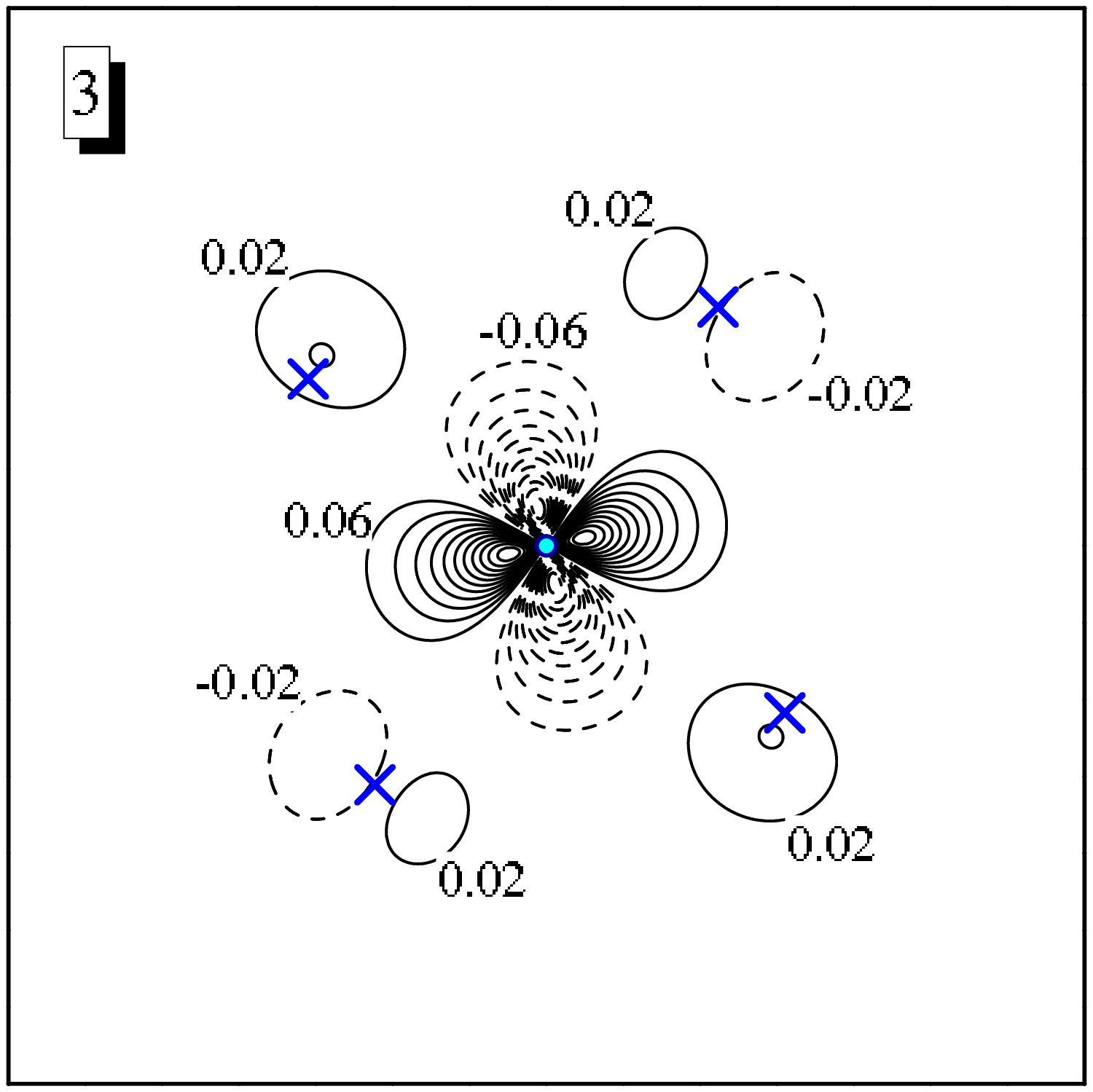}}
\end{center}
\begin{center}
\resizebox{5cm}{!}{\includegraphics{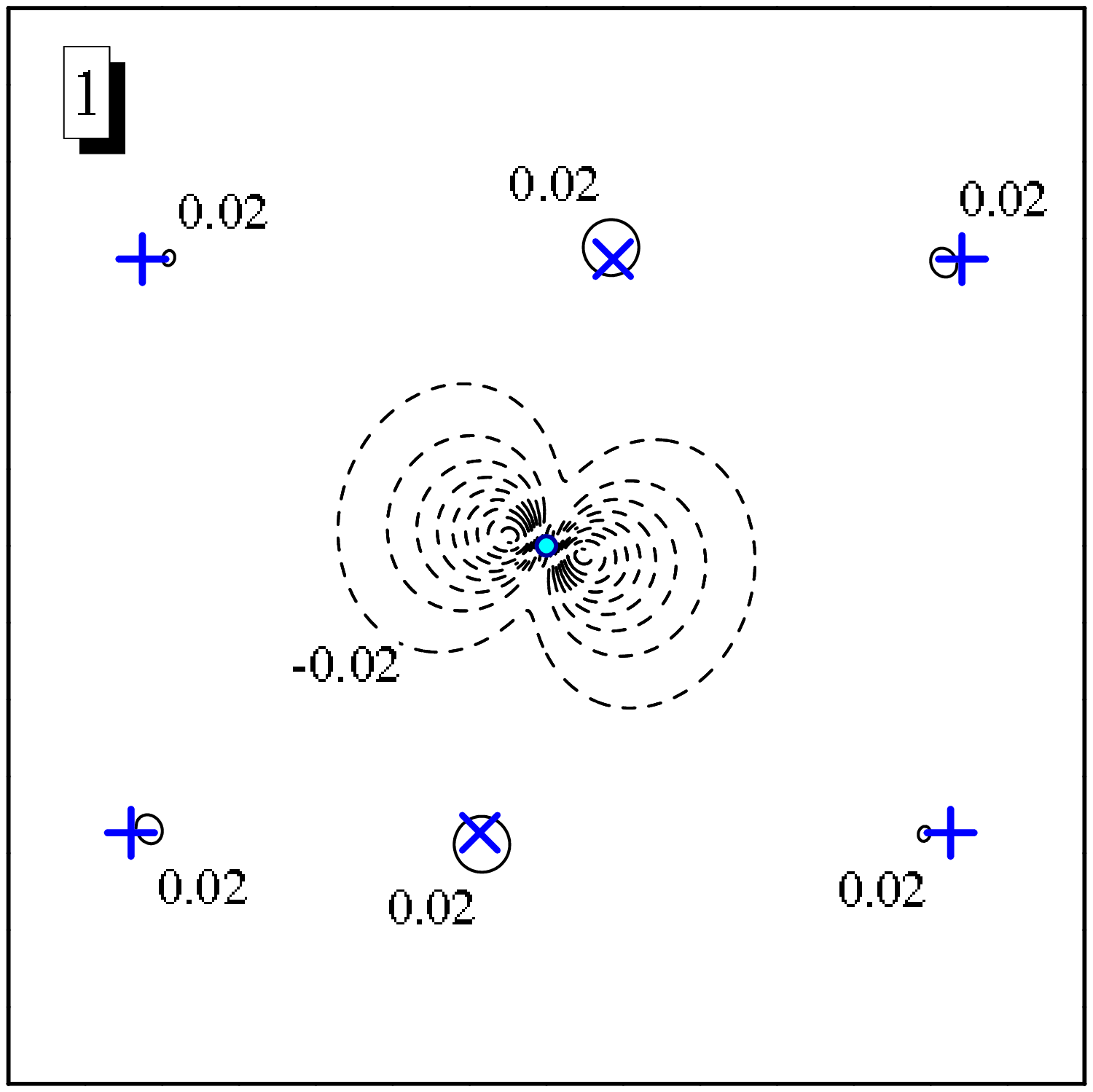}}
\resizebox{5cm}{!}{\includegraphics{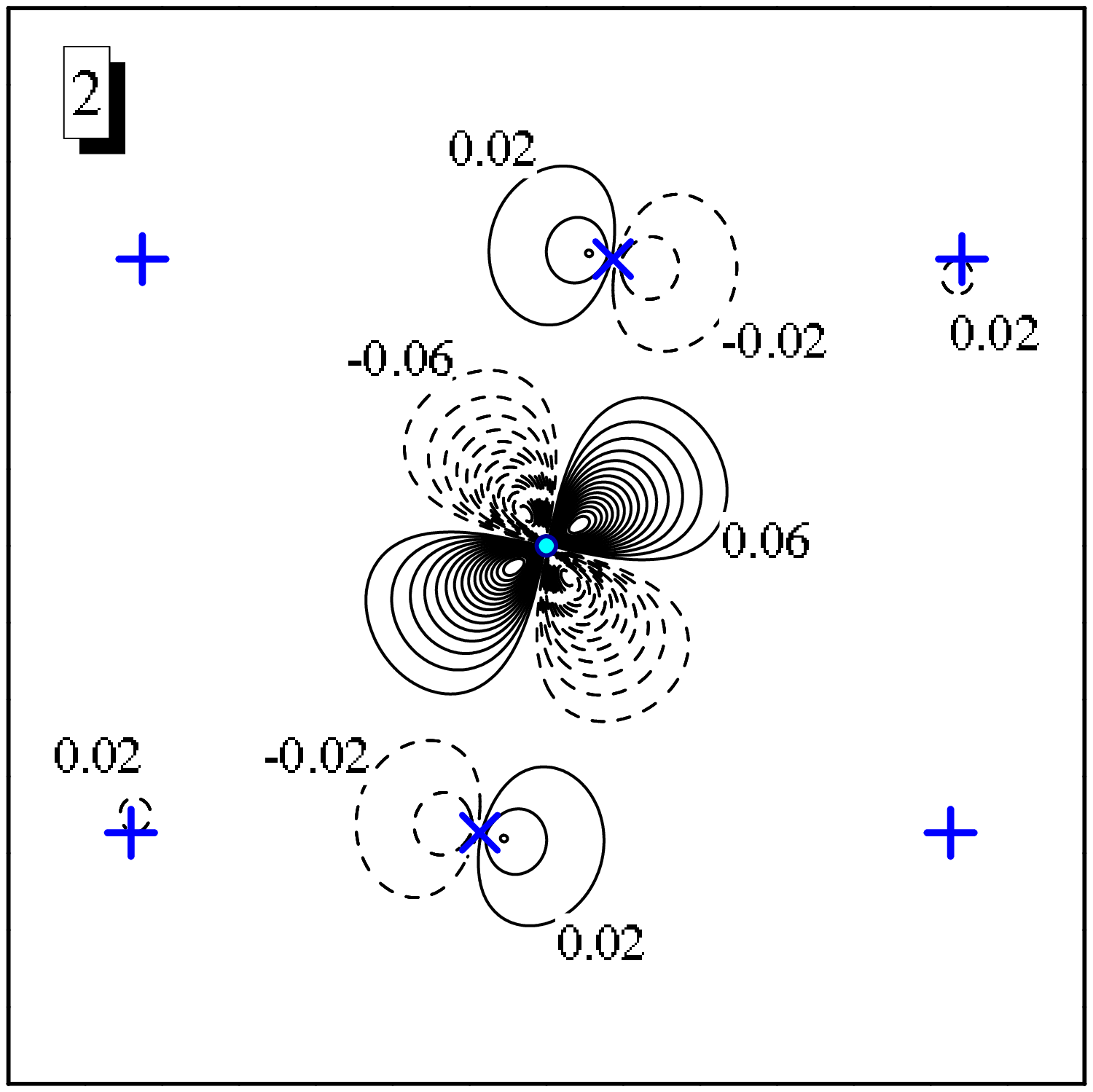}}
\resizebox{5cm}{!}{\includegraphics{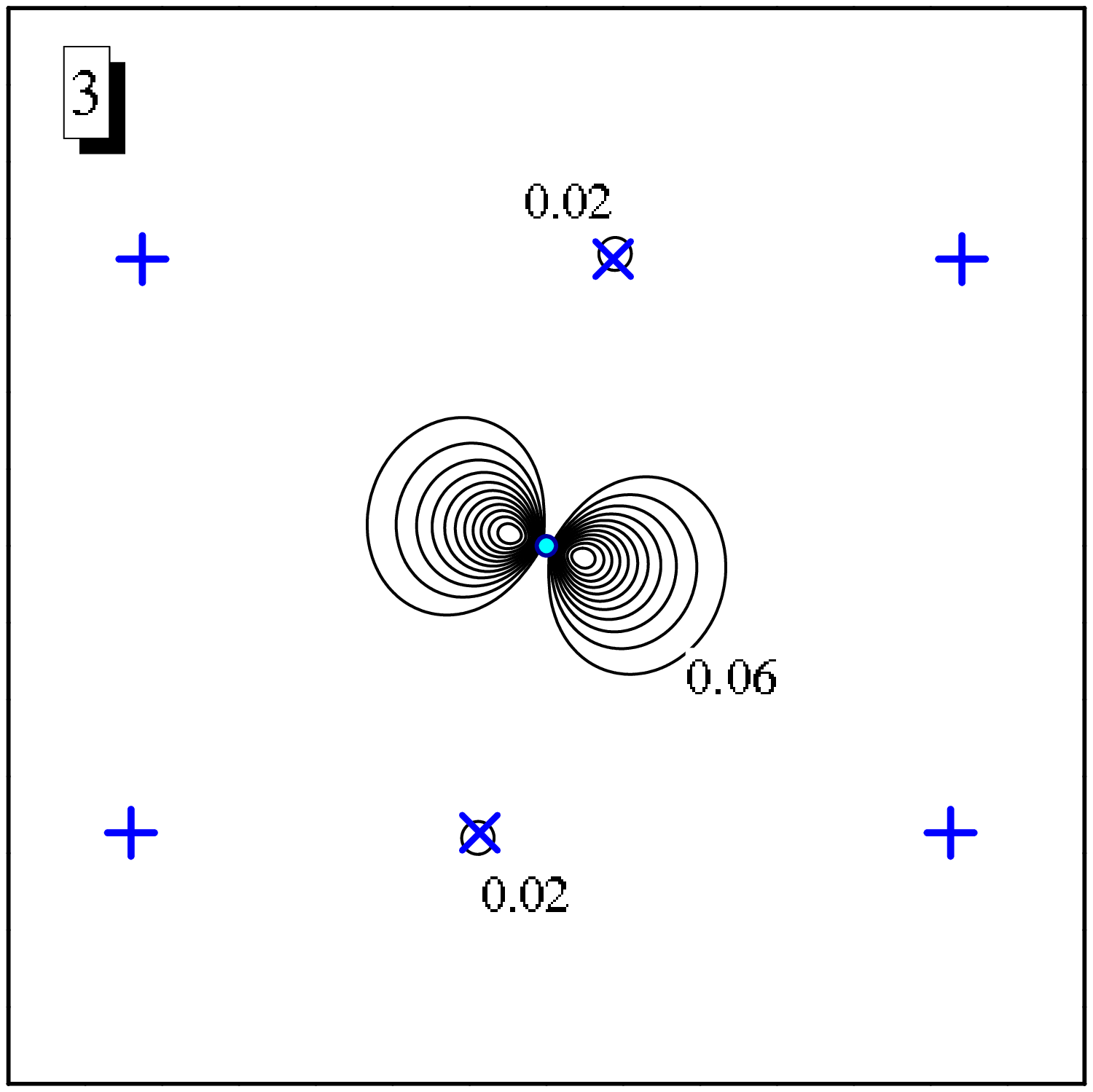}}
\end{center}
\begin{center}
\resizebox{5cm}{!}{\includegraphics{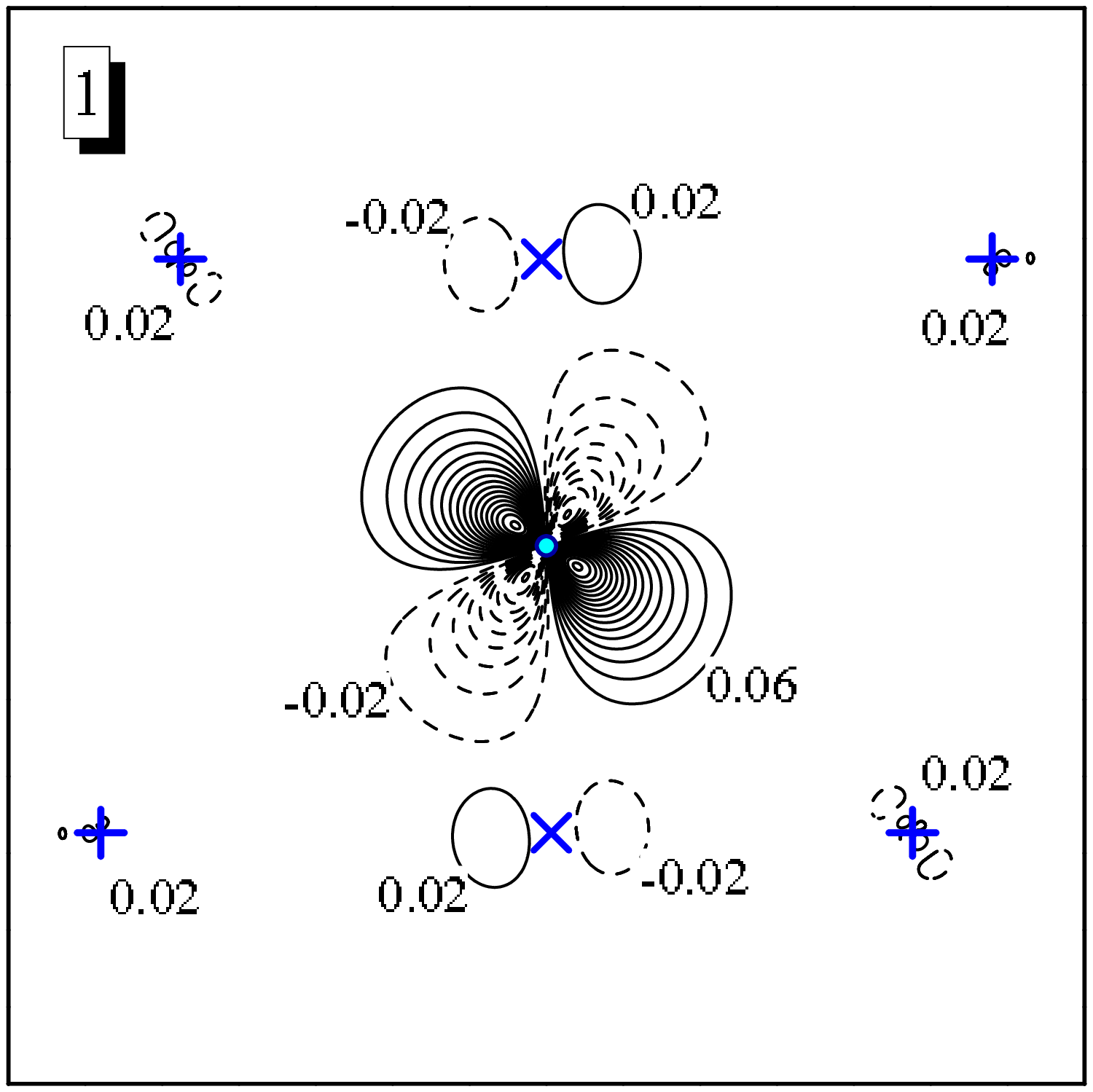}}
\resizebox{5cm}{!}{\includegraphics{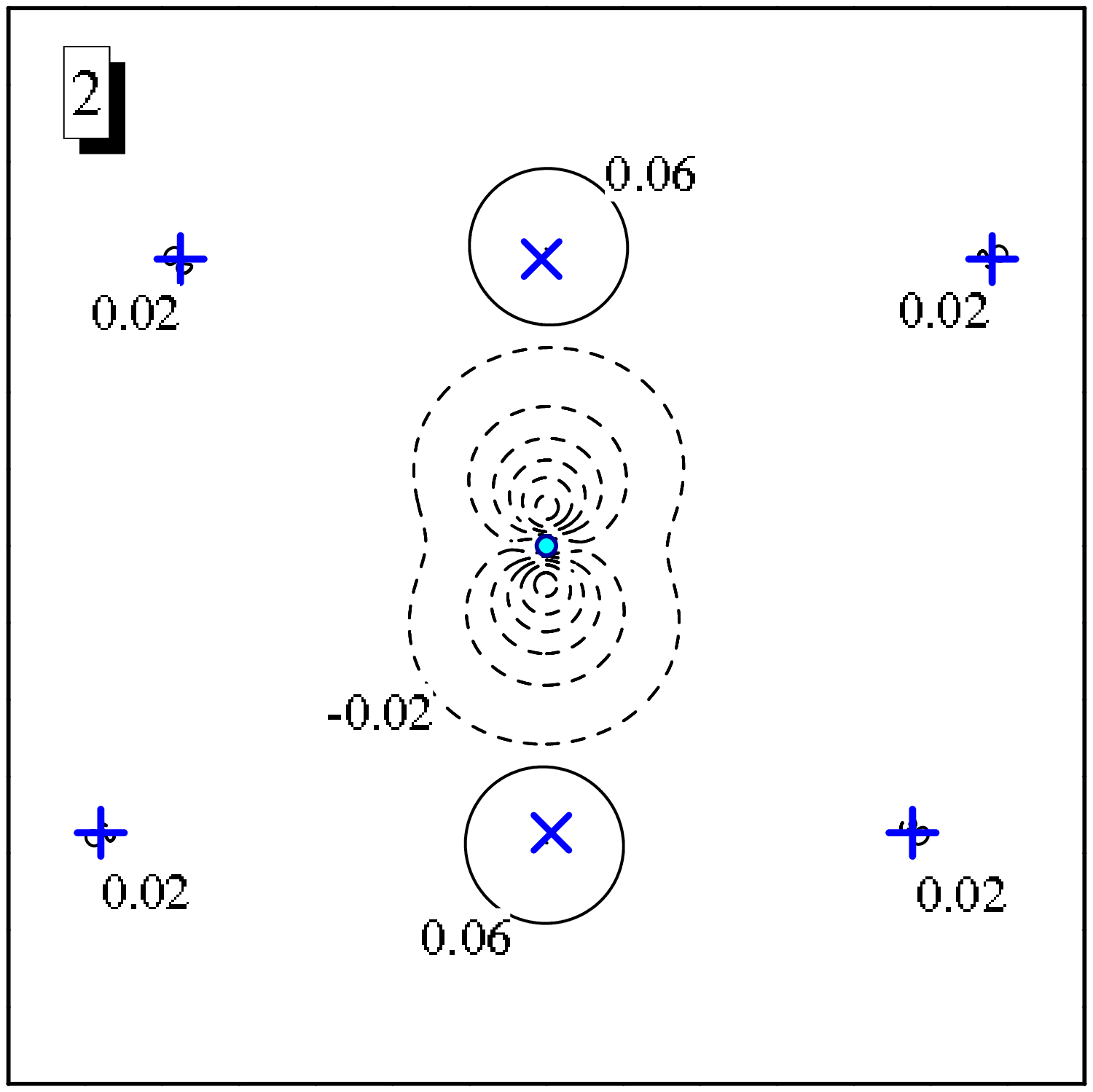}}
\resizebox{5cm}{!}{\includegraphics{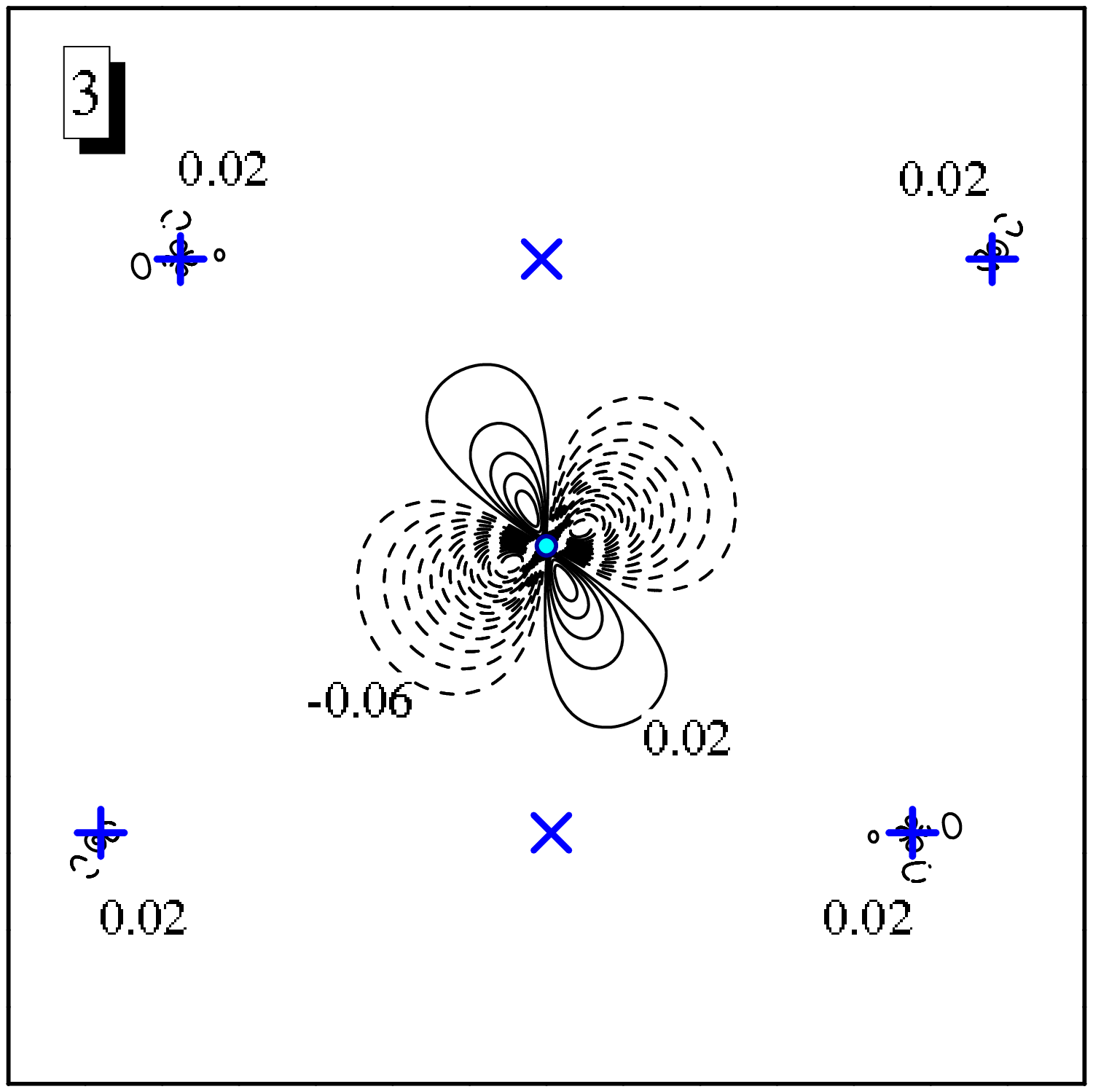}}
\end{center}
\caption{\label{fig.LaTiO3WF}
(Color online)
Contour plot of the Wannier functions in the case of LaTiO$_3$ in three
orthorhombic planes: ${\bf ab}$ (top), ${\bf ac}$ (middle), and
${\bf bc}$ (bottom).
The solid and dashed line correspond to the positive and negative values of the Wannier functions.
The projections of different atoms on the planes are denoted by the following symbols:
$+$ (La), $\circ$ (Ti), and $\times$ (O).
Around each site, the Wannier function increases/decreses with the step $0.04$
from the values indicated on the graph.}
\end{figure}

\begin{figure}[h!]
\begin{center}
\resizebox{10cm}{!}{\includegraphics{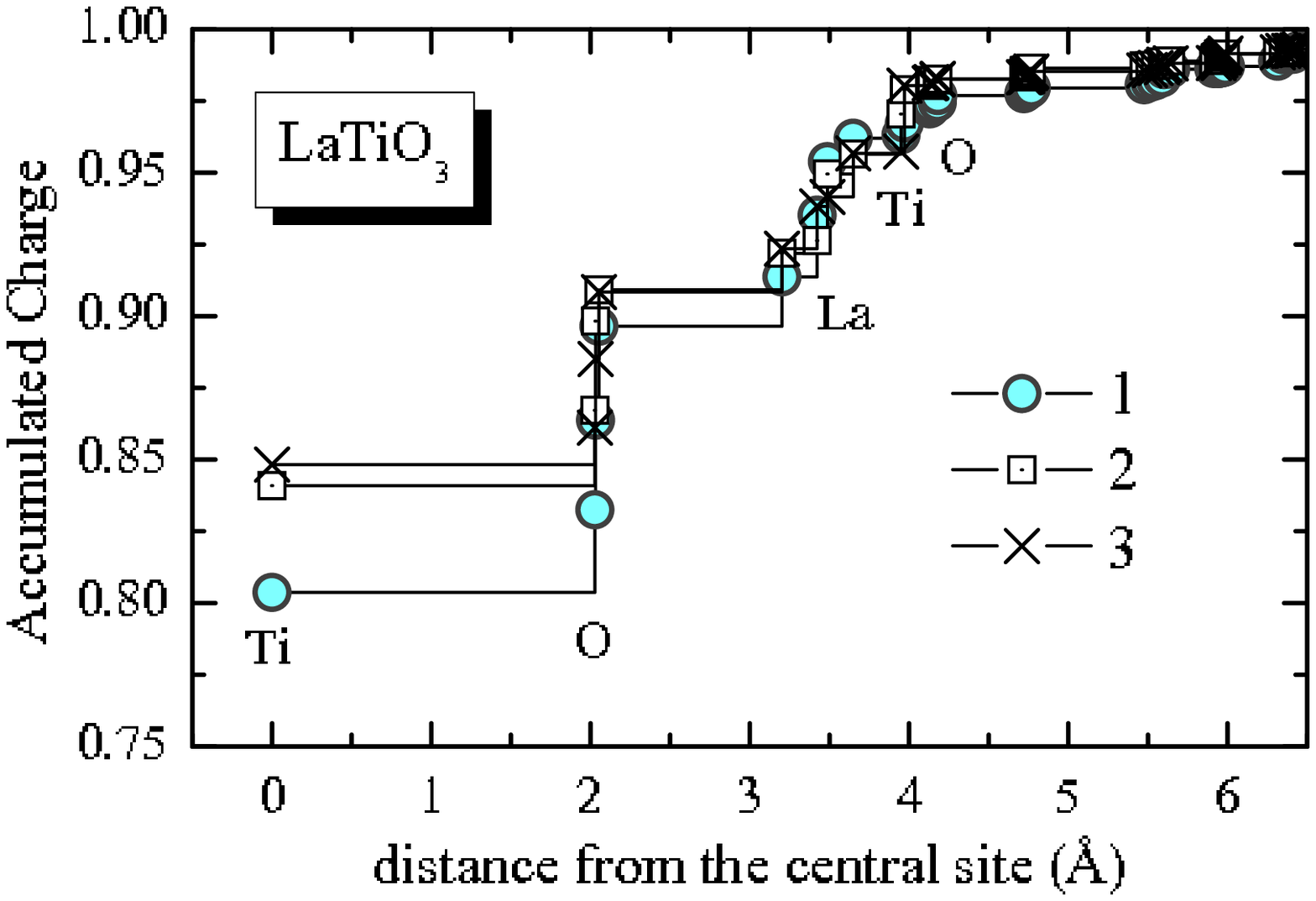}}
\end{center}
\caption{\label{fig.LaTiO3WFextension}
Spacial extension of the Wannier functions in the case of LaTiO$_3$: the electronic charge
accumulated around the central Ti site after adding every new sphere of
neighboring atomic sites.}
\end{figure}

  The (characteristic) example of
such Wannier functions constructed for LaTiO$_3$ is shown in Fig.~\ref{fig.LaTiO3WF},
and their extension in the real space is illustrated in Fig.~\ref{fig.LaTiO3WFextension}.
The functions are well localized: about 80-85\% of their total weight
is concentrated at the central
Ti site, 5-9 \% belong to
neighboring oxygen sites, and about 10 \% is distributed over
La, Ti, and O sites located in next coordination spheres.
Another measure of localization of the Wannier functions is
the expectation value of the square of the position operator:
$\langle {\bf r}^2 \rangle_\alpha$$=$$\langle \tilde{W}_{\bf R}^\alpha |
({\bf r}$$-$${\bf R})^2 | \tilde{W}_{\bf R}^\alpha \rangle$,\cite{MarzariVanderbilt}
which
yields
$\langle {\bf r}^2 \rangle_\alpha$$=$ $2.68$, $2.36$, and
$2.37$~\AA$^2$ for $\alpha$$=$ 1, 2, and 3, respectively.
The Wannier functions for LaTiO$_3$ are less localized in comparison
with the more distorted YTiO$_3$, where
$\langle {\bf r}^2 \rangle$
is of the order of $1.90$-$2.28$~\AA$^2$.\cite{condmat05}
However, this is to be expected.

  The parameters $\{ h_{{\bf R}{\bf R}'}^{\alpha \beta} \}$
include all kinds of hybridization (or covalent mixing)
effects between transition-metal $t_{2g}$ and other
atomic orbitals. However, there are other effects,
which are not yet included in
$\{ h_{{\bf R}{\bf R}'}^{\alpha \beta} \}$.
They come from the nonsphericity (n-s)
of the Madelung potential for the electron-ion interactions, and
contribute to the CF splitting.
The proper correction to $\{ h_{{\bf R}{\bf R}'}^{\alpha \beta} \}$ can be
easily calculated
in the basis of Wannier functions
$\{ \tilde{W}_{\bf R}^\alpha \}$:
\begin{equation}
\Delta^{\rm n-s} h_{\bf RR}^{\alpha \beta} = \sum_{{\bf R}'\neq{\bf R}}
\langle \tilde{W}_{\bf R}^{\alpha} |
\frac{- Z^{*}_{{\bf R}'} e^2}{|{\bf R}+{\bf r}-{\bf R}'|}
| \tilde{W}_{\bf R}^{\beta} \rangle,
\label{eqn:CFEI}
\end{equation}
where $Z^{*}_{{\bf R}'}$ is the total charge associated with the
site ${\bf R}'$
(namely, the nuclear charge minus the screening electronic charge encircled by the
atomic sphere),
and ${\bf r}$ is the position of the electron in the sphere ${\bf R}$.

  The main idea behind this treatment of the CF splitting is based on
certain hierarchy of interactions in solids. It implies that the strongest
interaction, which leads to the energetic separation of the $t_{2g}$ band from other
bands (Fig.~\ref{fig.DOSsummary}), is the covalent mixing. For example, in many
transition-metal oxides this interaction is mainly responsible for the famous splitting
between the transition-metal $t_{2g}$ and $e_g$ states.\cite{Kanamori1}
The nonsphericity of the Madelung potential is considerably weaker than this splitting.
However, it can be comparable with the effects
of covalent mixing
in the narrow
$t_{2g}$ band. Therefore, the basic idea is to treat this nonsphericity as a
pseudo-perturbation,\cite{LMTO} and calculate the matrix elements of the
Madelung potential
in the basis of Wannier functions constructed
for spherically averaged ASA potential.

  The same strategy can be used for the spin-orbit (s-o) interaction,
which yields the following correction to the kinetic part of the model Hamiltonian:
$$
\Delta^{\rm s-o} h_{\bf RR}^{\alpha \beta} =
\langle \tilde{W}_{\bf R}^{\alpha} | \frac{\hbar}{4m^2c^2} ({\bm \nabla}V \times {\bf p})\cdot {\bm \sigma}|
\tilde{W}_{\bf R}^{\alpha} \rangle
$$
($V$ being the LDA potential and
${\bm \sigma}$ being the vector of Pauli matrices).

  One of the most controversial issues, which is actively discussed
in the literature, is the values and the directions of the CF splitting in
the distorted $t_{2g}$ perovskite oxides. Therefore, we would like to discuss
this problem more in details. Basically, there are two sources
of discrepancies, which largely affect the conclusions about the orbital
ordering and the magnetic ground state:\\
1. (The origin of the CF splitting: the nonsphericity of the Madelung potential versus
the hybridization effects) The importance of nonsphericity of the Madelung potential has
been emphasized by several authors. The original idea is due to Mochizuki and Imada,
who considered the $t_{2g}$-level splitting in the Ti-compounds
associated with the displacements of
Y and La sites.\cite{MochizukiImada} It was paraphrased by Cwik~\textit{et al.},\cite{Cwik}
who suggested the main effect comes from the
deformation of the TiO$_6$ octahedra. A more general
picture has been considered by Schmitz~\textit{et al.},\cite{Schmitz}
who summed up all contributions in the Madelung potential.
A weak point of all these approaches is the approximate treatment of the
hybridization effects, which largely relied on the model parameters.
Moreover, the result depends crucially on the value of the dielectric constant,
which is treated as an adjustable parameter.
On the other hand, the parameters of the model Hamiltonians extracted from the
first-principles electronic structure calculations
using either downfolding
(Refs.~\onlinecite{PRB04}, \onlinecite{Pavarini1}, and
\onlinecite{Pavarini2}) or Wannier function (Ref.~\onlinecite{Streltsov})
methods
automatically include all effects of the covalent mixing.
In this sense, these are more rigorous techniques.
However, all these calculations were supplemented with the additional
atomic-spheres-approximation and neglected the nonspherical part of the
Madelung potential.
The Madelung term has been also neglected in our previous work (Ref.~\onlinecite{PRB04}).
As we shall see in Secs.~\ref{sec:YTO} and \ref{sec:LTO}, it will
definitely revise several statements
of Ref.~\onlinecite{PRB04}. However, the final conclusion about the magnetic
ground state of YTiO$_3$ and LaTiO$_3$ remains valid.
\\
2. (The nonuniqueness of the Wannier functions)
Different calculations yield different parameters of the
CF splitting. For example, for LaTiO$_3$ different authors reported the
following parameters of the CF splitting (between lowest and highest
$t_{2g}$-levels): 93 meV (Ref.~\onlinecite{PRB04}),
200 meV (Ref.~\onlinecite{Pavarini1}), and 270 meV (Ref.~\onlinecite{Streltsov}).
There is a particularly bad custom to criticize
Ref.~\onlinecite{PRB04},\cite{Schmitz,Streltsov,Haverkort}
which reports the smallest value, even with certain hints at the
accuracy of calculations.\cite{Streltsov}
It is also premature to think that the small CF splitting will
inevitably lead to the orbital liquid scenario,\cite{KhaliullinMaekawa,Streltsov}
because the transfer interactions are also strongly
affected by the lattice distortion, which makes a big difference from
the idealized cubic perovskites.\cite{condmat05}

  First, we would like to consider the second part of problem and argue that
different
values of the CF splitting are most likely related with the different
choice of the Wannier functions, which by no means is unique. This is \textit{not}
a problem of accuracy of calculations.

  In the downfolding method employed in
Refs.~\onlinecite{PRB04}, \onlinecite{Pavarini1}, and \onlinecite{Pavarini2},
all basis states were divided in two groups:
the ``$t_{2g}$'' part $\{ \tilde{\chi}_t \}$,
and the ``rest'' of the basis functions $\{ \tilde{\chi}_r \}$. The effective Hamiltonian
is constructed by eliminating the ``rest'' part.\cite{condmat05} A similar
idea (although formulated in slightly different way) is employed in
projections scheme for the construction of
the Wannier
functions,\cite{Streltsov} where $\{ \tilde{\chi}_t \}$
play a role of trial orbitals. The basic difficulty here is that, in the
distorted perovskites, the set of atomic ``$t_{2g}$'' orbitals cannot be defined
in an unique fashion: since the local symmetry is not cubic, the abbreviations like
``$t_{2g}$'' and ``$e_g$'' will always reflect some bad quantum numbers for the state, which
are mixed by the crystal field and/or the transfer interactions.
In numerical calculations, the set of ``$t_{2g}$'' orbitals is always specified in
some local coordinate frame, and the choice of this frame
appeared to be different in different calculations.
For example, Pavarini~\textit{et~al.} (Refs.~\onlinecite{Pavarini1} and \onlinecite{Pavarini2})
and Streltsov~\textit{et~al.} (Ref.~\onlinecite{Streltsov}) selected their
local coordinate frames from some geometrical considerations. A completely
different strategy was pursued by the present author in
Ref.~\onlinecite{PRB04}, where the atomic ``$t_{2g}$'' orbitals
were determined from the diagonalization of the local density matrix
constructed from the $t_{2g}$ bands and projected onto all five $3d$-orbitals
of the transition-metal sites.

  Then, we are ready to argue that different choice of the local coordinate frame naturally
explains the difference in the parameters of the CF splitting reported by different
authors. For these purposes we consider two different setups in the downfolding scheme,
and consider LaTiO$_3$ as an example.
The first scheme is absolutely identical to that proposed in Ref.~\onlinecite{PRB04}, where
the atomic ``$t_{2g}$'' orbitals have been defined as three most populated orbitals
obtained from the diagonalization of the density matrix for the $t_{2g}$ bands.
In the second scheme, we first construct a more general $40$$\times$$40$
tight-binding Hamiltonian, comprising the Ti($3d$) and La($5d$) states, and reproducing
the behavior of $40$ overlapping Ti($3d$)-La($5d$) bands.
Other orbitals have been eliminated using the downfolding method. Then, we
diagonalize the site-diagonal part of this Hamiltonian and define three
lowest eigenstates at the Ti-site as the atomic ``$t_{2g}$'' orbitals.
After that we eliminate the rest of the Ti($3d$) and La($5d$) states using
the downfolding method and obtain the minimal $12$$\times$$12$ Hamiltonian
for the $t_{2g}$ bands.

\begin{figure}[t!]
\begin{center}
\resizebox{12cm}{!}{\includegraphics{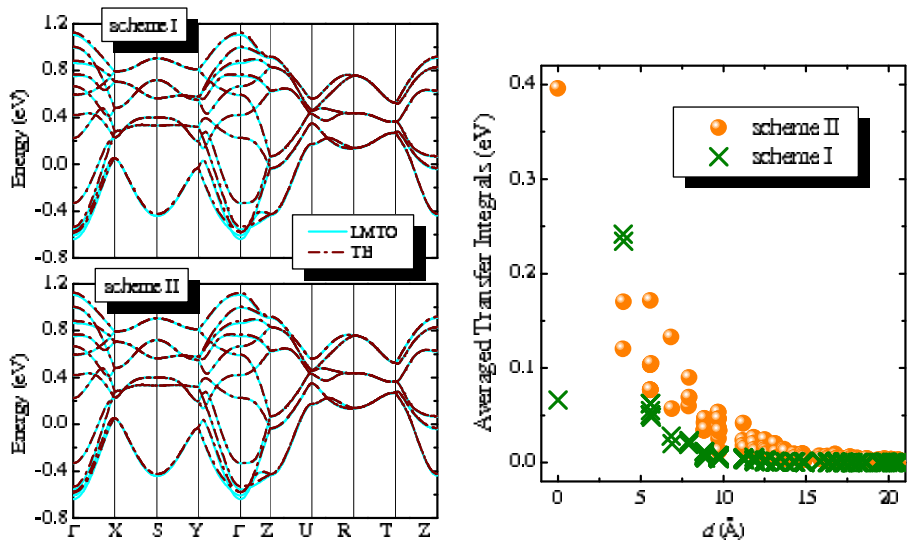}}
\end{center}
\caption{\label{fig.LaTiO3twoschemes}
(Color online) Left panel:
LDA energy bands for LaTiO$_3$ obtained in LMTO calculations and
after the tight-binding (TB) parametrization using
two different schemes of downfolding.
In the scheme I, the local coordinate frame has been obtained
from the diagonalization of
the local density matrix. In the scheme II, the local coordinate frame
has been obtained from the diagonalization of the site-diagonal part
of a more general $40$$\times$$40$ Ti($3d$)-La($5d$) tight-binding Hamiltonian.
Notations of the high-symmetry points of the Brillouin zone are taken from
Ref.~\protect\onlinecite{BradlayCracknell}.
Right panel: distance-dependence of averaged parameters of the kinetic energy
$\bar{h}_{{\bf RR}'}(d)$$=$$\left( \sum_{\alpha \beta} h^{\alpha \beta}_{{\bf RR}'}
h^{\beta \alpha}_{{\bf R}'{\bf R}} \right)^{1/2}$, where
$d$ is the distance between transition-metal sites ${\bf R}$ and ${\bf R}'$.
The data for
$d$$=$$0$ correspond to the crystal-field splitting of the covalent type, and
$d$$\sim$$4$~\AA~is the distance between nearest transition-metal sites.
Two schemes generate very similar electronic structure in the reciprocal
space, which is well consistent
with results of original LMTO calculations. However, the transfer integrals and
the crystal-field splitting are different. Scheme II yields larger
crystal-field splitting, but less localized transfer integrals.}
\end{figure}

  Both downfolding schemes are nearly perfect
and well reproduce the behavior of
the $t_{2g}$ bands in the reciprocal space (Fig.~\ref{fig.LaTiO3twoschemes}).
However, they yield very different parameters after the Fourier transformation
to the real space.
For example, the splitting of atomic $t_{2g}$ levels (in meV) obtained in the schemes
I and II is ($-$$49$,$4$,$44$) and ($-$$320$,$123$,$197$), respectively.
Moreover, the eigenvectors corresponding to the lowest ``$t_{2g}$'' levels
appear to be also different.
In the orthorhombic coordinate frame, specified by the vectors ${\bf a}$, ${\bf b}$, and ${\bf c}$,
the eigenvectors have the following form (referred to the site 1):
$|\Psi_{\rm I} \rangle$$=$$
0.32|{\rm xy}\rangle$$-$$0.73|{\rm yz}\rangle$$-$$0.10|{\rm z^2}\rangle$$-$$0.18|{\rm zx}\rangle
$$+$$0.57|{\rm x^2}$$-$${\rm y^2}\rangle$
and
$|\Psi_{\rm II} \rangle$$=$$
0.31|{\rm xy}\rangle$$-$$0.20|{\rm yz}\rangle$$+$$0.15|{\rm z^2}\rangle$$+$$0.55|{\rm zx}\rangle
$$+$$0.73|{\rm x^2}$$-$${\rm y^2}\rangle$,
for the scheme I and II, respectively. Thus, the small value of the CF splitting reported in
Ref.~\onlinecite{PRB04} is related with the particular choice of the Wannier functions
(or the parameters of the downfolding scheme).
Had we changed our definition of the local coordinate frame, our conclusion would have been
also different, and we could easily obtain the CF splitting of the order of $500$ meV
(and even larger).

  Then, it is of course right to ask which scheme is better? In principle, the physics should
not depend on the choice of the Wannier functions, and as long as we are dealing only with
the kinetic-energy part of the model Hamiltonian, both schemes are totally equivalent as they equally
well reproduce the behavior of the $t_{2g}$ bands. However, what we want to do next is to
combine this kinetic-energy part with the Coulomb interactions,
and \textit{to use only the site-diagonal part of these interactions}.
This is of course an approximation, and
in order to justify it one should guarantee that the
Wannier functions, which are used as the basis for the
matrix elements of the Coulomb interactions,
were sufficiently well localized in the real space, so that all inter-site
matrix elements could be neglected.

  The degree of localization of the Wannier orbitals is
related with the spread of transfer integrals. Loosely speaking, in order to contribute to the
transfer integral between $N$-$th$ neighbors, the Wannier function should have a finite weight
at this neighbor. The distance-dependence of transfer integrals
calculated in two different scheme
is shown in Fig.~\ref{fig.LaTiO3twoschemes}.
One can clearly see that the transfer integrals obtained in the scheme I are indeed well
localized and basically restricted by the nearest neighbors.
The transfer integrals obtained in the scheme II are less localized and spread far
beyond the nearest neighbors. Therefore, the Wannier functions,
corresponding to the scheme I, should be more
localized. This is not surprising, because the scheme I guarantee that the
density matrix (or the integrated density of states) at the transition-metal site is already
well described by the central parts (or ``heads'') of the Wannier functions, given by
the atomic orbitals $\{ \tilde{\chi}_t \}$.\cite{condmat05}
Therefore, the tails of the Wannier functions, coming from the neighboring sites, should be small
and cancel each other. In the scheme II, the local density matrix is composed by both
``heads'' of the Wannier functions as well as their tails coming from the
neighboring sites. Intuitively, this means that the tail part of the Wannier functions
is larger for the scheme II and these functions are less localized.

\begin{figure}[t!]
\begin{center}
\resizebox{10cm}{!}{\includegraphics{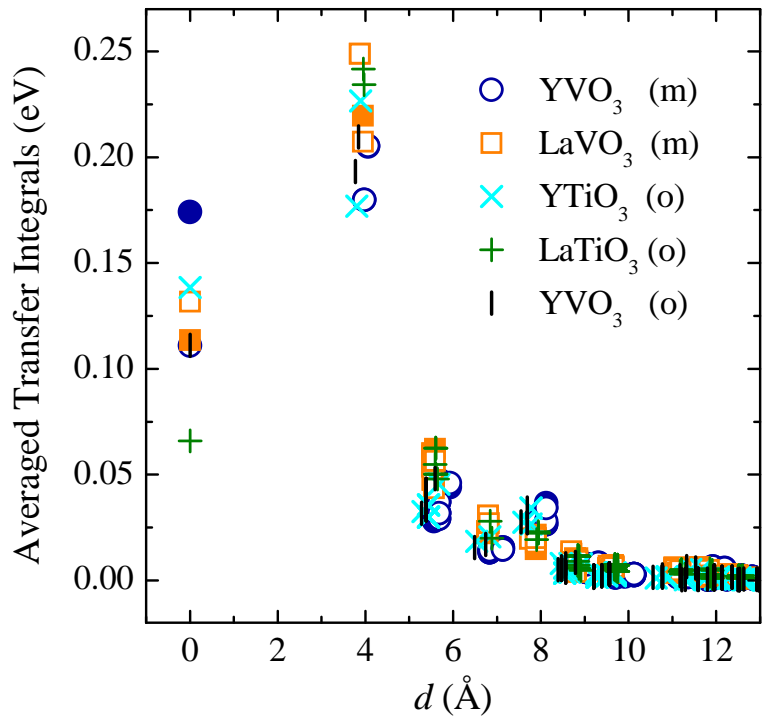}}
\end{center}
\caption{\label{fig.avrtransfer} (Color online) Behavior of
averaged parameters of the kinetic energy
for various perovskite compounds.
In the orthorhombic (o) structure, all four sublattices of the transition-metal
sites are equivalent as they can be transformed
to each other by the symmetry operations of the $D_{2h}^{16}$ group.
In the monoclinic (m) structure, the transfer interactions corresponding to two
different sublattices are shown by closed and open symbols.
For other notations, see Fig.~\protect\ref{fig.LaTiO3twoschemes}.}
\end{figure}

  Thus, we believe that the scheme I is more suitable for the purposes of our work
and we apply it to all perovskite compounds. The transfer integrals, obtained in such a way,
are indeed well localized and restricted by the nearest neighbors (Fig.~\ref{fig.avrtransfer}).

\begin{figure}[t!]
\begin{center}
\resizebox{15cm}{!}{\includegraphics{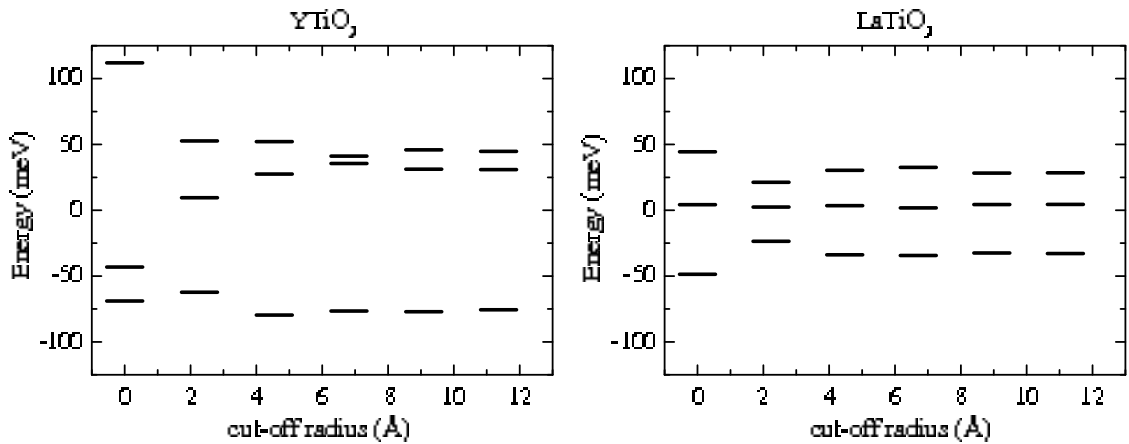}}
\end{center}
\caption{\label{fig.CFconvergence}
Convergence of the crystal-field splitting as the function of
cut-off radius in Eq.~(\protect\ref{eqn:CFEI}).}
\end{figure}

  The next important contribution to the CF splitting comes from the nonsphericity
of the Madelung potential. This effect is \textit{comparable} with the CF splitting
of the covalent type. We have also found that there are several different contributions
to the CF splitting,
which tend to cancel each other. For example, the CF splitting of the
covalent type in YTiO$_3$ and LaTiO$_3$ is largely compensated by the nonsphericity
of the Madelung potential coming from the
region encircling the the
neighboring oxygen sites
and corresponding to the cut-off radius
$d$$\sim$$2$~\AA~in Fig.~\ref{fig.CFconvergence}.\cite{comment.8}
The next important contribution comes from the Y/La and Ti sites, located in the
next coordination spheres ($d$$\sim$$4$~\AA). In some compounds, even longer-range
interactions spreading up to $d$$\sim$$10$~\AA~can have a relatively large
weight in the sum (\ref{eqn:CFEI}).

\begin{figure}[t!]
\begin{center}
\resizebox{10cm}{!}{\includegraphics{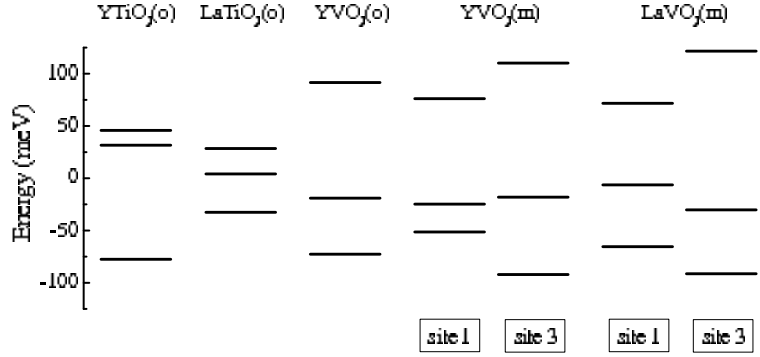}}
\end{center}
\caption{\label{fig.CFsummary}
$t_{2g}$-level splitting in various compounds. The notations ``site 1''
and ``site 3'' correspond to two nonequivalent transition-metal sites in
the monoclinic structure (shown in Fig.~\protect\ref{fig.structure}).}
\end{figure}

  The final scheme of the ``$t_{2g}$''-level splitting, which takes into
account all these effects, is summarized in Fig.~\ref{fig.CFsummary}.
The splitting is not particularly large. Nevertheless, as we shall see below,
it appears to be sufficient in order to explain the experimentally observed
orbital ordering and the magnetic ground state
\textit{in all considered compounds, except LaTiO$_3$}.
The Madelung potential changes not only the magnitude of the crystal-field splitting,
but also the \textit{type} of the atomic ``$t_{2g}$''-orbitals,
which are split off by the distortion (Table~\ref{tab:CForbitals}).
\begin{table}[h!]
\caption{Lowest (for YTiO$_3$ and LaTiO$_3$) and highest (for YVO$_3$ and LaVO$_3$)
atomic orbital obtained from the diagonalization of
the site-diagonal part of the model Hamiltonian
in the downfolding method (denoted as the ``covalent part'') and and after including
the
nonsphericity of the Madelung potential (denoted as ``total'').
The symbols `o' and `m' stand for orthorhombic and monoclinic phases, respectively.
The order of the basis orbitals is ${\rm xy}$, ${\rm yz}$, ${\rm z^2}$, ${\rm zx}$,
and ${\rm x^2}$$-$${\rm y^2}$, in the orthorhombic coordinate frame.
The positions of the transition-metal sites are shown in Fig.~\protect\ref{fig.structure}.}
\label{tab:CForbitals}
\begin{ruledtabular}
\begin{tabular}{ccccc}
 compound  & phase & site & covalent part & total           \\
\hline
 YTiO$_3$  &  o    & 1    & ($\phantom{-}$$0.32$,$\phantom{-}$$0.78$,$\phantom{-}$$0.36$,$-$$0.33$,$-$$0.22$)
                          & ($-$$0.13$,$\phantom{-}$$0.45$,$\phantom{-}$$0.38$,$-$$0.61$,$\phantom{-}$$0.50$)           \\
 LaTiO$_3$ &  o    & 1    & ($-$$0.32$,$\phantom{-}$$0.73$,$\phantom{-}$$0.10$,$\phantom{-}$$0.18$,$-$$0.57$)
                          & ($-$$0.06$,$\phantom{-}$$0.85$,$\phantom{-}$$0.15$,$\phantom{-}$$0.34$,$\phantom{-}$$0.37$) \\
 YVO$_3$   &  o    & 1    & ($\phantom{-}$$0.04$,$\phantom{-}$$0.84$,$\phantom{-}$$0.19$,$-$$0.22$,$\phantom{-}$$0.47$)
                          & ($\phantom{-}$$0.29$,$\phantom{-}$$0.90$,$\phantom{-}$$0.25$,$-$$0.22$,$-$$0.06$)           \\
 YVO$_3$   &  m    & 1    & ($-$$0.04$,$\phantom{-}$$0.47$,$\phantom{-}$$0.19$,$-$$0.61$,$\phantom{-}$$0.60$)
                          & ($\phantom{-}$$0.12$,$\phantom{-}$$0.54$,$\phantom{-}$$0.18$,$-$$0.74$,$\phantom{-}$$0.33$) \\
           &       & 3    & ($-$$0.01$,$\phantom{-}$$0.04$,$-$$0.09$,$\phantom{-}$$0.79$,$\phantom{-}$$0.60$)
                          & ($\phantom{-}$$0.28$,$\phantom{-}$$0.24$,$-$$0.28$,$\phantom{-}$$0.89$,$-$$0.05$)           \\
 LaVO$_3$  &  m    & 1    & ($\phantom{-}$$0.04$,$-$$0.31$,$\phantom{-}$$0.11$,$\phantom{-}$$0.80$,$-$$0.49$)
                          & ($\phantom{-}$$0.16$,$\phantom{-}$$0.09$,$\phantom{-}$$0.01$,$\phantom{-}$$0.98$,$-$$0.04$) \\
           &       & 3    & ($-$$0.01$,$\phantom{-}$$0.37$,$\phantom{-}$$0.10$,$\phantom{-}$$0.75$,$\phantom{-}$$0.54$)
                          & ($-$$0.09$,$\phantom{-}$$0.55$,$\phantom{-}$$0.14$,$\phantom{-}$$0.78$,$\phantom{-}$$0.24$) \\
\end{tabular}
\end{ruledtabular}
\end{table}
The type of these orbitals is extremely important in the analysis of the
magnetic properties.

  Finally, both magnitude and form of the CF splitting can be different
for the nonequivalent transition-metal sites in the
monoclinic structure. Therefore, one can generally expect rather different magnetic behavior
in two nonequivalent ${\bf ab}$-planes of the monoclinic phase.

\subsection{\label{sec:Coulomb}Effective Coulomb Interactions}

  The effective Coulomb interaction in
the $t_{2g}$ band is defined as the energy cost for moving
an electron between two Wannier orbitals, $\tilde{W}_{\bf R}^\alpha$ and
$\tilde{W}_{{\bf R}'}^\beta$,
which have been initially populated by $\mathfrak{n}_{{\bf R}\alpha}$ and
$\mathfrak{n}_{{\bf R}'\beta}$ electrons:
$$
U_{\alpha \alpha \beta \beta} = E\left[ \mathfrak{n}_{{\bf R}\alpha} + 1,
\mathfrak{n}_{{\bf R}'\beta} - 1 \right] -
E\left[ \mathfrak{n}_{{\bf R}\alpha} , \mathfrak{n}_{{\bf R}'\beta} \right].
$$
More generally, one can consider the electron transfer from any
linear combination of the Wannier orbitals at the site ${\bf R}'$ to
any linear combination at the site ${\bf R}$. This define the full
matrix of screened Coulomb interactions $\hat{U}$$=$$\| U_{\alpha \beta \gamma \delta}\|$.
It can be calculated under certain approximations,
which have been discussed in details
in Ref.~\onlinecite{condmat05}.
The method consists of two parts.\\
1. First, we perform the standard constraint-LDA (c-LDA) calculations
and artificially switch off
all matrix elements
of hybridization involving the atomic $3d$-states.\cite{cLDA}
This part takes into
account the screening of Coulomb interactions
caused by the relaxation of the $3d$-atomic basis functions and the
redistribution of the rest of the charge density.
Typical values of
on-site Coulomb interactions ($u$)
obtained in this approach for the Ti- and V-based perovskites vary between $8.5$ and $9.3$ eV.
Using a similar approach, one can calculate the intraatomic exchange coupling constant ($j$),
which is about $0.9$ eV for all considered compounds.
Finally, from the obtained parameters $u$ and $j$, one can restore the full
$5$$\times$$5$$\times$$5$$\times$$5$ matrix
$\hat{u}$$\equiv$$\| u_{\alpha \beta \gamma \delta} \|$ of screened Coulomb interactions
between atomic $3d$-orbitals with the same spin, as it is typically done in the
LDA$+$$U$ method.\cite{PRB94}\\
2. Then, we switch on the hybridization and evaluate the screening caused by the
change of this hybridization between the
atomic $3d$-orbitals and the rest of the basis states
in the random-phase approximation (RPA):
\begin{equation}
\hat{U} = \left[ 1 - \hat{u}\hat{P}(0) \right]^{-1}\hat{u}.
\label{eqn:Dyson}
\end{equation}
This scheme implies that different channels of screening can be included
consecutively. Namely, the $\hat{u}$-matrix derived from c-LDA is used as the
bare Coulomb interaction in the Dyson equation (\ref{eqn:Dyson}), and the
$5$$\times$$5$$\times$$5$$\times$$5$
polarization matrix
$\hat{P}$$\equiv$$\| P_{\alpha \beta \gamma \delta } \|$
describes solely the effects of hybridization
of the $3d$-states with O($2p$), and either Y($4d$) or La($5d$) states, which lead to
the formation of the distinct oxygen-$2p$, transition-metal $t_{2g}$, and a hybrid $e_g$ band
in Fig.~\ref{fig.DOSsummary}. The matrix elements
$\hat{P}$ are given by
\begin{equation}
P_{\alpha \beta \gamma \delta}(\omega) = \sum_{n {\bf k}} \sum_{n' {\bf k}'}
\frac{(\mathfrak{n}_{n {\bf k}}-\mathfrak{n}_{n' {\bf k}'})
d^\dagger_{\alpha n' {\bf k}'} d_{\beta {n \bf k}}
d^\dagger_{\gamma n {\bf k}} d_{\delta n' {\bf k}'}}
{\omega - \varepsilon_{n' {\bf k}'} + \varepsilon_{n {\bf k}} +
i\delta (\mathfrak{n}_{n {\bf k}}-\mathfrak{n}_{n' {\bf k}'})},
\label{eqn:Polarization}
\end{equation}
where $\{ \varepsilon_{n {\bf k}} \}$ and $\{ \mathfrak{n}_{n {\bf k}} \}$
are LDA eigenvalues and occupation numbers
for the band $n$ and momentum ${\bf k}$ in the first
Brillouin zone (the spin index is already included in the definition of $n$) and
$d_{\gamma {n {\bf k}}}$$=$$\langle \tilde{\chi}_\gamma | \psi_{n {\bf k}} \rangle$
is the projection of LDA eigenstate $\psi_{n {\bf k}}$ onto the
atomic $3d$ orbital $\tilde{\chi}_\gamma$.

  Hence, we obtain a
$5$$\times$$5$$\times$$5$$\times$$5$ matrix of screened Coulomb
interactions in the basis of all
five $3d$ orbitals. This matrix is then transformed into the local
coordinate frame spanned by three ``$t_{2g}$'' orbitals with the same spin.
The obtained
$3$$\times$$3$$\times$$3$$\times$$3$ matrix is expanded in the spin subspace
using the orthogonality condition
between Wannier orbitals with different spins. This yields a
$6$$\times$$6$$\times$$6$$\times$$6$ matrix
$\hat{U}$, which is used
in the actual calculations.

  Only for explanatory purposes, we fit $\hat{U}$ in terms of
two Kanamori parameters: the intra-orbital Coulomb interaction ${\cal U}$
and the exchange interaction ${\cal J}$.\cite{Kanamori2}
The results of such fitting are shown in Table~\ref{tab:Kanamori}.
There is the clear dependence of the parameter ${\cal U}$ on the
local environment in solid, which is captured by the RPA calculations.\cite{condmat05}
Generally, the value of ${\cal U}$ is larger for more distorted Y-compounds.
There is also a clear correlation between the value of ${\cal U}$ and the magnitude of
local distortion around two nonequivalent transition-metal sites in the monoclinic phases:
the sites experiencing larger distortion (according to the magnitude
of the CF splitting in Fig.~\ref{fig.CFsummary}) have larger ${\cal U}$.
On the other hand, ${\cal J}$ is practically insensitive to the local environment in solids.
\begin{table}[h!]
\caption{Results of fitting of the
effective Coulomb interactions
in the $t_{2g}$ band obtained in the hybrid c-LDA+RPA approach
in terms of two Kanamori parameters: the intra-orbital Coulomb interaction ${\cal U}$
and the exchange interaction ${\cal J}$ (in eV).\protect\cite{Kanamori2}
The symbols `o' and `m' stand for the orthorhombic and monoclinic phases, respectively.
The positions of the transition-metal sites are shown in Fig.~\protect\ref{fig.structure}.}
\label{tab:Kanamori}
\begin{ruledtabular}
\begin{tabular}{ccccc}
 compound  & phase & site & ${\cal U}$ & ${\cal J}$ \\
\hline
 YTiO$_3$  &  o    & 1    & $3.45$     & $0.62$     \\
 LaTiO$_3$ &  o    & 1    & $3.20$     & $0.61$     \\
 YVO$_3$   &  o    & 1    & $3.27$     & $0.63$     \\
 YVO$_3$   &  m    & 1    & $3.19$     & $0.63$     \\
           &       & 3    & $3.26$     & $0.63$     \\
 LaVO$_3$  &  m    & 1    & $3.11$     & $0.62$     \\
           &       & 3    & $3.12$     & $0.62$     \\
\end{tabular}
\end{ruledtabular}
\end{table}

\section{\label{sec:Method}Solution of Model Hamiltonian}

  In this section we briefly discuss the methods of solution of the
model Hamiltonian (\ref{eqn:Hmanybody}). We start with the simplest
HF approach, which totally neglect the correlation effects.
Then, we consider two simple corrections to the HP approximation, which
allow to include some of these effects. One is the
second-order perturbation theory for the total energy. It shares common
problems of the regular (nondegenerate) perturbation theory and allows
to calculate easily the correction to the total energy, starting from
the single-Slater-determinant HF approximation
for the wavefunctions. Therefore, we expect this
method to work well for the systems where the orbital degeneracy is already lifted
and the ground state is described reasonably well
by a single Slater determinant, so that other corrections can be treated as a perturbation.
The second scheme is the variational superexchange theory for the $d^1$ perovskites,
which takes into account the multiplet structure of the excited atomic states.
It allows to study the effect of electron correlations on the orbital ordering.
However, it is limited by typical approximations made in the
theory of superexchange interactions, which treat all transfer integrals as
a perturbation.

  All calculations have been performed in the basis of Wannier functions
$\{ \tilde{W}_{\bf R}^\alpha \}$, which have a finite weight
at the central transition-metal sites
as well as the oxygen and other atomic sites located in the
nearest neighborhood to the transition-metal atoms. In order to calculate the local quantities,
associated with the transition-metal atoms, such as the spin or orbital magnetic
moments as well as the distributions of the charge densities,
the Wannier
functions have been expanded over the standard LMTO basis, and the
aforementioned quantities have been obtained by the integration over the atomic spheres
of the transition-metal sites.

\subsection{\label{sec:HFApproximation}Hartree-Fock Approximation}

  The Hartree-Fock approximation provides the simplest solution of the many-body
problem described by the model Hamiltonian (\ref{eqn:Hmanybody}). In this case, the trial
wavefunction for the many-electron ground state is searched in the form of a single Slater determinant
$|S\{ \varphi_{n {\bf k}} \} \rangle$, which is constructed from the one-electron orbitals
$\{ \varphi_{n {\bf k}} \}$. The latter are subjected to the variational principle and requested to minimize
the total energy
$$
E_{\rm HF}= \min_{\{ \varphi_{n {\bf k}}\}}
\langle S\{ \varphi_{n {\bf k}} \}|\hat{\cal{H}}| S\{ \varphi_{n {\bf k}} \} \rangle
$$
for a given number of particles $\cal{N}$,
yielding the set of well-known HF equations:
\begin{equation}
\left( \hat{h}_{\bf k} + \hat{V} \right) | \varphi_{n {\bf k}} \rangle =
\varepsilon_{n {\bf k}} | \varphi_{n {\bf k}} \rangle,
\label{eqn:HFeq}
\end{equation}
where
$\hat{h}_{\bf k}$$\equiv$$\| h_{\bf k}^{\alpha \beta} \|$
is the kinetic part of the model Hamiltonian (\ref{eqn:Hmanybody}) in the reciprocal space:
$h_{\bf k}^{\alpha \beta}$$=
$$\frac{1}{N} \sum_{{\bf R}'} h^{\alpha \beta}_{{\bf R}{\bf R}'} e^{-i {\bf k} \cdot ({\bf R}-{\bf R}')}$
($N$ being the number of sites),
and $\hat{V}$$\equiv$$\| V_{\alpha \beta} \|$ is the HF potential:
\begin{equation}
V_{\alpha \beta} = \sum_{\gamma \delta}
\left( U_{\alpha \beta \gamma \delta} - U_{\alpha \delta \gamma \beta} \right)
n_{\gamma \delta}.
\label{eqn:HFpot}
\end{equation}
Eq.~(\ref{eqn:HFeq}) is solved self-consistently together with the equation
$$
\hat{n} = \sum_{n {\bf k}}^{occ} | \varphi_{n {\bf k}} \rangle \langle \varphi_{n {\bf k}} |
$$
for the
density matrix $\hat{n}$$\equiv$$\|n_{\alpha \beta}\|$ in the basis
of Wannier orbitals.

  After the self-consistency,
the total energy can be calculated as
$$
E_{\rm HF} = \sum_{n {\bf k}}^{occ} \varepsilon_{n {\bf k}} -\frac{1}{2}
\sum_{\alpha \beta} V_{\beta \alpha} n_{\alpha \beta}.
$$

  Using $\{ \varepsilon_{n {\bf k}} \}$ and $\{ \varphi_{n {\bf k}}  \}$,
one can calculate the Green function in the HF approximation:
$$
\hat{\cal G}_{{\bf RR}'}(\omega) = \sum_{n {\bf k}}
\frac{ | \varphi_{n {\bf k}} \rangle \langle \varphi_{n {\bf k}} |}
{ \omega - \varepsilon_{n {\bf k}} + i\delta } e^{i {\bf k} \cdot ({\bf R}-{\bf R}')}.
$$
The latter is widely used for the analysis of interatomic magnetic
interactions:\cite{JHeisenberg}
\begin{equation}
J_{{\bf RR}'} = \frac{1}{2 \pi} {\rm Im} \int_{-\infty}^{\varepsilon_{\rm F}}
d \varepsilon {\rm Tr}_L \left\{ \hat{\cal G}_{{\bf RR}'}^\uparrow (\omega)
\Delta \hat{V} \hat{\cal G}_{{\bf R}'{\bf R}}^\downarrow (\omega)
\Delta \hat{V} \right\},
\label{eqn:JHeisenberg}
\end{equation}
where
$\hat{\cal G}_{{\bf RR}'}^{\uparrow, \downarrow}$$=$$\frac{1}{2} {\rm Tr}_S
\{ (\hat{1}$$\pm$$\hat{\sigma}_z ) \hat{\cal G}_{{\bf RR}'} \}$
is the projection of the Green function onto the majority ($\uparrow$)
and minority ($\downarrow$) spin states,
$\Delta \hat{V}$$=$${\rm Tr}_S \{ \hat{\sigma}_z \hat{V} \}$
is the magnetic (spin) part of the HF potential,
${\rm Tr}_S$ (${\rm Tr}_L$) denotes the trace over the spin (orbital) indices,
$\hat{1}$ and $\hat{\sigma}_z$ are correspondingly unity and Pauli matrices
of the dimension $6$,
and $\varepsilon_{\rm F}$ is the Fermi energy.

  The interatomic magnetic interactions $\{ J_{{\bf RR}'} \}$
characterize
the spin stiffness of the magnetic phase.
Therefore, they can be directly compared with the experimental spin-wave spectra
derived from the inelastic neutron scattering measurement.

  According to Eq.~\ref{eqn:JHeisenberg},
$J_{{\bf RR}'}$$>$$0$ ($<$$0$)
means that for a
given magnetic
structure, the spin arrangement in the bond $\langle {\bf RR}' \rangle$
corresponds to the local minimum (maximum) of the total energy.
However, in the following we will use the universal notations,
according to which $J_{{\bf RR}'}$$>$$0$ and $<$$0$ will
stand
the ferromagnetic and antiferromagnetic
coupling, respectively.
This corresponds to the \textit{local mapping} onto the Heisenberg model of the
form
\begin{equation}
E_{\rm Heis} = - \sum_{\langle {\bf RR}' \rangle}
J_{{\bf RR}'} {\bf e}_{\bf R} \cdot {\bf e}_{{\bf R}'},
\label{eqn:EHeisenberg}
\end{equation}
where ${\bf e}_{\bf R}$ is the direction of the spin magnetic
moment at the site ${\bf R}$.
The ``local mapping'' means that, strictly speaking, Eq.~\ref{eqn:EHeisenberg}
is justified only for the infinitesimal rotations of the spin magnetic moment near
an equilibrium.

  The magnetic interactions are extremely sensitive to the orbital ordering.
Therefore, they can be regarded as the local prob of the orbital
ordering in each magnetic state.

\subsection{\label{sec:2ndorder}Second Order Perturbation Theory for Correlation Energy}

  The simplest way to go beyond the HF approximation is to include the correlation
interactions
in the second order of perturbation theory for the total energy.\cite{cor2ndorder}
The correlation interaction (or a fluctuation)
is defined as the difference between true many-body
Hamiltonian~(\ref{eqn:Hmanybody}), and its one-electron
counterpart, obtained at the level of HF approximation:
\begin{equation}
\hat{\cal{H}}_{\rm C} = \sum_{\bf R} \left(
\frac{1}{2} \sum_{\alpha \beta \gamma \delta}
U_{\alpha \beta \gamma \delta}
\hat{c}^\dagger_{{\bf R}\alpha} \hat{c}^\dagger_{{\bf R}\gamma}
\hat{c}^{\phantom{\dagger}}_{{\bf R}\beta} \hat{c}^{\phantom{\dagger}}_{{\bf R}\delta} -
\sum_{\alpha \beta} V_{\alpha \beta}
\hat{c}^\dagger_{{\bf R}\alpha} \hat{c}^{\phantom{\dagger}}_{{\bf R}\beta} \right).
\label{eqn:H2ndorder}
\end{equation}
By treating $\hat{\cal{H}}_{\rm C}$ as a perturbation, the correlation energy can be easily
estimated as:\cite{cor2ndorder}
\begin{equation}
E_{\rm C} = - \sum_{S} \frac{
\langle G | \hat{\cal{H}}_{\rm C} | S \rangle \langle S | \hat{\cal{H}}_{\rm C} | G \rangle }
{E_{\rm HF}(S) - E_{\rm HF}(G)},
\label{eqn:dE2ndorder}
\end{equation}
where $|G \rangle$ and $|S \rangle$ are the Slater determinants
corresponding to the low-energy ground state
in the HF approximation, and the
excited state, respectively.
Due to the variational properties of the
Hartree-Fock approach, the only processes which contribute to
$E_{\rm C}$ correspond to the
two-particle excitations, for which each
$|S \rangle$ is obtained from $|G \rangle$ by replacing
two one-electron orbitals, say $\varphi_{n_1 {\bf k}_1}$
and $\varphi_{n_2 {\bf k}_2}$, from the occupied part of the spectrum
by two unoccupied orbitals, say $\varphi_{n_3 {\bf k}_3}$ and $\varphi_{n_4 {\bf k}_4}$.
Hence, using the notations of Sec.~\ref{sec:Model}, the matrix elements take the form:
\begin{equation}
\langle S | \hat{\cal{H}}_{\rm C} | G \rangle =
\langle \varphi_{n_3 {\bf k}_3} \varphi_{n_4 {\bf k}_4} | v_{\rm scr} |
\varphi_{n_1 {\bf k}_1} \varphi_{n_2 {\bf k}_2} \rangle -
\langle \varphi_{n_3 {\bf k}_3} \varphi_{n_4 {\bf k}_4} | v_{\rm scr} |
\varphi_{n_2 {\bf k}_2} \varphi_{n_1 {\bf k}_1} \rangle.
\label{eqn:dHmelement}
\end{equation}
These matrix elements
satisfy the following condition:
$\langle S | \hat{\cal{H}}_{\rm C} | G \rangle$$\sim$$\frac{1}{N}
\sum_{\bf R} e^{i({\bf k}_3+{\bf k}_4-{\bf k}_1-{\bf k}_2) \cdot {\bf R}}$,
provided that
the screened Coulomb interactions are diagonal with respect to the site
indices.
In the following we will retain only the ${\bf R}$$=$$0$ part in this sum.
This corresponds to the single-site approximation for the correlation interactions,
which is known to be good for three-dimensional systems
and becomes exact in the limit of infinite spacial dimensions.\cite{DMFT}

  Finally, we employ a common approximation
of noninteracting quasiparticles
and replace the denominator of Eq.~(\ref{eqn:dE2ndorder})
by the linear combination of HF eigenvalues:
$E_{\rm HF}(S)$$-$$E_{\rm HF}(G)$$\approx$$\varepsilon_{n_3 {\bf k}_3}$$+$$\varepsilon_{n_4 {\bf k}_4}$$-
$$\varepsilon_{n_1 {\bf k}_1}$$-$$\varepsilon_{n_2 {\bf k}_2}$.\cite{cor2ndorder}

  The form of Eq.~(\ref{eqn:dE2ndorder}) implies that the HF ground state is
\textit{nondegenerate}, and the correlation effects can be systematically
included by considering the regular perturbation theory expansion.
It does not apply to the cubic systems, where the ground state is infinitely
degenerate (with respect to different orbital configurations), and where
a suitable approach
for the correlation energy
should be based
on the degenerate perturbation
theory.\cite{KhaliullinMaekawa}
Thus, the use of Eq.~(\ref{eqn:dE2ndorder}) implies that
the orbital degeneracy
is already lifted by the
crystal distortion.
As we shall see below, this approximation can be
justified for a number of systems.

\subsection{\label{sec:SEmethod}Effects of Multiplet Structure in the
Theory of Superexchange Interactions}

  Another method, which allows to treat some correlation effects beyond the
mean-field HF approximation is based on the generalization of the theory of
superexchange interactions in order to describe the correct multiplet structure
of the atomic states.
Similar idea has been discussed in the context of colossal magnetoresistive
perovskite manganites.\cite{SEmultiplet} The formulation is extremely simple for the
$d^1$ compounds, like YTiO$_3$ and LaTiO$_3$.

\begin{figure}[t!]
\begin{center}
\resizebox{8cm}{!}{\includegraphics{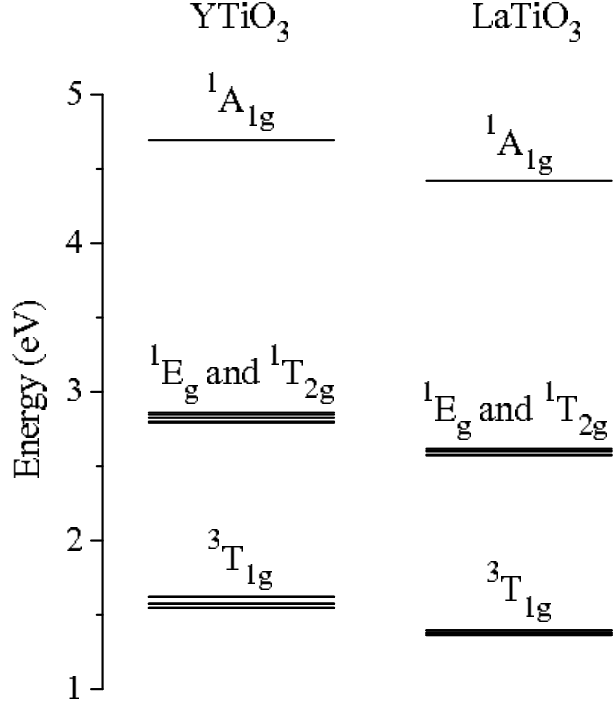}}
\end{center}
\caption{\label{fig.multiplet}
The multiplet structure of the excited atomic configuration $t_{2g}^2$
in YTiO$_3$ and LaTiO$_3$.}
\end{figure}

  The superexchange interaction in the bond $\langle {\bf RR}' \rangle$ is basically the
gain of the kinetic energy, which is acquired by an electron occupying the atomic
orbital $\phi_{\bf R}$ of site ${\bf R}$ in the process of virtual hoppings into the
subspace of unoccupied orbitals of the (neighboring) site ${\bf R}'$, and vice versa.\cite{PWA,KugelKhomskii}
In the atomic limit
for the $d^1$ compounds, there is only one
$t_{2g}$ electron at each Ti site. This is essentially an one-electron problem, where the
form of the atomic orbital $\phi_{\bf R}$ is determined by the site-diagonal part of
kinetic-energy $\| h_{\bf RR}^{\alpha \beta} \|$, incorporating the effects
of the spin-orbit interaction and the CF splitting. Therefore, in the pure
atomic limit, the ground-state wavefunction for each bond
$\langle {\bf RR}' \rangle$ can be described by a single Slater determinant:
$$
| G_{{\bf RR}'} \rangle = \frac{1}{\sqrt{2}} \left\{
\phi_{\bf R}(1) \phi_{{\bf R}'}(2) - \phi_{{\bf R}'}(1) \phi_{\bf R}(2) \right\}.
$$
Then,
$\phi_{\bf R}$ and $\phi_{{\bf R}'}$
are expanded over the Wannier orbitals associated with the site ${\bf R}$ and ${\bf R}'$
and the transfer integrals connecting different Wannier orbitals are treated as a perturbation.
The excited states at the sites ${\bf R}$ and ${\bf R}'$, which appear in the process of
the virtual hoppings are the two-electron states and subjected to the multiplet splitting.
This is exactly the point where the electron correlations, beyond the HF approximation,
enter the problem. In order to incorporate these effects we note that
from
$m$$=$$6$ Wannier spin-orbitals $\{ \tilde{W}_{\bf R}^\alpha \}$ at each Ti site, one can construct
$\frac{1}{2}m(m$$-$$1)$$=$$15$ antisymmetric two-electron Slater's determinants $\{ |S \rangle \}$
($S$$=$$1,$$\dots,$$15$), which can be used as the basis for the screened Coulomb interactions
in the excited state: $U_{SS'}$$=$$\langle S|v_{\rm scr}|S' \rangle$. The diagonalization of
this $15$$\times$$15$ matrix yield the complete set of eigenvalues $\{ E_{{\bf R}M} \}$ and
eigenfunctions $\{ | {\bf R} M \rangle \}$ of the two-electron states at the site
${\bf R}$ ($M$$=$$1,$$\dots,$$15$).
An example of such a multiplet structure for YTiO$_3$ and
LaTiO$_3$ is shown in Fig.~\ref{fig.multiplet}.
According to the first Hund rule, the lowest-energy configuration corresponds to the spin-triplet
state ${\rm ^3T_{1g}}$.
The degeneracy of the
${\rm ^3T_{1g}}$, ${\rm ^1E_g}$, and ${\rm ^3T_{2g}}$ levels is lifted
by the orthorhombic distortion, which affects the
matrix elements of the effective
$\hat{U}$ via the RPA channel of
screening. The splitting is larger for the
more distorted YTiO$_3$.

  In order to calculate the energy gain caused by the superexchange interactions,
the eigenfunctions $\{ | {\bf R} M \rangle \}$
shall be projected onto the physical subspace of
two-electron states which can be created by transferring an electron from
the neighboring sites. The corresponding projector operators have the form:
${\hat{\cal P}}_{\bf R}$$=$$\sum_{\alpha} | P_{\bf R}^\alpha \rangle
\langle P_{\bf R}^\alpha |$, where
$| P_{\bf R}^\alpha \rangle$$=
$$\frac{1}{\sqrt{2}}\{ \phi_{\bf R}(1)\tilde{W}_{\bf R}^\alpha(2)$$-$$
\tilde{W}_{\bf R}^\alpha(1)\phi_{\bf R}(2) \}$ is the Slater determinant
constructed from the occupied orbital $\phi_{\bf R}$ and one of the
basis Wannier orbitals $\tilde{W}_{\bf R}^\alpha$.
In the other words,
the projection ${\hat{\cal P}}_{\bf R}$ imposes an additional constraint,
which guarantees that one of the orbitals in the two-electron state
is
$\phi_{\bf R}$.
Then, the energy gain caused by the virtual hoppings in the bond $\langle {\bf RR}' \rangle$
is given by
\begin{equation}
\Delta E_{{\bf RR}'} = - \left\langle G_{{\bf RR}'} \left| \hat{h}_{{\bf RR}'}
\left( \sum_M \frac{{\hat{\cal P}}_{{\bf R}'}|{\bf R}' M \rangle
\langle {\bf R}' M| {\hat{\cal P}}_{{\bf R}'}}{E_{{\bf R}'M}} \right)
\hat{h}_{{\bf R}'{\bf R}} +
({\bf i} \leftrightarrow {\bf j}) \right| G_{{\bf RR}'} \right\rangle.
\label{eqn:Esuperexchange}
\end{equation}
The total energy of the system
in the superexchange approximation is obtained after summation over all bonds,
which should be combined with the site-diagonal elements, incorporating the
effects of the
CF splitting and the relativistic spin-orbit interaction:
$$
E_{\rm SE}= \sum_{\bf R}
\langle \phi_{\bf R} | \hat{h}_{\bf RR} | \phi_{\bf R} \rangle +
\sum_{\langle {\bf RR}' \rangle}
\Delta E_{{\bf RR}'}.
$$
Finally, the set of occupied orbitals $\{ \phi_{\bf R} \}$ is obtained
by minimizing
$E_{\rm SE}$, e.g., using the steepest descent method.

  For a given orbital ordering,
the multiplet effects are expected to be more important in the case of the
AFM spin ordering, where an electron comes to the neighboring site
with the opposite direction of spin. In this case,
the excited configuration is
subjected to the multiplet splitting into the spin-singlet and spin-triplet
states, which additionally stabilizes the AFM spin state.\cite{comment.5}

\section{\label{sec:results}Results and Discussions}

  In this section we present results of solution the model
Hamiltonian (\ref{eqn:Hmanybody})
for the distorted perovskite compounds. We start with Y-based perovskites, where
the orbital ordering is largely controlled by the CF splitting coming
from the experimental lattice distortion. We will show that this distortion
imposes a severe constraint on the magnetic properties
and predetermines the type of the
magnetic ground state.
Then, we turn to the La-based perovskites, where the situation is less clear:
while the magnetic structure of LaVO$_3$ can be still understood
on the basis of its experimental lattice structure, LaTiO$_3$ poses
many open and unresolved questions.

  First, we will consider the impact of the crystal structure without the spin-orbit interaction.
The effects of the spin-orbit interaction will be discussed separately in Sec.~\ref{sec:spinorbit}.

\subsection{\label{sec:YVO}YVO$_3$}

  YVO$_3$ exhibits two structural phase transitions.\cite{Blake,Tsvetkov}
The first one
is the second-order transition from orthorhombic ($D_{2h}^{16}$)
to monoclinic (apparently $D_{2h}^5$) phase, which takes places at
$200$ K and is believed to coincide with the onset of the
orbital ordering. The second one is the first-order transition at $77$ K from
monoclinic to another orthorhombic phase.
The magnetic transition temperature
is
$116$ K, which lies in the
monoclinic region and does not coincide with any structural phase
transition.
On the other hand,
the change of the crystal structure at $77$ K coincides with the
magnetic phase transition.
The magnetic structure in the interval $77$ K $<$ T $<$ $116$ K
is C-type AFM, while below $77$ K it becomes G-type AFM.

\subsubsection{Low-temperature orthorhombic phase $({\rm T < 77~K})$}

  YVO$_3$ has the largest CF splitting amongst
perovskite compounds, which crystallize in the orthorhombic
phase (Fig.~\ref{fig.CFsummary}).
It lowers the energies of two $t_{2g}$ levels --
just enough to accommodate
two $d$ electrons.
The highest level is
separated from the middle one
by a $111$~meV gap. Therefore, the orbital ordering in this
$d^2$ compound is expected to be quenched (at least partially)
by the crystal distortion.

\begin{figure}[t!]
\begin{center}
\resizebox{5cm}{!}{\includegraphics{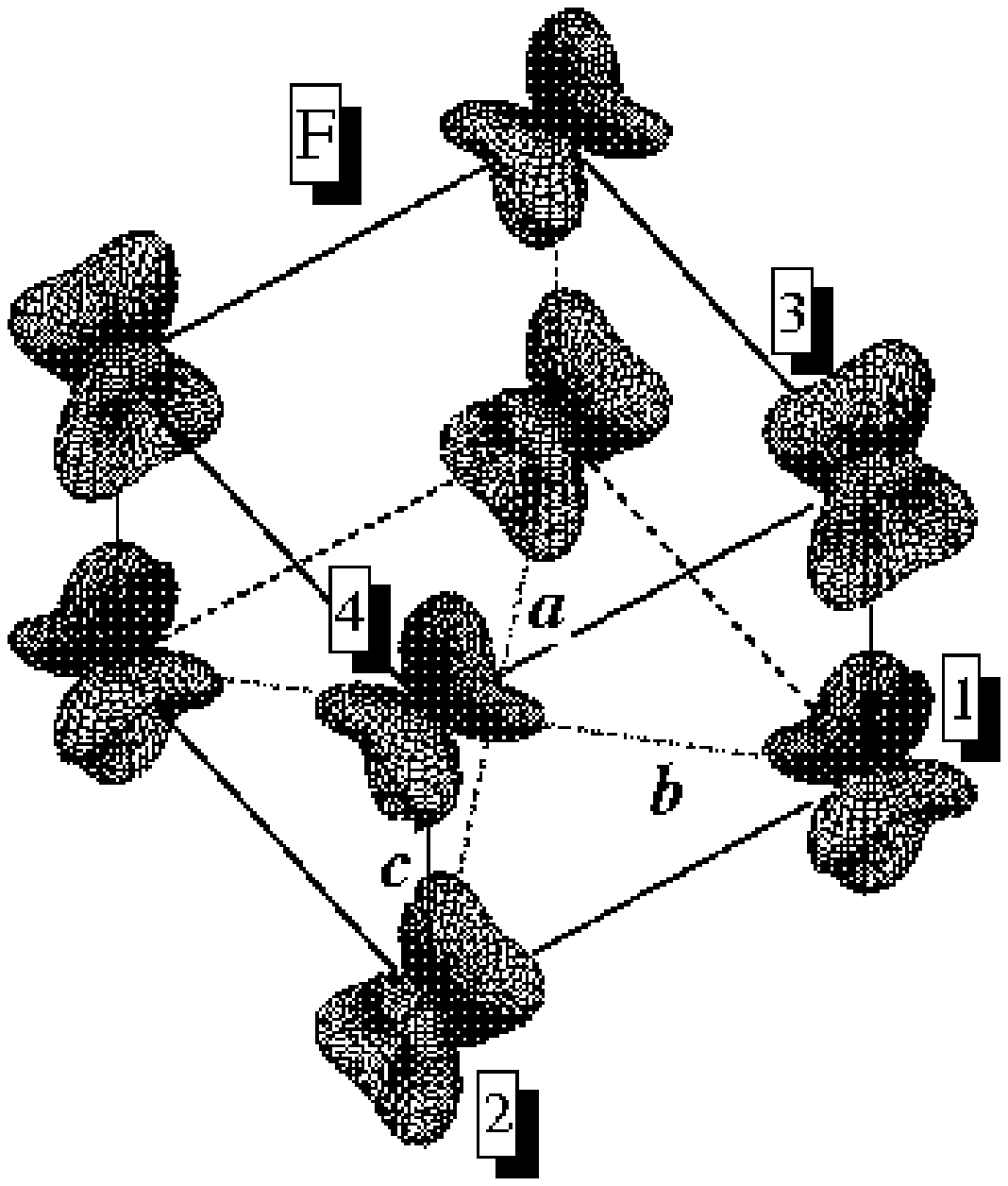}}
\resizebox{5cm}{!}{\includegraphics{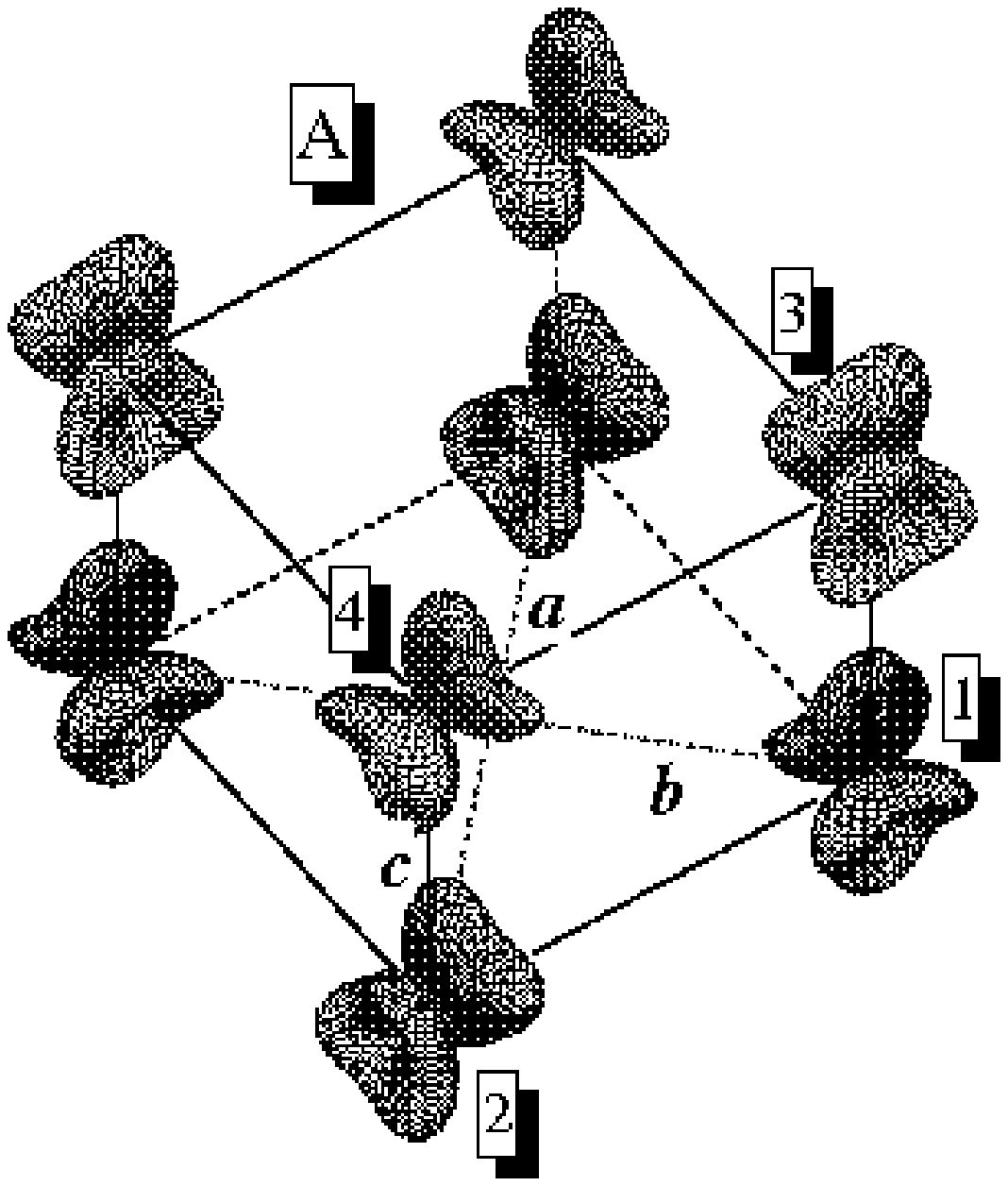}}
\resizebox{5cm}{!}{\makebox{ }}
\end{center}
\begin{center}
\resizebox{5cm}{!}{\includegraphics{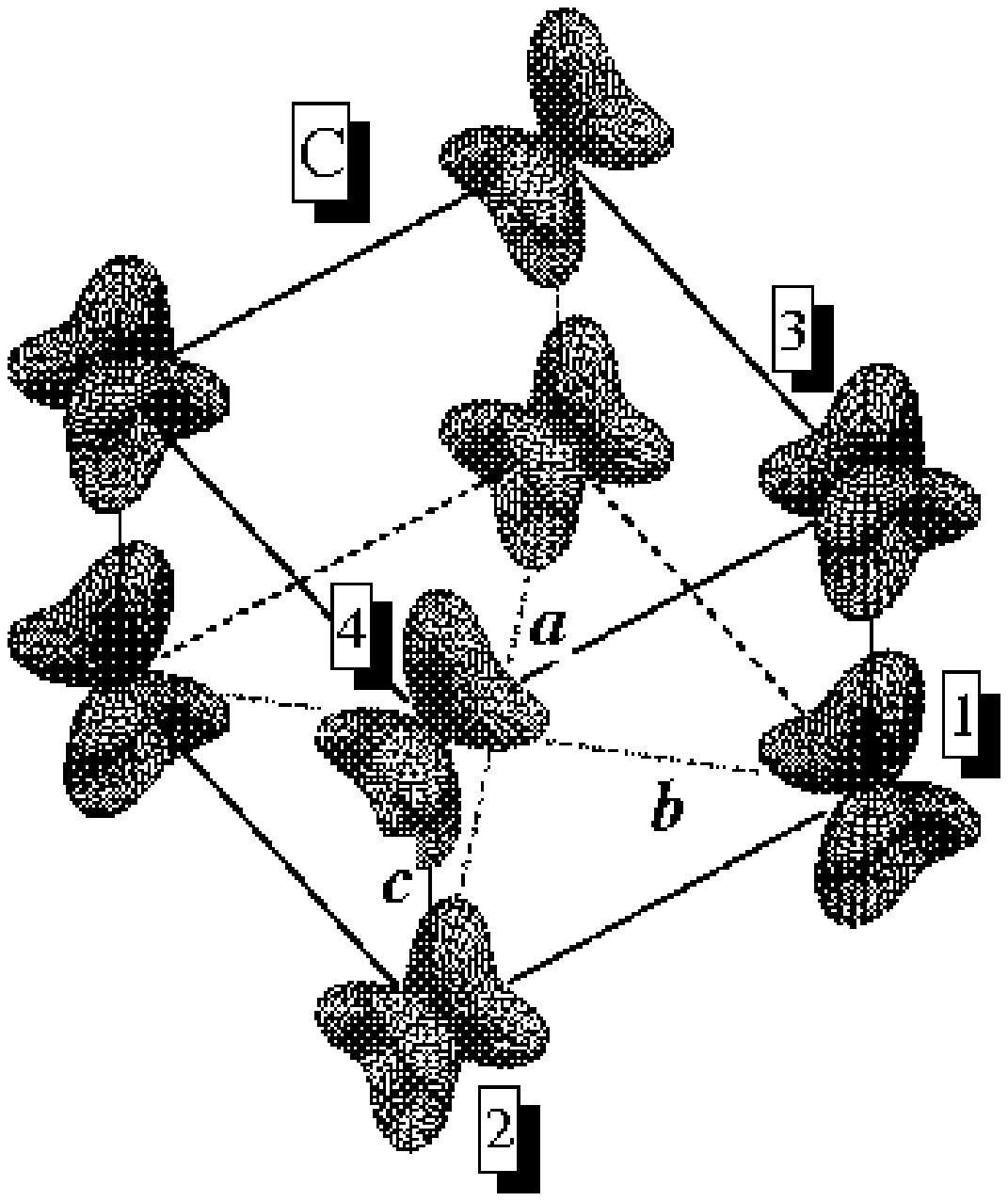}}
\resizebox{5cm}{!}{\includegraphics{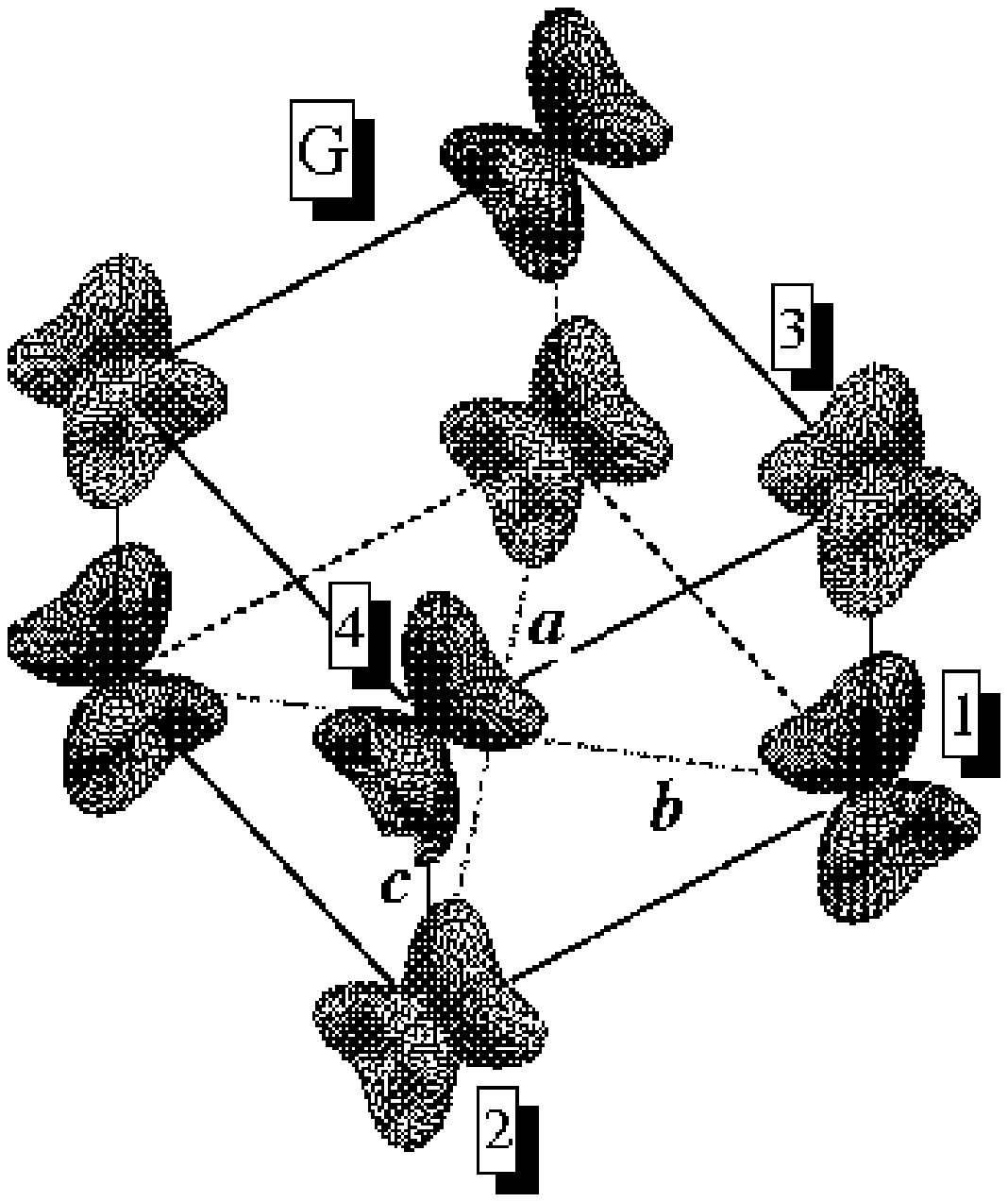}}
\resizebox{5cm}{!}{\includegraphics{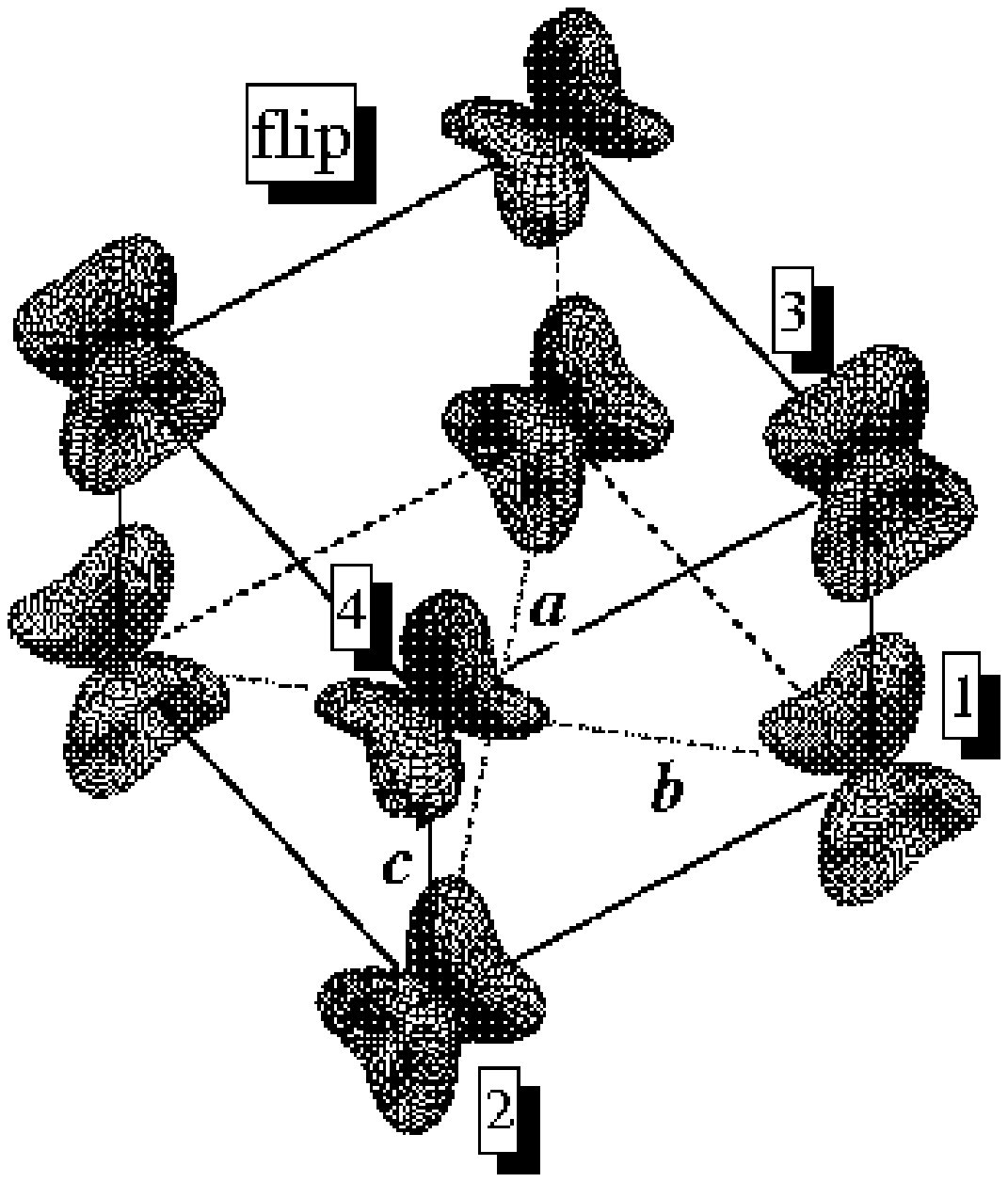}}
\end{center}
\caption{\label{fig.OOYVO} (Color online)
Distribution of the charge density around V-sites
in various magnetic phases
of orthorhombically distorted YVO$_3$ (${\rm T}$$<$$77$ K),
as obtained in Hartree-Fock calculations.
Different magnetic sublattices
are shown by different colors.}
\end{figure}

  This is clearly seen in our Hartree-Fock calculations. The orbital ordering
depends on the magnetic state. However, this dependence is
weak and can be hardly seen on the plot (Fig.~\ref{fig.OOYVO}).
The form of the orbital ordering, which can be schematically viewed as an
alternation of the ($xy$,$yz$) and ($xy$,$zx$) orbitals in the cubic
coordinate frame associated with the VO$_6$ octahedra,
is in an excellent agreement with the theoretical prediction of Sawada and Terakura
based on the semi-empirical LDA$+$$U$ method,\cite{SawadaTerakura}
which was later on confirmed by the synchrotron x-ray-diffraction
measurement.\cite{Noguchi}

\begin{figure}[t!]
\begin{center}
\resizebox{5cm}{!}{\includegraphics{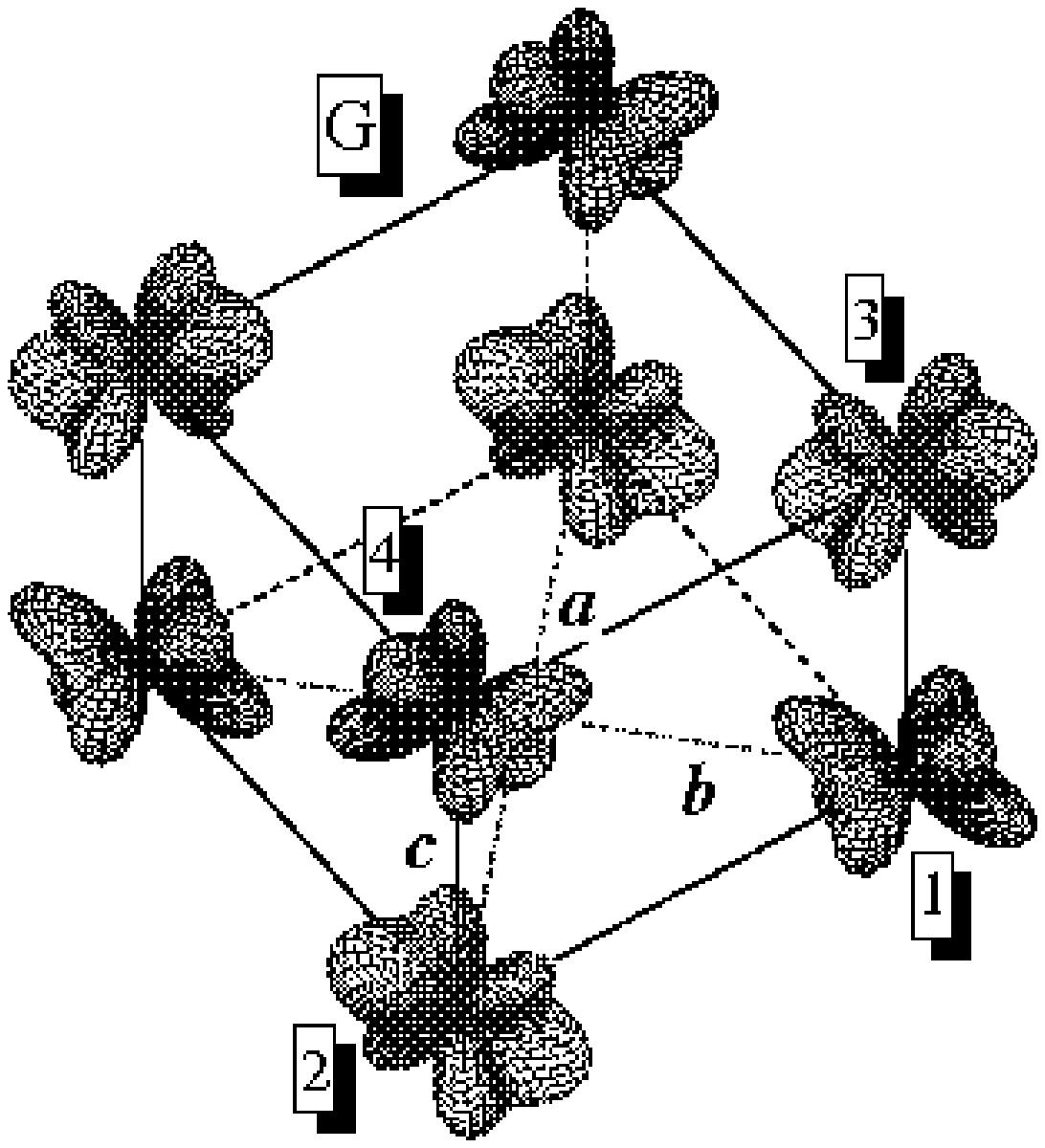}}
\resizebox{5cm}{!}{\includegraphics{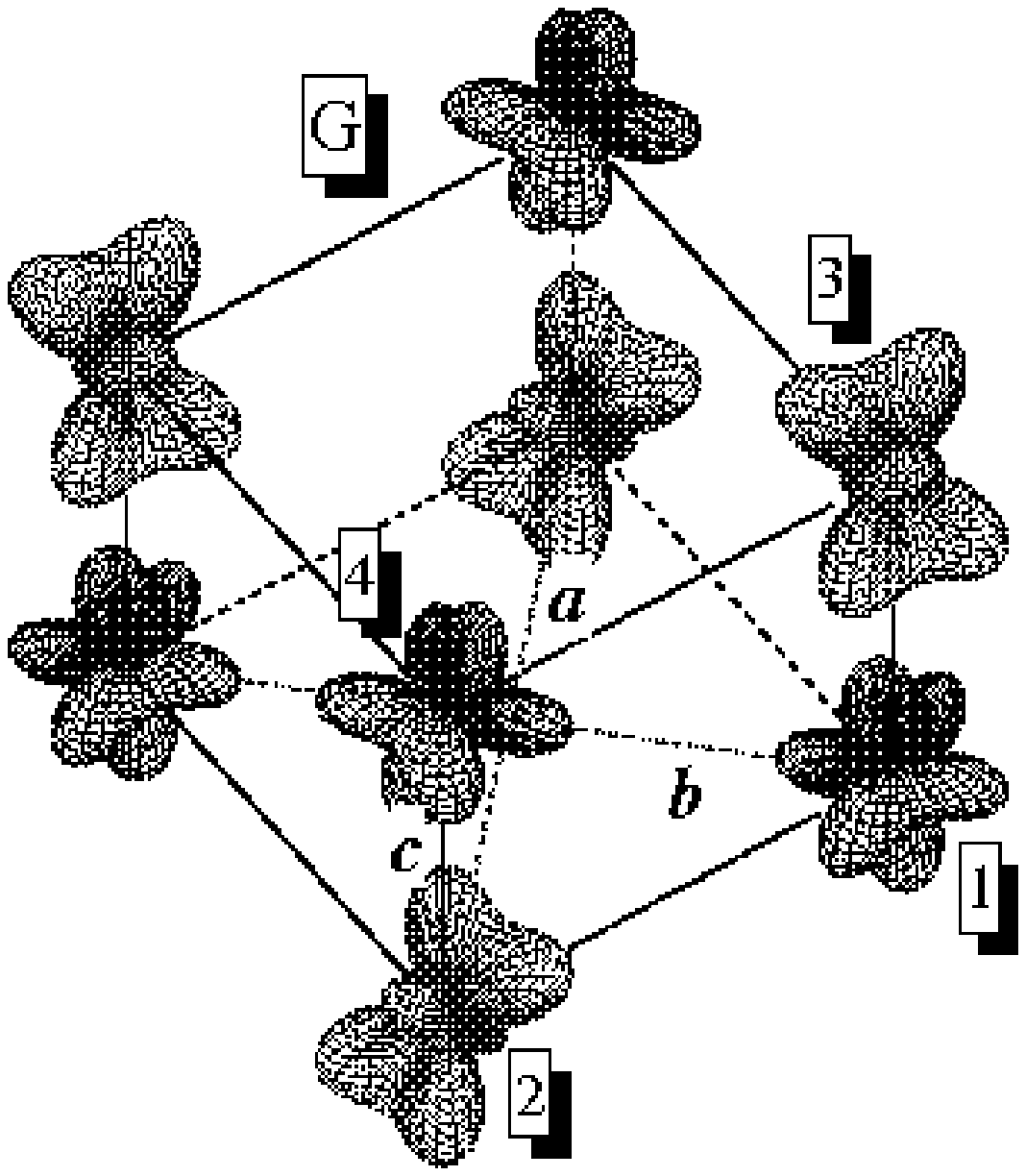}}
\resizebox{5cm}{!}{\includegraphics{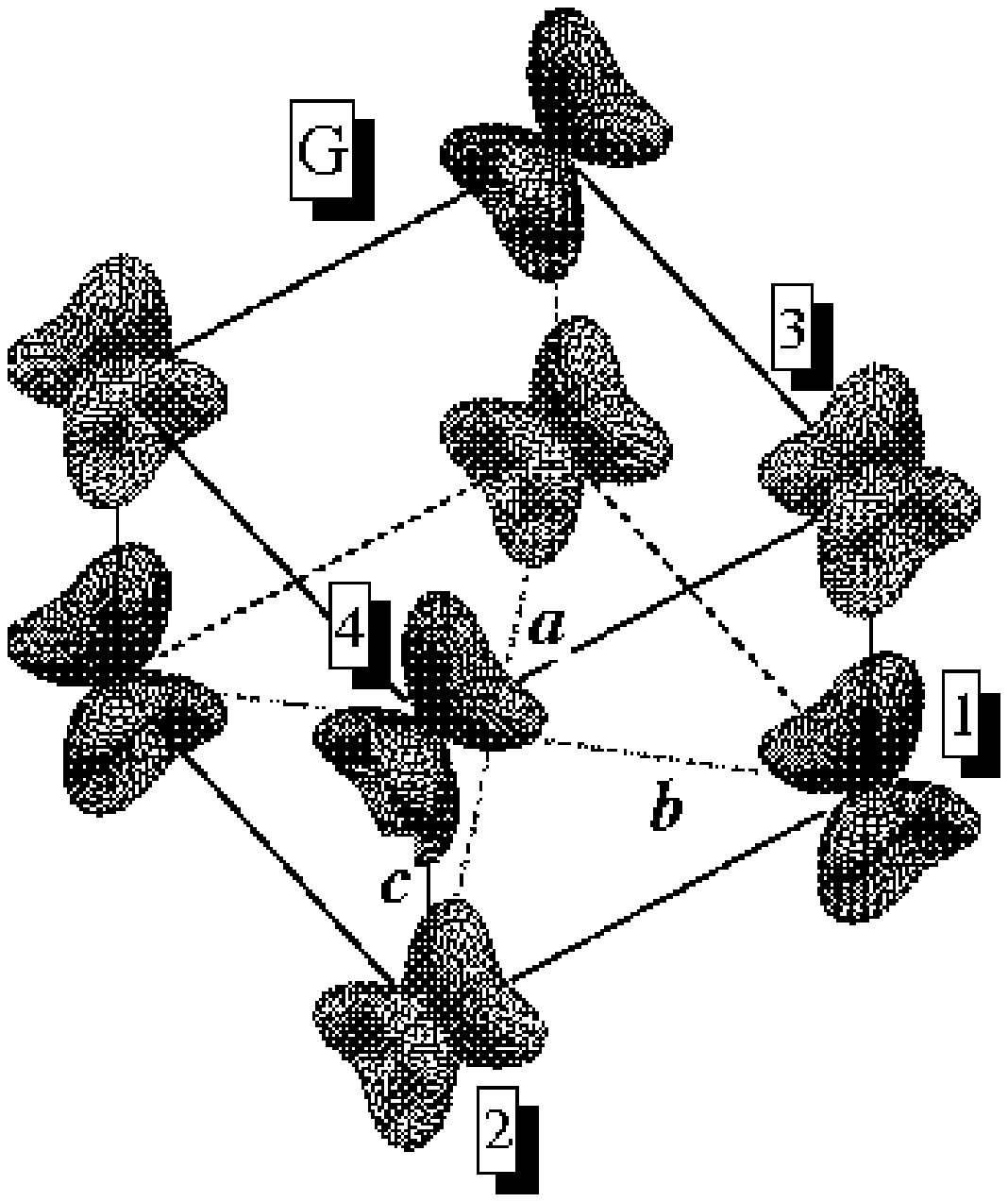}}
\end{center}
\caption{\label{fig.OOYVOcomparison} (Color online)
Orbital ordering in the G-type antiferromagnetic phase of orthorhombically
distorted YVO$_3$ computed without crystal field (left),
including the crystal field of the covalent type (center), and the
crystal field of both covalent and Madelung types (right).}
\end{figure}

  The first important question, which we would like to address here is where
does this orbital ordering come from?
In Fig.~\ref{fig.OOYVOcomparison} we show results of calculations obtained using
three different settings for the site-diagonal part of kinetic-energy part of
the model Hamiltonian: (i) $h_{\bf RR}^{\alpha \beta}$$=$$0$ (no CF splitting);
(ii) the parameters $\{ h_{\bf RR}^{\alpha \beta} \}$ extracted from the downfolding method,
which takes into account only the covalent type of the CF splitting; and
(iii) the parameters $\{ h_{\bf RR}^{\alpha \beta} \}$ obtained in
the downfolding method and corrected
for the nonsphericity of the Madelung potential (\ref{eqn:CFEI}).
One can clearly see that in order to reproduce the correct orbital ordering,
\textit{all} contributions to the CF splitting appear to be important.
Had we neglected some of these contributions, not only
the orbital ordering but also the magnetic ground state
would have been different. For example, without the Madelung term,
the magnetic ground state is expected to be of the A-type, in clear disagreement
with the experimental data.

  As it has been already discussed by other
authors,\cite{SawadaTerakura,MizokawaFujimori,FangNagaosa}
the C-type orbital ordering shown in Fig.~\ref{fig.OOYVO} favors the
G-type AFM spin ordering, which
emerges as the ground state already at the level of HF calculations (Table~\ref{tab:YVOo}).
The order of the magnetic states, corresponding to
the increase of the total energy, is
G$\rightarrow$C$\rightarrow$flip$\rightarrow$A$\rightarrow$F, which is well
consistent with results of all-electron LDA$+$$U$
calculations.\cite{SawadaTerakura,FangNagaosa,comment.3}
\begin{table}[h!]
\caption{Magnetic interactions ($J_{{\bf RR}'}$),
Hartree-Fock energies ($E_{\rm HF}$), and total energies ($E_{\rm tot}$)
in the orthorhombic phase of YVO$_3$ (${\rm T}$$<$$77$ K).
The energies are measured from the most stable
magnetic state in meV per one formula unit. The magnetic interactions
are measured in meV. The total energy is defined as Hartree-Fock energy plus
correlation energies ($E_{\rm C}$) in the second order of perturbation theory:
$E_{\rm tot}$$=$$E_{\rm HF}$$+$$E_{\rm C}$.
}
\label{tab:YVOo}
\begin{ruledtabular}
\begin{tabular}{ccccccc}
 phase  & $J_{12}$  & $J_{13}$  & $J_{24}$  & $J_{34}$  & $E_{\rm HF}$ &  $E_{\rm tot}$    \\
\hline
   F    & $-$$1.4$  & $-$$3.6$  & $-$$3.6$  & $-$$1.4$  & $21.7$       &          $27.1$   \\
   A    & $-$$2.5$  & $-$$3.7$  & $-$$3.7$  & $-$$2.5$  & $14.6$       &          $17.3$   \\
   C    & $-$$4.7$  & $-$$4.9$  & $-$$4.9$  & $-$$4.7$  & $10.1$       &          $11.7$   \\
   G    & $-$$4.4$  & $-$$4.8$  & $-$$4.8$  & $-$$4.4$  & $0$          &          $0$      \\
  flip  & $-$$4.0$  & $-$$4.8$  & $-$$3.9$  & $-$$1.8$  & $11.6$       &          $14.0$   \\
\end{tabular}
\end{ruledtabular}
\end{table}

  However, the orbital ordering \textit{is not} fully quenched by the crystal distortion
and to certain extent
can adjust the change of the magnetic state through the
Kugel-Khomskii mechanism.\cite{KugelKhomskii}
This is seen particularly well in the behavior of interatomic magnetic
interactions, which reveal an appreciable dependence on the magnetic state.
For example, by going from the G-state to the F-state,
the in-plane interaction $J_{12}$ ($J_{34}$)
changes by nearly 70\%,
and the
inter-plane interaction
$J_{13}$ ($J_{24}$) changes by 25\%.

  In agreement with the experimental finding,\cite{Ulrich2003})
the magnetic interactions in the G-type AFM ground state are nearly isotropic.
However, this isotropic behavior can be easily destroyed
by a small change of the orbital ordering, which is realized for example
in other magnetic states.

  The absolute values of $J_{12}$ and $J_{13}$ obtained in the HF calculations
for the G-type AFM phase
are underestimated by about $1$ meV  in comparison with the
experimental parameters extracted from the fit
of the spin-wave spectra
($J_{12}$$=$$J_{13}$$=-$$5.7$$\pm$$0.3$ meV).
This seems to be reasonable because the HF method is a
single-Slater-determinant approach, which does not include
the correlation effects. The magnitude of the correlations energy
depends on the magnetic state and
is expected to be larger in the case of the AFM spin alignment,
where
the net magnetization is zero and the choice of the many-electron wave function
in the form of a
single Slater determinant
is always an approximation.\cite{Imai} On the other hand, the saturated
ferromagnetic state can be described relatively well by a single Slater determinant.
All these trends are clearly seen in the total energy calculations,
which take into account the correlation effect in the second order of
perturbation theory (Table~\ref{tab:YVOo}). The correlations additionally
stabilize the G-type AFM ground state relative to other magnetic states.
However, it does not change the order of the magnetic states.

  Unfortunately, it is not possible to estimates the effects of correlations
on the interatomic magnetic interactions directly, using Eq.~(\ref{eqn:JHeisenberg}).
Nevertheless, one can try to use the total energy differences between different
magnetic states and map them onto the Heisenberg model.\cite{comment.4}
This is a cruder approximation, which implies that the orbital ordering
is completely quenched by the crystal distortion and does not depend on the
magnetic state. We will use it here only in order to get a qualitative
idea about the impact of correlation effects on the interatomic magnetic interactions.
Then, the standard HF approximation yields
$J_{12}$$=-$$3.3$ meV and $J_{13}$$=-$$4.3$ meV,
the second order perturbation theory for the correlation energy
yields $J_{12}$$=-$$4.1$ meV and $J_{13}$$=-$$5.4$ meV.
Therefore, the $1$ meV difference between parameters of magnetic interactions obtained
in the HF calculations and the experimental spin-wave dispersion data can be naturally
attributed to the correlation effects.
This example also clearly shows at the importance of more rigorous treatment
of the correlation effects and necessity to go
beyond the single-Slater-determinant HF approximation
for the $t_{2g}$ perovskite oxides.

\subsubsection{High-temperature monoclinic phase $({\rm 77~K < T < 116~K)}$}

  The monoclinic distortion
creates two inequivalent types of V-sites,
which lie in different ${\bf ab}$-planes:
(1,2) and (3,4) in Fig.~\ref{fig.structure}. This leads to the new scheme of the
$t_{2g}$-level splitting (Fig.~\ref{fig.CFsummary}).
Energetically, the new scheme
is rather similar to the previous one, observed in
the orthorhombic phase. In both planes it lowers the energies of two $t_{2g}$ levels.
The energy gap, which separates the highest level from the middle one is
$101$ and $128$ meV for the sites 1 and 3, respectively. However, the type of
the orbitals which are split off by the distortion is different (Table~\ref{tab:CForbitals}).
This leads to the new type of stacking between the planes, which reminiscent
the G-type orbital ordering.

\begin{figure}[t!]
\begin{center}
\resizebox{5cm}{!}{\includegraphics{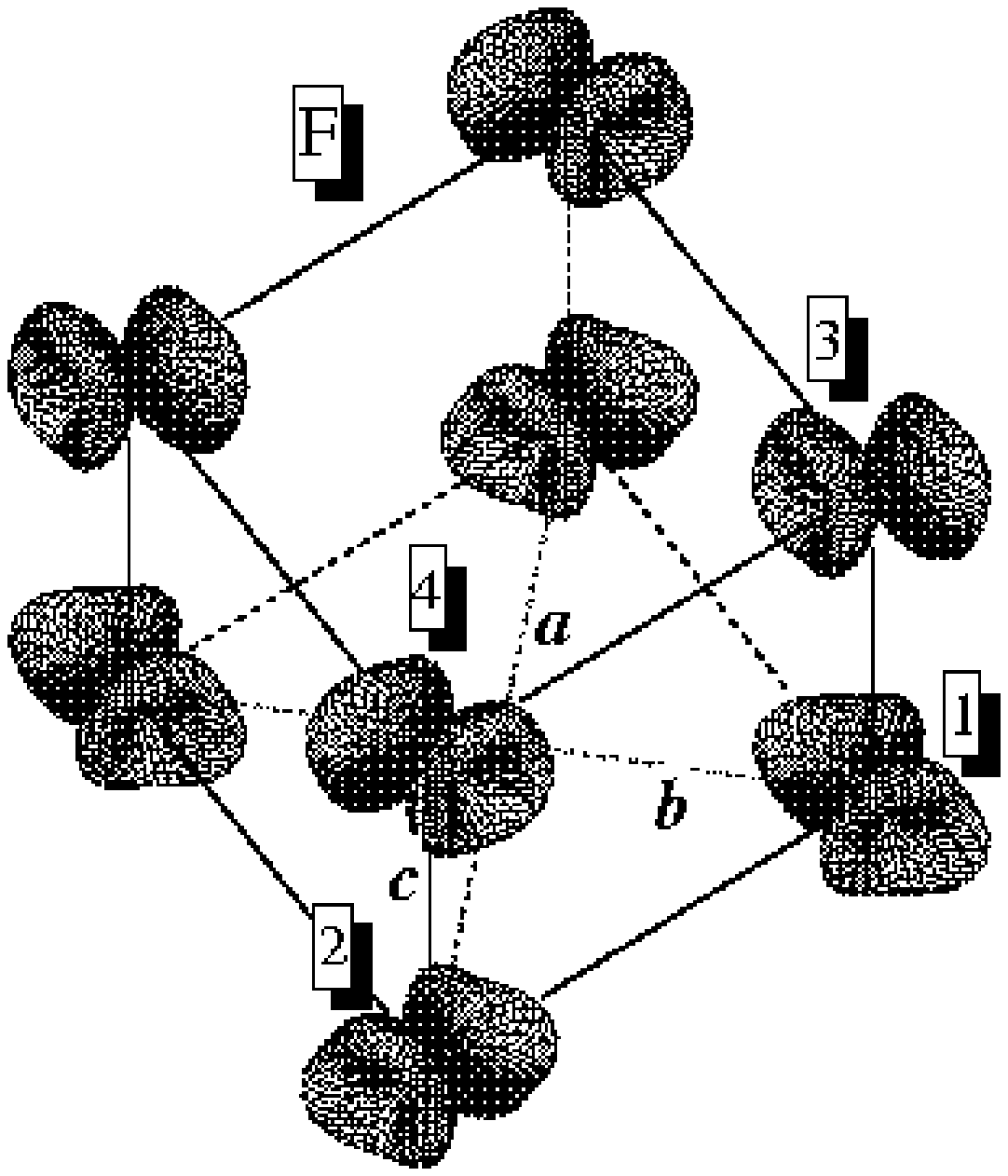}}
\resizebox{5cm}{!}{\includegraphics{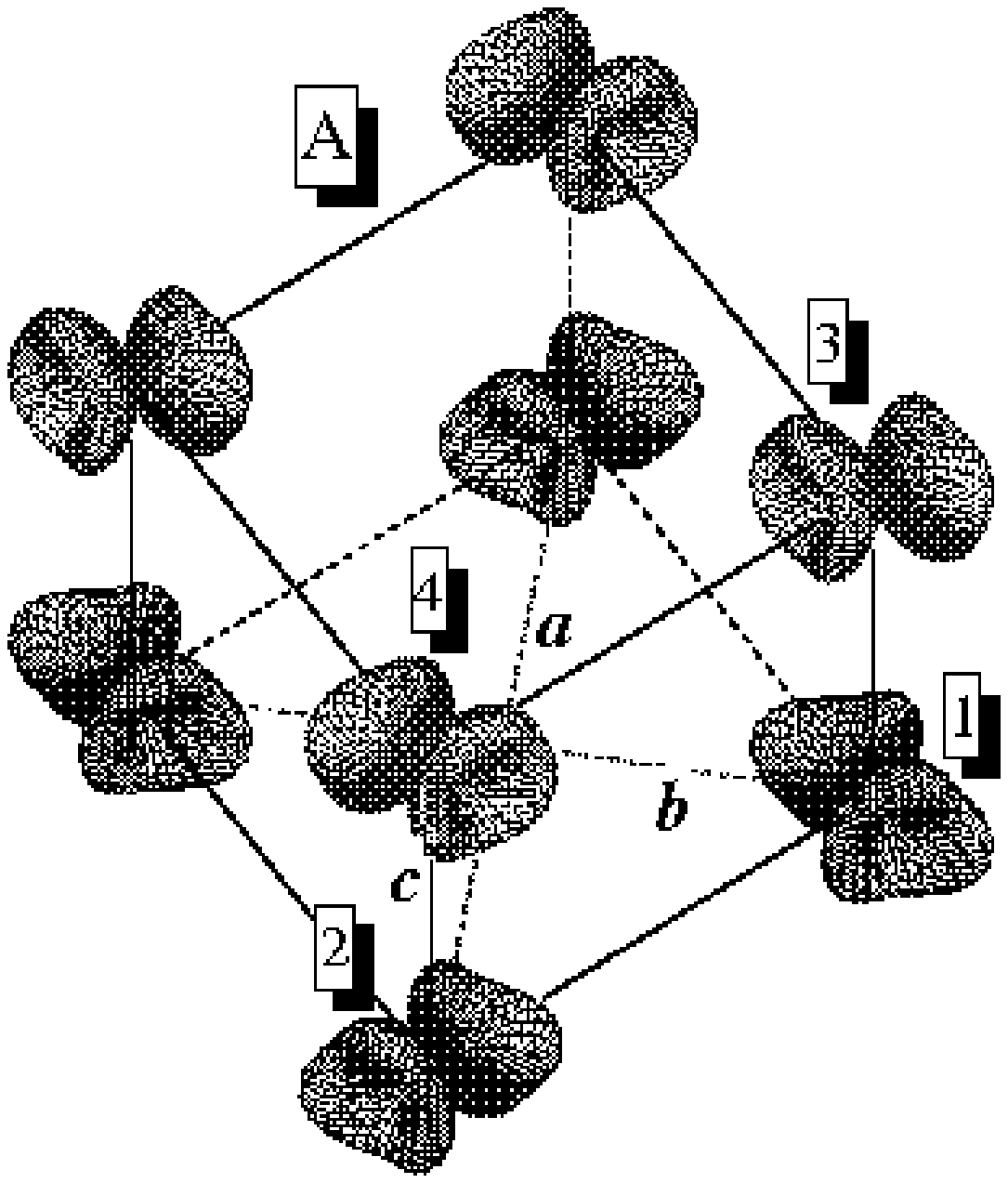}}
\resizebox{5cm}{!}{\includegraphics{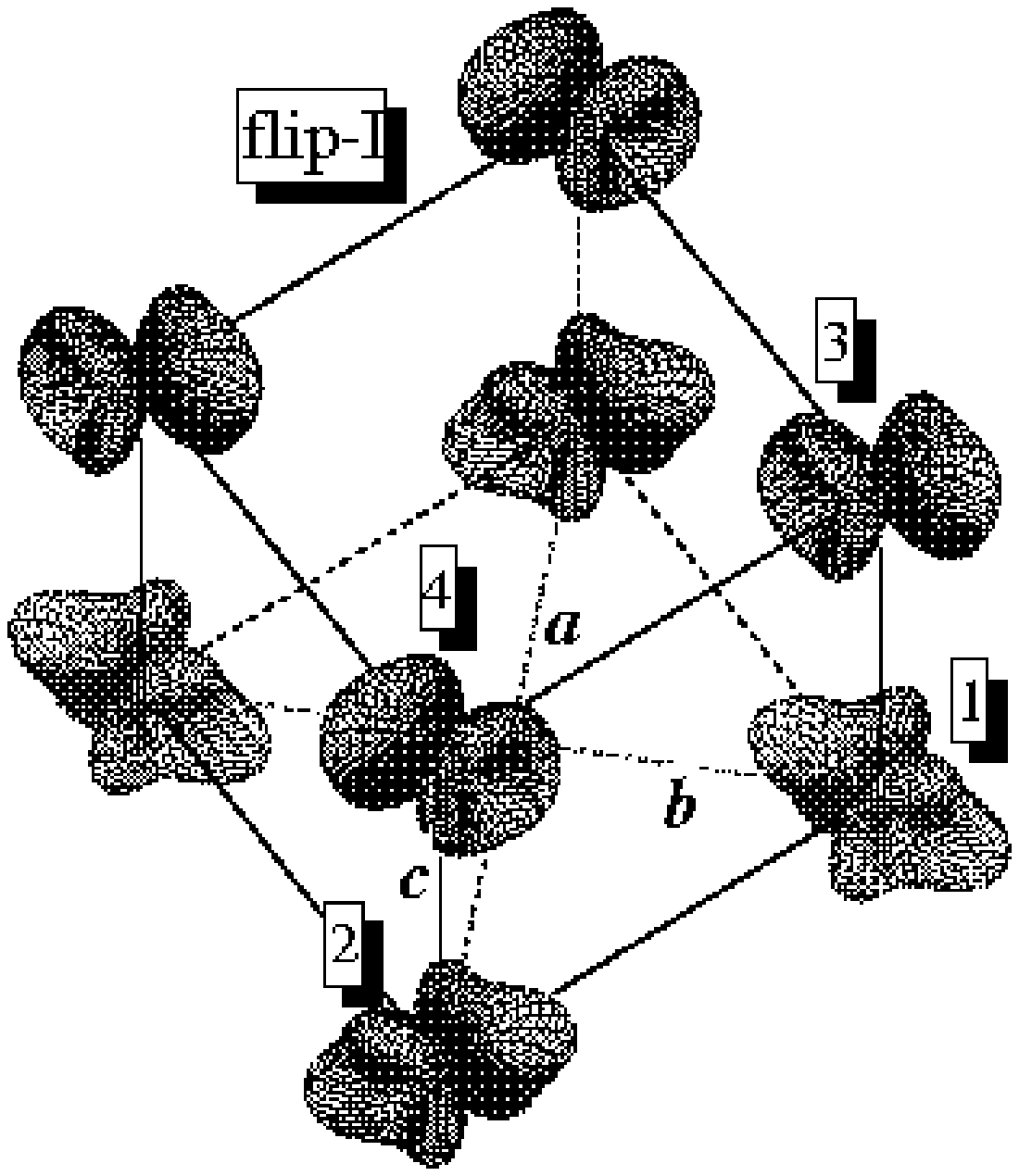}}
\end{center}
\begin{center}
\resizebox{5cm}{!}{\includegraphics{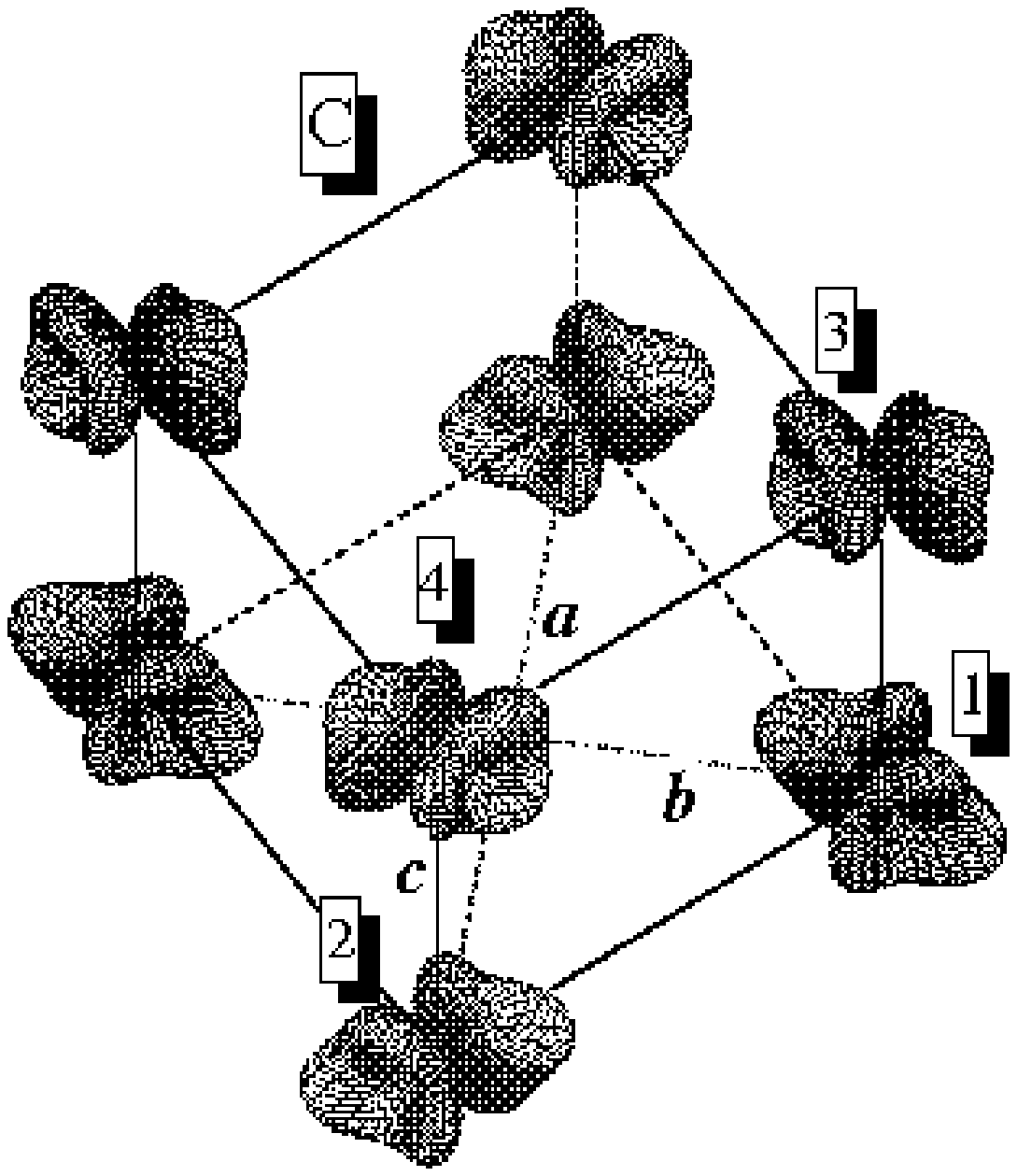}}
\resizebox{5cm}{!}{\includegraphics{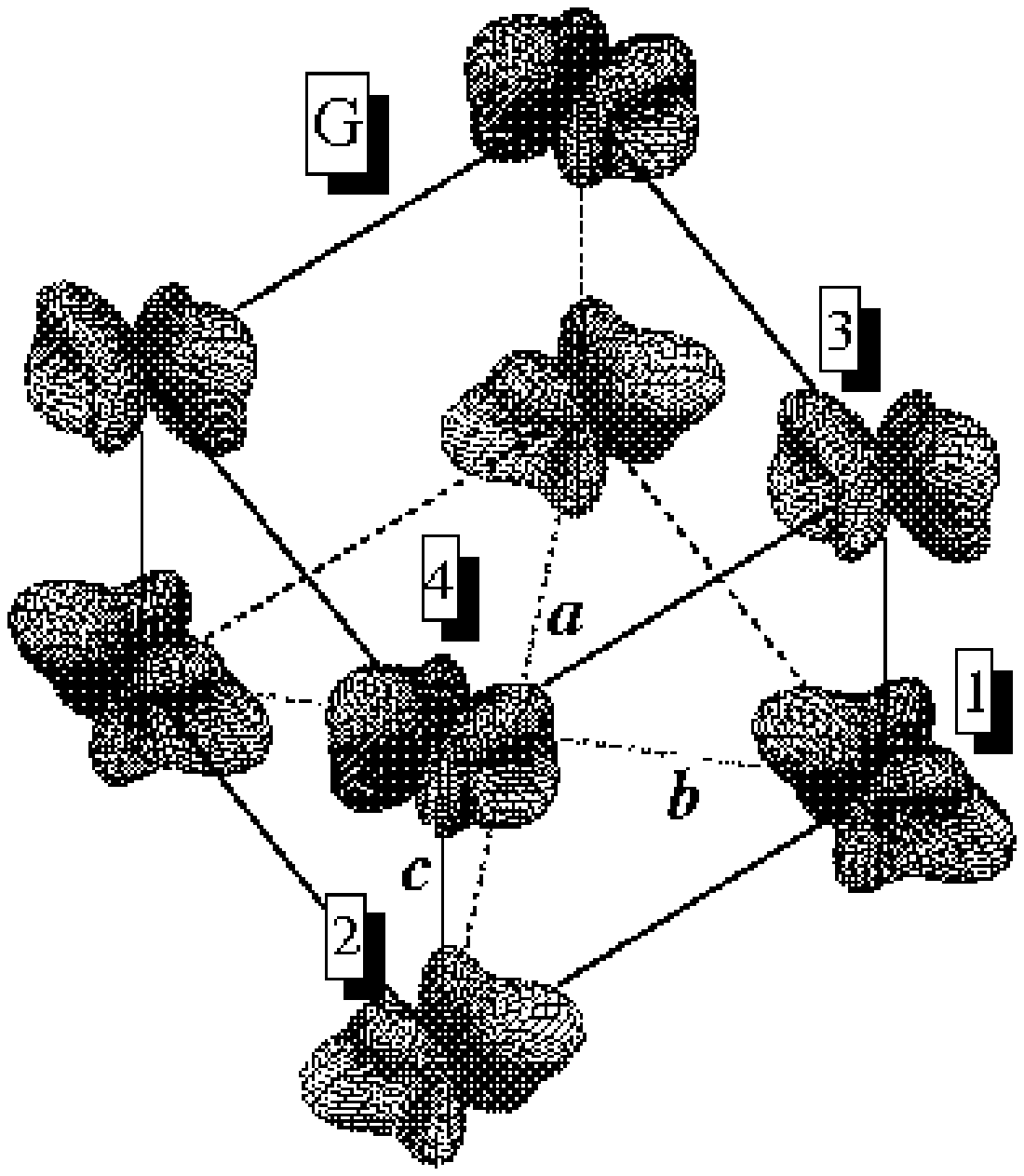}}
\resizebox{5cm}{!}{\includegraphics{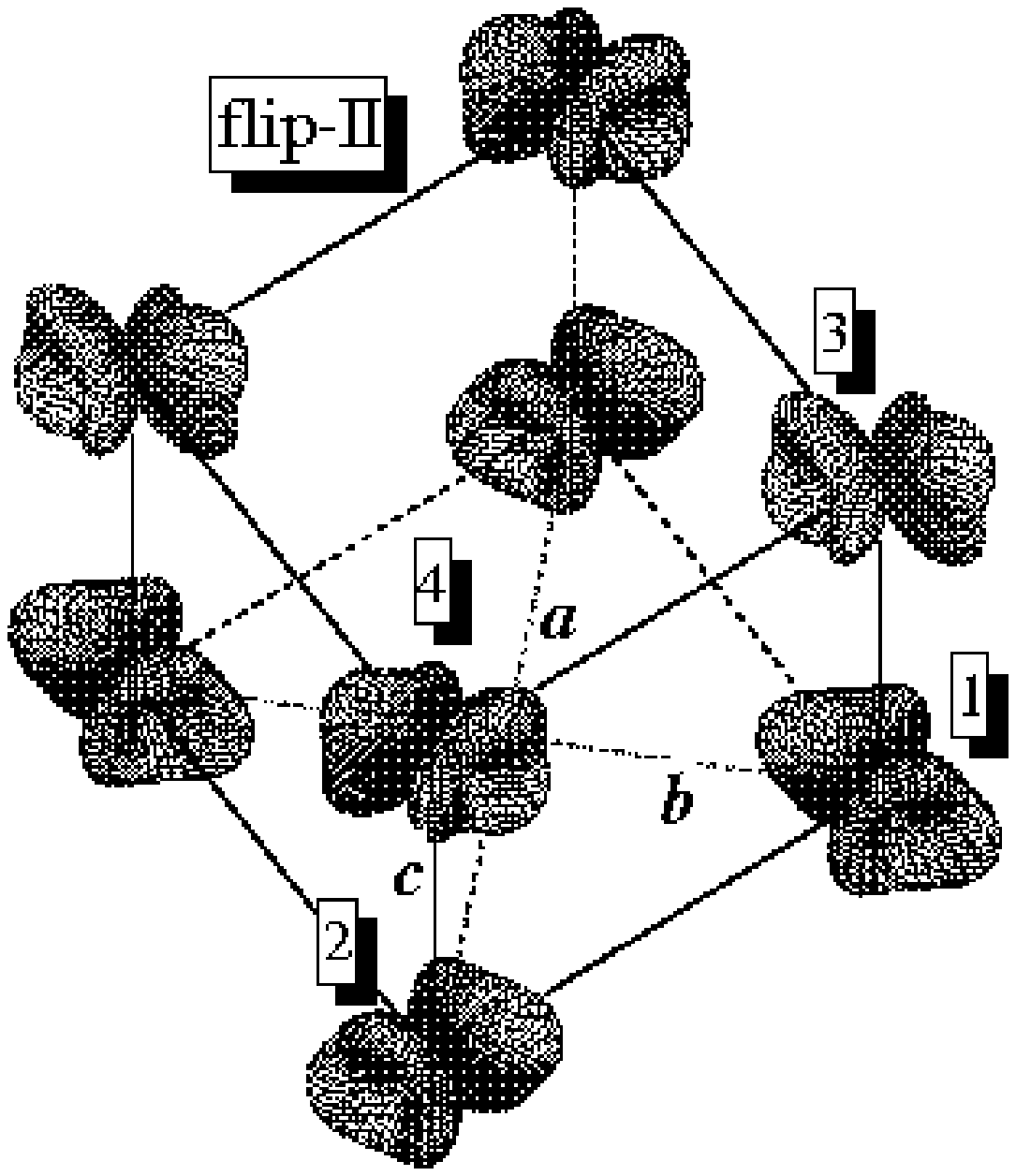}}
\end{center}
\caption{\label{fig.OOYVOm} (Color online)
Distribution of the charge density around V-sites
in various magnetic phases
of monoclinically distorted YVO$_3$, as obtained in Hartree-Fock calculations.
Different magnetic sublattices
are shown by different colors.}
\end{figure}

  However, the orbital ordering is not completely quenched by the crystal
distortion and there is a substantial variation of the orbital structure
depending on the magnetic state, which can be seen already in the
distribution of the charge density
in Fig.~\ref{fig.OOYVOm}.
Nevertheless,
the basic G-type orbital ordering pattern is clearly seen in all magnetic structures.
This appears to be sufficient to stabilize the
experimentally observed
C-type AFM phase, which becomes the ground state already in the HF approach
(Table~\ref{tab:YVOm}). The correlation effects play a very important role.
They additionally stabilize the C-type AFM ground state and reverse the
order of other magnetic states (e.g., F and A in Table~\ref{tab:YVOm}).
\begin{table}[h!]
\caption{Magnetic interactions ($J_{{\bf RR}'}$),
Hartree-Fock energies ($E_{\rm HF}$), and total energies ($E_{\rm tot}$)
in the monoclinic phase of YVO$_3$ ($77$ K $< {\rm T} <$ $116$ K).
See Table~\protect\ref{tab:YVOo} for other notations.}
\label{tab:YVOm}
\begin{ruledtabular}
\begin{tabular}{ccccccc}
 phase  & $J_{12}$           & $J_{13}$  & $J_{24}$  & $J_{34}$  & $E_{\rm HF}$     &  $E_{\rm tot}$   \\
\hline
   F    & $\phantom{-}$$0.1$ & $2.7$     & $2.7$     & $-$$3.5$  & $11.7$           & $17.5$           \\
   A    & $-$$0.3$           & $2.5$     & $2.5$     & $-$$3.4$  & $14.0$           & $17.1$           \\
   C    & $-$$0.9$           & $2.2$     & $2.2$     & $-$$4.5$  & $0$              & $0$              \\
   G    & $-$$1.6$           & $1.8$     & $1.8$     & $-$$4.8$  & $\phantom{1}6.6$ & $\phantom{1}7.2$ \\
 flip-I & $-$$1.8$           & $1.6$     & $2.5$     & $-$$3.4$  & $11.5$           & $14.2$           \\
flip-II & $-$$0.4$           & $2.3$     & $2.8$     & $-$$4.8$  & $\phantom{1}4.4$ & $\phantom{1}6.5$ \\
\end{tabular}
\end{ruledtabular}
\end{table}

  The orbital ordering in the plane (3,4) clearly reminiscent the one
observed in the orthorhombic phase (Fig.~\ref{fig.OOYVO}).
The shape of orbitals in the plane (1,2) appears to be more distorted.

 The behavior of interatomic magnetic interactions
in the high-temperature phase of YVO$_3$ has attracted a
considerable attention recently.
The experimental
spin-wave spectrum shows a clear splitting into
acoustic and optical branches, which are separated
by a 5 meV gap in the middle,
${\bf q}_m$$=$$(0,0,\pi/2c)$,
point of the first Brillouin zone
along the [001] direction.\cite{Ulrich2003}
The splitting has been initially attributed to the
dimerization effect associated with an
orbital Peierls
state, which should lead to the alternation of the strong and weak ferromagnetic bonds along the
${\bf c}$ axis.\cite{Ulrich2003}
However, more recently the puzzling features of the experimental
spectra have be naturally explained by the difference of the exchange coupling
constants in the adjacent ${\bf ab}$-planes,
which is expected for the monoclinic $C^5_{2h}$
symmetry.\cite{FangNagaosa}
This effect is clearly seen in our HF calculations:
while the AFM exchange coupling in the plane (3,4)
does not change so much in comparison with the orthorhombic phase,
the one in the plane (1,2) drops by almost $4$ meV (referring to
the C-type AFM state in Table~\ref{tab:YVOm}).
The value of the gap in the point ${\bf q}_m$
can be estimated in the linear spin-wave theory as
$\Delta_{\rm SW}$$=$$2 J_{13} |\sqrt{1-4J_{12}/J_{13}}$$-$$
\sqrt{1-4J_{34}/J_{13}}|$.
Then, using results of HF calculations
we obtain $\Delta_{\rm SW}$$\approx$$6.2$ meV, which is
in fair agreement with the experimental finding.
We can further speculate that the ferromagnetic coupling $J_{13}$
is overestimated in the HF approximation due to the
lack of the correlations effects (in the next section we
shall see that this is indeed the case for the totally
ferromagnetic YTiO$_3$).
Therefore, the correlation effects will probably yield a
smaller value of $\Delta_{\rm SW}$, which is proportional to $J_{13}$.

\subsection{\label{sec:YTO}YTiO$_3$}

  YTiO$_3$ is a $d^1$ compound. The lattice distortion splits off
one $t_{2g}$-level to the low-energy part of the spectrum (Fig.~\ref{fig.CFsummary}).
This is just enough for trapping one $d$ electron at each Ti-site.
Therefore, the situation is similar to YVO$_3$.
The lowest $t_{2g}$-level is separated from the middle one by a
109 meV gap, which is comparable with the magnitude of the
CF splitting in YVO$_3$.

\begin{figure}[t!]
\begin{center}
\resizebox{5cm}{!}{\includegraphics{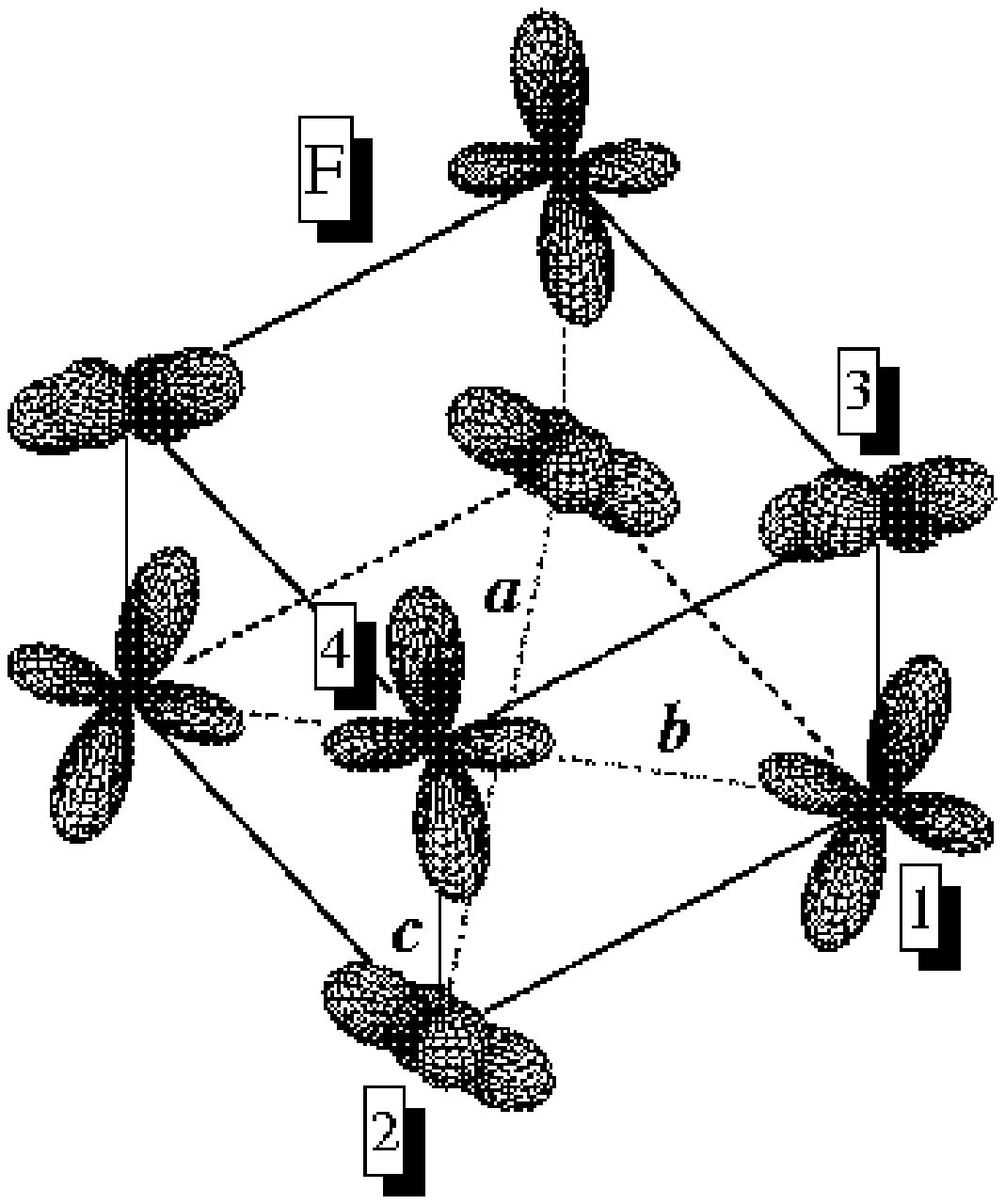}}
\resizebox{5cm}{!}{\includegraphics{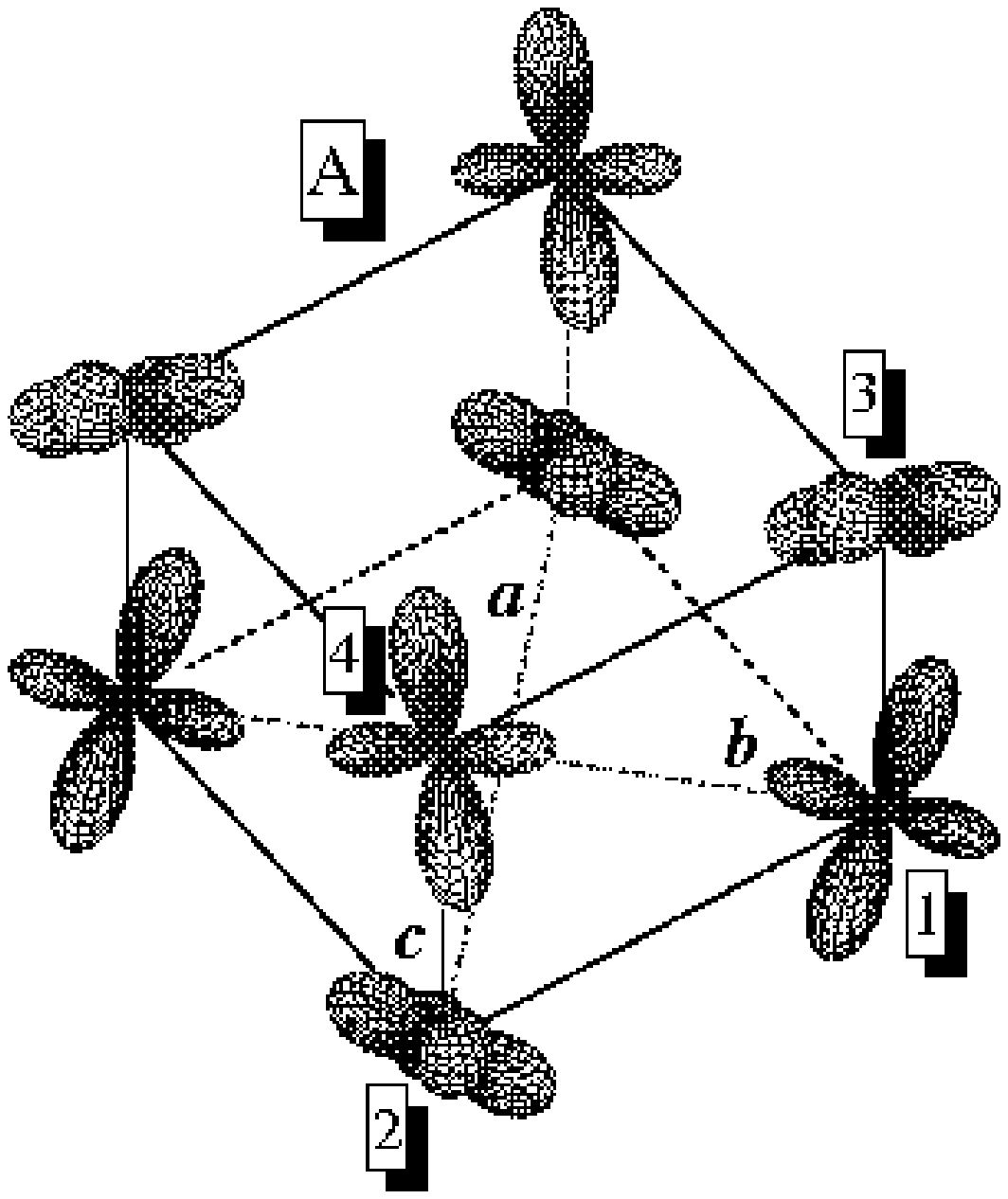}}
\resizebox{5cm}{!}{\makebox{ }}
\end{center}
\begin{center}
\resizebox{5cm}{!}{\includegraphics{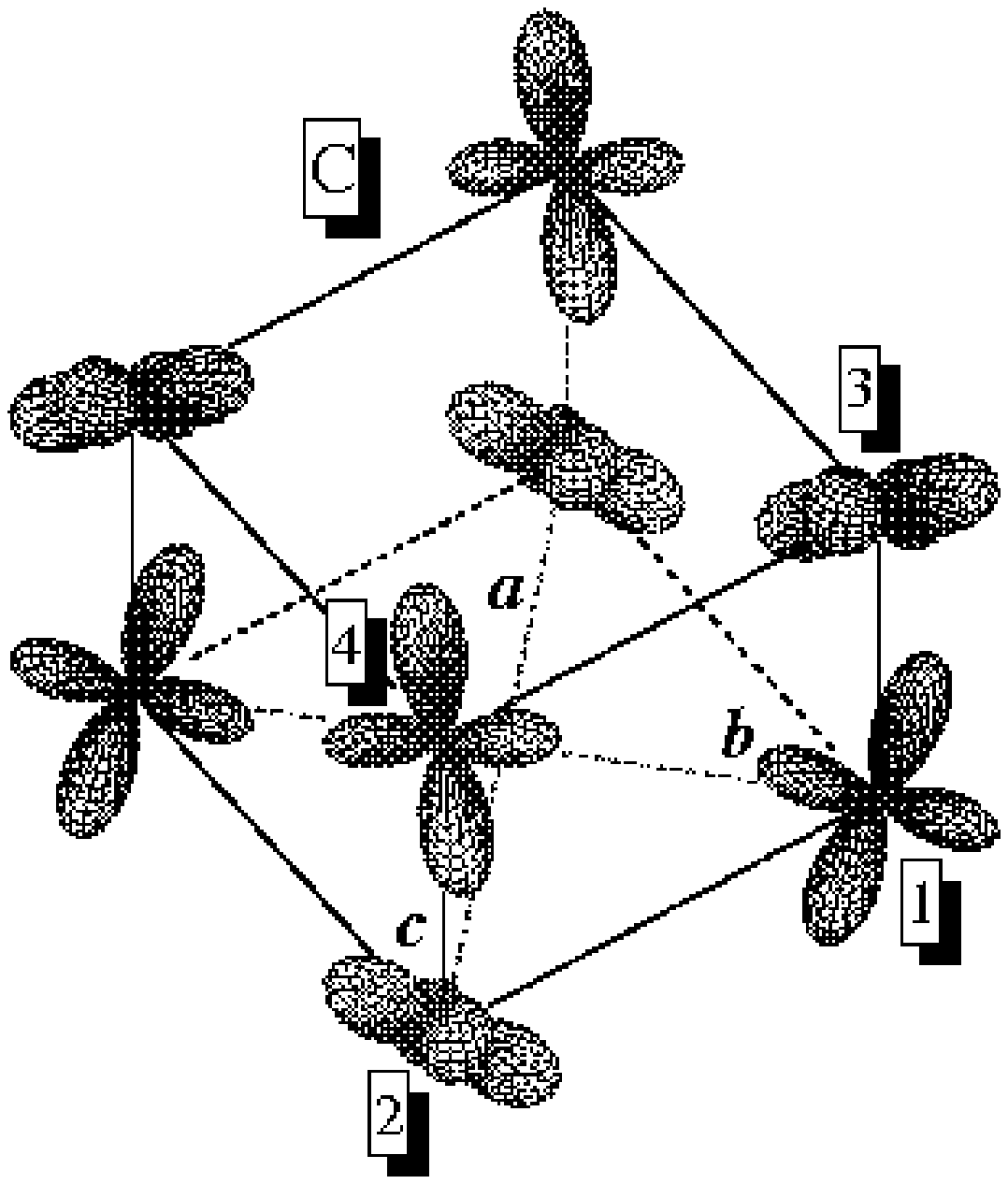}}
\resizebox{5cm}{!}{\includegraphics{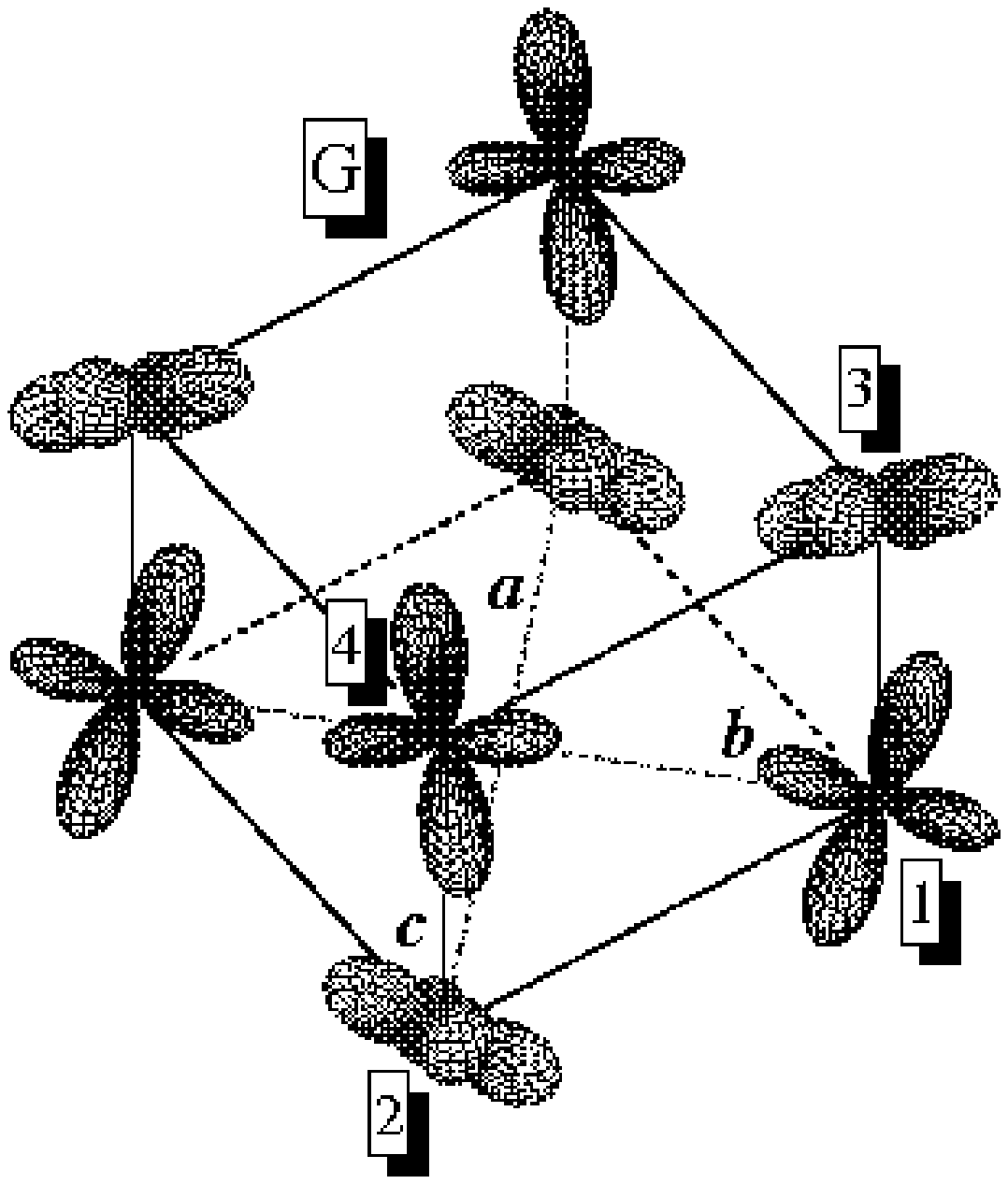}}
\resizebox{5cm}{!}{\includegraphics{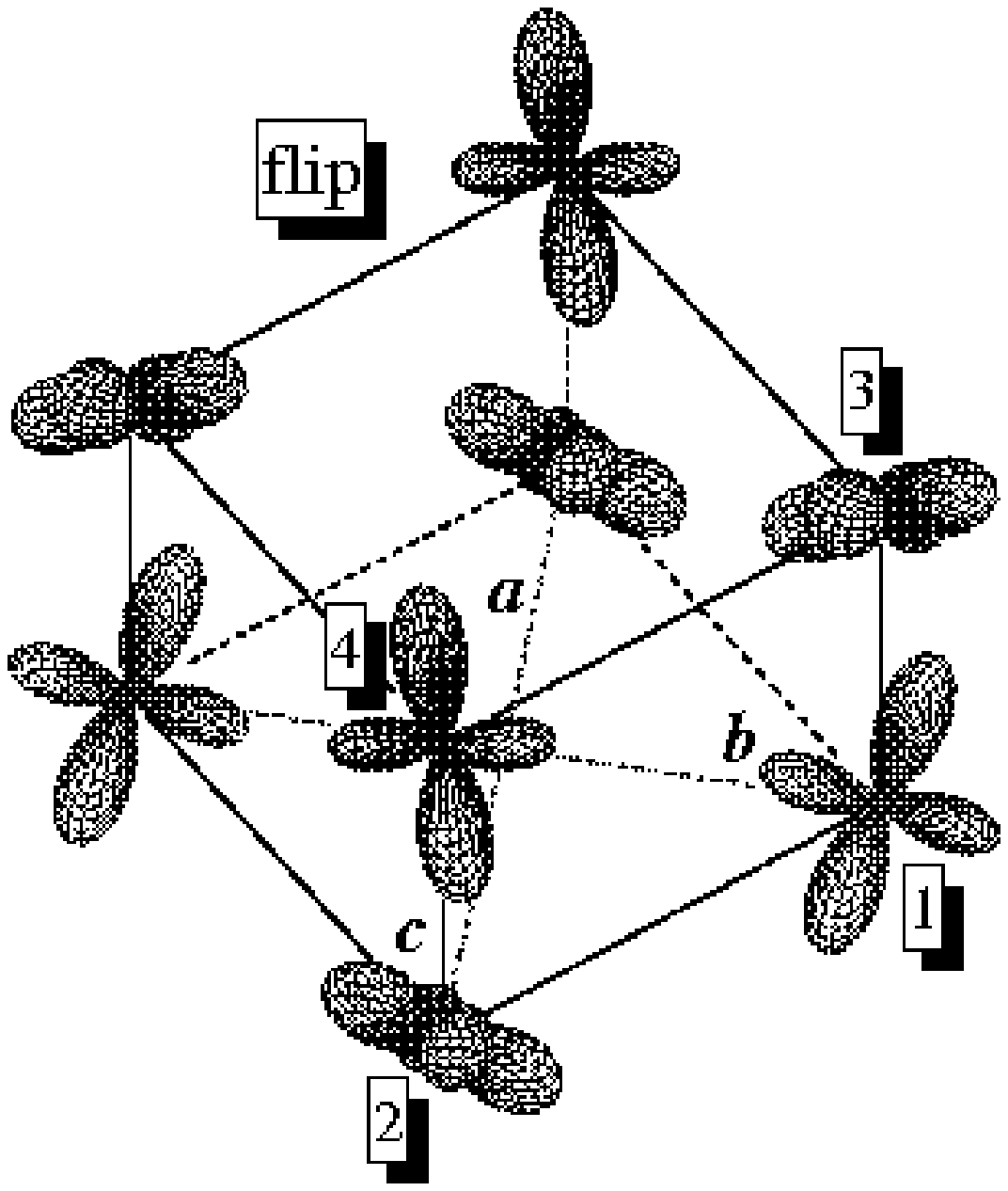}}
\end{center}
\caption{\label{fig.OOYTO} (Color online)
Distribution of the charge density around Ti-sites
in various magnetic phases
of YTiO$_3$, as obtained in Hartree-Fock calculations.
Different magnetic sublattices
are shown by different colors.}
\end{figure}

  The lattice distortion stabilizes the orbital ordering, which is
shown in Fig.~\ref{fig.OOYTO}. In this case, the orbital ordering
is strongly quenched by the distortion and not only the
charge density but also the parameters of the magnetic interactions
(Table~\ref{tab:YTO}),
which are sensitive to the orbital ordering, only weakly depend
on the magnetic state.
\begin{table}[h!]
\caption{Magnetic interactions ($J_{{\bf RR}'}$),
Hartree-Fock energies ($E_{\rm HF}$), total energies ($E_{\rm tot}$), and
superexchange energies ($E_{\rm SE}$)
in YTiO$_3$. $E_{\rm SE}$ is defined as the total energy in the
superexchange approximation.
Other definitions are explained in Table~\protect\ref{tab:YVOo}.}
\label{tab:YTO}
\begin{ruledtabular}
\begin{tabular}{cccccccc}
 phase  & $J_{12}$ & $J_{13}$ & $J_{24}$ & $J_{34}$ & $E_{\rm HF}$     & $E_{\rm tot}$    & $E_{\rm SE}$     \\
\hline
   F    & $3.9$    & $1.2$    & $1.2$    & $3.9$    & $0$              & $0$              & $0$              \\
   A    & $3.9$    & $1.1$    & $1.1$    & $3.9$    & $\phantom{1}2.0$ & $\phantom{1}0.8$ & $\phantom{1}0.1$ \\
   C    & $3.2$    & $1.1$    & $1.1$    & $3.2$    & $14.4$           & $10.9$           & $12.1$           \\
   G    & $3.2$    & $1.0$    & $1.0$    & $3.2$    & $16.2$           & $12.5$           & $13.9$           \\
 flip   & $3.2$    & $1.1$    & $1.1$    & $3.9$    & $\phantom{1}8.2$ & $\phantom{1}6.1$ & $\phantom{1}6.7$ \\
\end{tabular}
\end{ruledtabular}
\end{table}
Partly, this is because YTiO$_3$ has the largest on-site Coulomb
interaction ${\cal U}$ amongst considered perovskite oxides
(Table~\ref{tab:Kanamori}). Therefore, the superexchange
contribution to the orbital ordering is expected to be smaller in
comparison with the CF splitting. The obtained orbital ordering is
in the excellent agreement with results of LDA$+$$U$ calculations by
Sawada and Terakura,\cite{SawadaTerakura} and the experimental
measurement using the nuclear magnetic resonance (Ref.~\onlinecite{Itoh}),
the polarized neutron diffraction
(Ref.~\onlinecite{Akimitsu}), and the soft x-ray linear dichroism
(Ref.~\onlinecite{Iga}).

  The observed orbital ordering patter is compatible with the ferromagnetic
ground state, which easily emerges at the level of HF
theories.\cite{SawadaTerakura,MizokawaFujimori}
The same trend is clearly seen in our calculations, and both the
order F$\rightarrow$A$\rightarrow$C$\rightarrow$flip$\rightarrow$G
and the
total-energies difference between different
magnetic states obtained in the HF approximation are well consistent with
the LDA$+$$U$ results of Sawada and Terakura.\cite{SawadaTerakura}

  The correlation effects are important. Similar to YVO$_3$, they
tend to additionally stabilize the AFM configurations and
destabilize the ferromagnetic ground state. The situation is very
fragile, and after taking into account the correlation effects,
the energy difference between
F state and the next A-type AFM state
becomes only $0.1$-$0.8$ meV per one formula unit.
For the $d^1$ compounds, we can apply two independent schemes
for studying
the correlation effects: one is the second order perturbation theory
and the other one is the theory of superexchange interactions, which takes into
account the multiplet structure of the excited states. For YTiO$_3$,
both scheme yields very consistent results, apart from the
small quantitative difference, which is inevitable for
two different approximations.

  Yet, the magnetic behavior of YTiO$_3$ poses several open questions,
which are not fully understood.

\begin{figure}[t!]
\begin{center}
\resizebox{5cm}{!}{\includegraphics{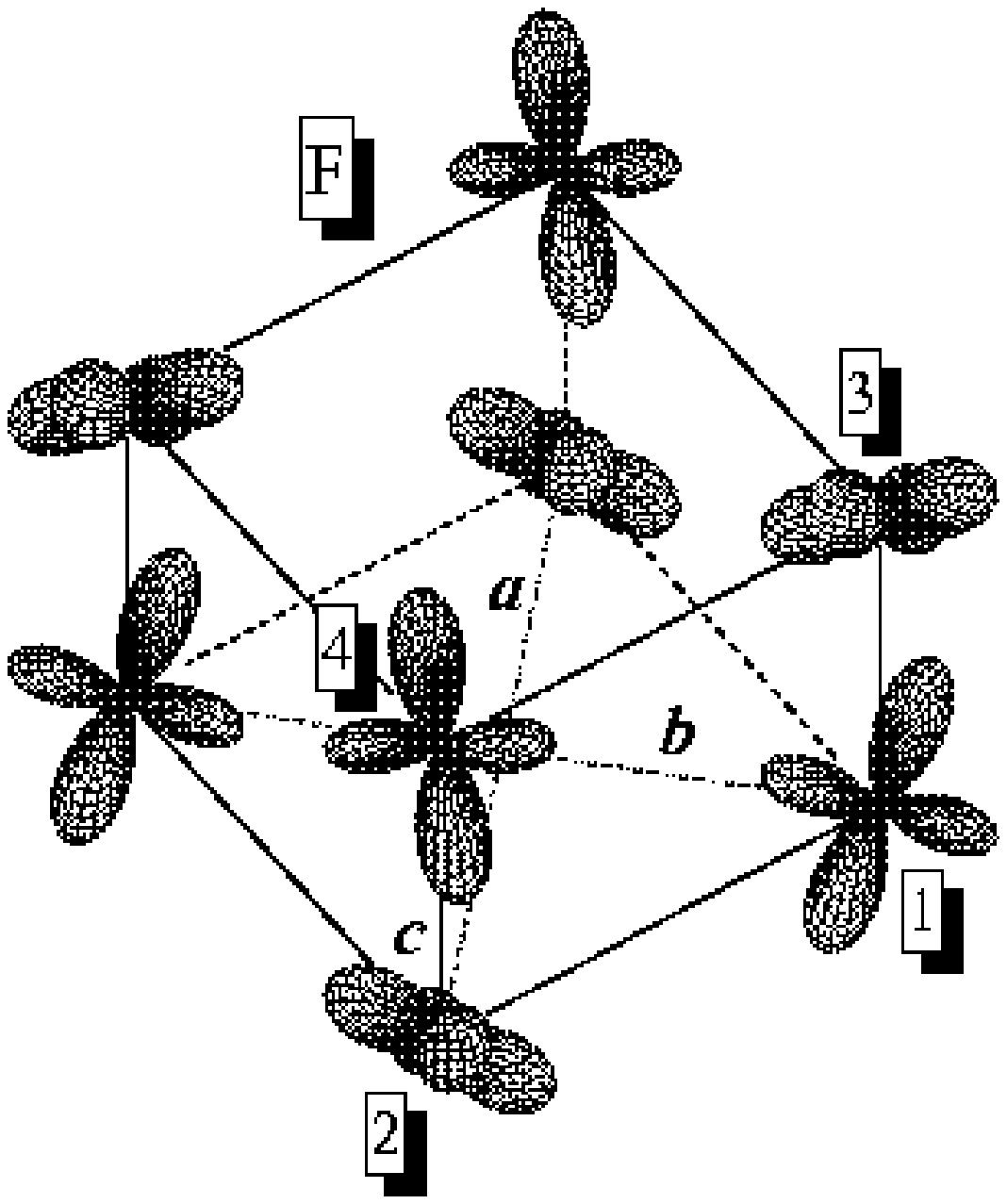}}
\resizebox{5cm}{!}{\includegraphics{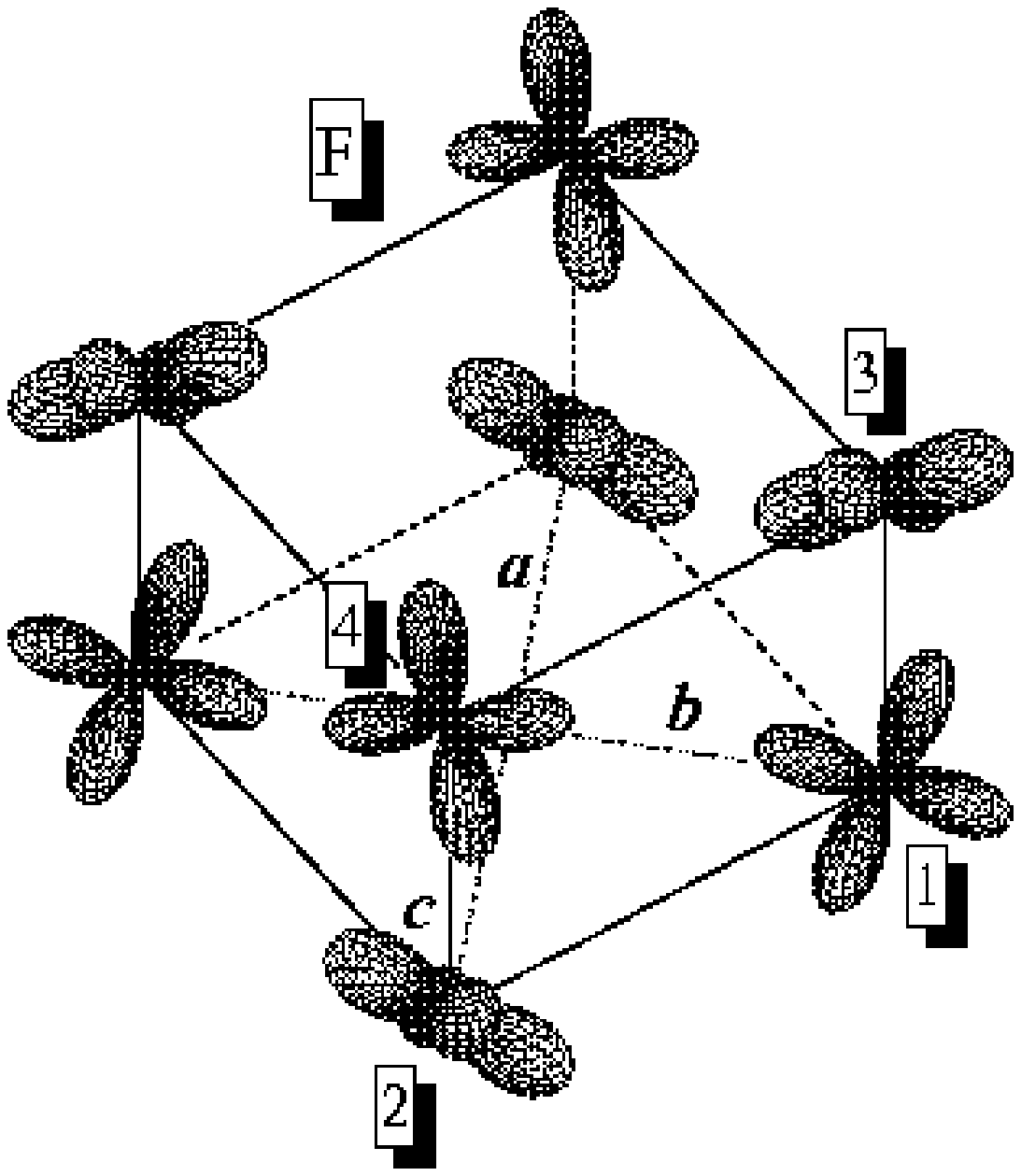}}
\end{center}
\caption{\label{fig.OOYTiOcomparison} (Color online)
Orbital ordering in the ferromagnetic phase of YTiO$_3$ computed with (left)
and without (right)
crystal-field splitting.}
\end{figure}

  First of all, YTiO$_3$ is an exceptional example amongst
$t_{2g}$ perovskite oxides, because the
ferromagnetic ground state can be
anticipated already on the basis of the canonical Goodenough-Kanamori-Anderson
rules for the superexchange interactions in the simple cubic structure.
This immediately revives the idea of Kugel and Khomskii about the
superexchange-driven orbital ordering and the concomitant
Jahn-Teller distortion, which was intensively
discussed in the context of KCuF$_3$.\cite{KugelKhomskii}
Then, one may ask whether the experimental orbital ordering in YTiO$_3$ can be stabilized
by the pure superexchange mechanism, without appealing to the CF splitting.
This can be easily checked by substituting $h_{\bf RR}^{\alpha \beta}$$=$$0$
into kinetic-energy part of the model Hamiltonian.
Surprisingly, the orbital ordering \textit{in the ferromagnetic state} is practically
the same with and without the CF splitting (Ref.~\ref{fig.OOYTiOcomparison}).
The interatomic magnetic interactions, $J_{12}$$=$$4.3$ meV and
$J_{13}$$=$$1.6$ meV, are also consistent with the data listed in
Table~\ref{tab:YTO}, and which include the effects of the CF splitting.
This naturally explain results of
our previous work (Ref.\onlinecite{PRB04}),
where similar orbital ordering and
interatomic magnetic interactions have been
obtained without the Madelung term in the CF splitting.
Therefore, it is tempting to conclude that the orbital ordering
in YTiO$_3$ is driven by the superexchange interactions, and the lattice
distortion simply follows the anisotropic distribution of the
charge density associated with this orbital state.
However, the situation is not so simple, because our calculations have been
performed in the room temperature structure
(T$=$$293$ K, Ref.~\onlinecite{Maclean}), which is much higher than the Curie
temperature (T$_{\rm C}$$\approx$$30$ K, Ref.~\onlinecite{Akimitsu}). Furthermore, the
orbital ordering shown in Fig.~\ref{fig.OOYTiOcomparison} in the absence of the
CF splitting takes places only in the ferromagnetic phase:
had we changed the magnetic state,
our orbital ordering would have been also different.
Therefore, a more plausible scenario for YTiO$_3$ is that the lattice distortion goes first
and sets up a particular form of the CF splitting, which
stabilizes the experimental
orbital ordering pattern and the ferromagnetic ground state.
Nevertheless, the good agreement between two orbital states shown in
Fig.~\ref{fig.OOYTiOcomparison} is really curious. Is it a simple coincidence
or there is some physical meaning behind this result?
We would like to emphasize again that the situation is totally
different from YVO$_3$ shown in Fig.~\ref{fig.OOYVOcomparison}.

  Another group of questions is related with the behavior of
interatomic magnetic interactions.
The first puzzling feature is the
nearly isotropic
experimental
spin-wave spectrum reported by Ulrich~\textit{et~al},\cite{Ulrich2002}
which cannot be explained in terms interatomic magnetic
interactions derived from the
first-principles
electronic structure calculations.
For example, the exchange parameters listed in Table~\ref{tab:YTO}
are clearly anisotropic, and the interactions along the ${\bf c}$-axis are much
weaker than in the ${\bf ab}$-plane.
A similar conclusion is expected from the analysis of the total-energy
differences reported
by Sawada and Terakura.\cite{SawadaTerakura}
Since the anisotropy of magnetic interactions is directly related
with the particular form of the orbital ordering, it seems that the inelastic neutron-scattering
data by Ulrich~\textit{et al.} are in an apparent disagreement not only with the
results of the first-principles
electronic structure calculations but also with a number of other experimental
data, which report the same type of the orbital ordering.\cite{Itoh,Akimitsu,Iga}
Therefore, what is so special about the
inelastic neutron-scattering measurements and why different experimental methods
probing the orbital state yield qualitatively different conclusions in the case of YTiO$_3$?

  The behavior of interatomic magnetic interactions
predetermines not only in the form of the spin-wave spectrum, but also in the
absolute value of T$_{\rm C}$. If the magnetic properties of YTiO$_3$
are indeed controlled by the large lattice distortion, which is set up far above T$_{\rm C}$,
it should to be a good Heisenberg ferromagnet. This is directly seen
in our HF calculations, where the parameters of interatomic magnetic
interactions practically do not depend on the magnetic state (Table~\ref{tab:YTO}).
Then,
the applicability of the Heisenberg model
is no longer restricted by infinitesimal rotations of the spin magnetic moments,
and
T$_{\rm C}$ can be easily evaluated using the standard expressions, which are
well known from the theory of Heisenberg magnets.\cite{Nagaev}
The simplest one is the mean-field formula:
$3k_{\rm B}{\rm T_C^{MF}}$$=$$4J_{12}$$+$$2J_{13}$,
where the prefactor $S(S$$+$$1)$ is already included in the definition of
our exchange parameters, although with some approximations for the
spin $1/2$.\cite{comment.1}
By combining this expression with the HF approximation for the exchange
interactions, one finds ${\rm T_C^{MF}}$$=$$64$ K, which exceeds the experimental
value by factor two.
However, ${\rm T_C^{MF}}$ does not include spontaneous fluctuations
and correlations between the motion of the neighboring spins.
This is exactly the point where the anisotropy of exchange interactions can help
to reduce the theoretical value of ${\rm T_C}$.
Indeed,
according to the Mermin-Wagner theorem,\cite{MerminWagner}
the two-dimensional Heisenberg model
does not support any long-range spin order
at any nonzero temperature.
Therefore,
since for $J_{13}/J_{12}$$\ll$$1$ the system will eventually approach the
two-dimensional limit,
it is reasonable to expect a
substantial reduction of ${\rm T_C}$.
In order to describe these effects quantitatively,
one can use
the spherical approximation for the Heisenberg model,\cite{Nagaev} according to which
$3k_{\rm B}{\rm T_C}$$=$$1/[ \sum_{\bf k} (J_0$$-$$J_{\bf k})^{-1} ]$,
$J_{\bf k}$$=$$\sum_{{\bf R}'} J_{{\bf RR}'} e^{i {\bf k} \cdot {\bf R}'}$,
and the summation over ${\bf k}$ is restricted by the first Brillouin zone.
Then, using parameters extracted from the HF calculations we obtain  ${\rm T_C}$$=$$36$ K,
which can be further reduced by taking into account the correlation effects.
For example, by mapping the total energies obtained in the second order of
perturbation theory and in the theory of superexchange interactions onto the
Heisenberg model, we obtain ${\rm T_C}$$=$ $27$ and $28$ K, respectively,
which is in fair agreement with the experimental data.

\subsection{\label{sec:LVO}LaVO$_3$}

  The monoclinic LaVO$_3$ has two inequivalent V-sites. Contrary to YVO$_3$,
these sites differ not only by the direction, but also by the magnitude of the
CF splitting (Table~\ref{fig.CFsummary}), which is $78$ and $152$ meV for the sites
1 and 3, respectively (referring to the splitting between middle and highest
$t_{2g}$ levels). Therefore, already from this very simple analysis of the CF splitting
one can expect very different behavior of the orbital degrees of freedom
in different ${\bf ab}$-planes: the strong quenching
in the plane
(3,4), and a relative flexibility in the plane (1,2)
(Fig.~\ref{fig.structure}). Thus, the situation is
qualitatively different from YVO$_3$.

\begin{figure}[t!]
\begin{center}
\resizebox{5cm}{!}{\includegraphics{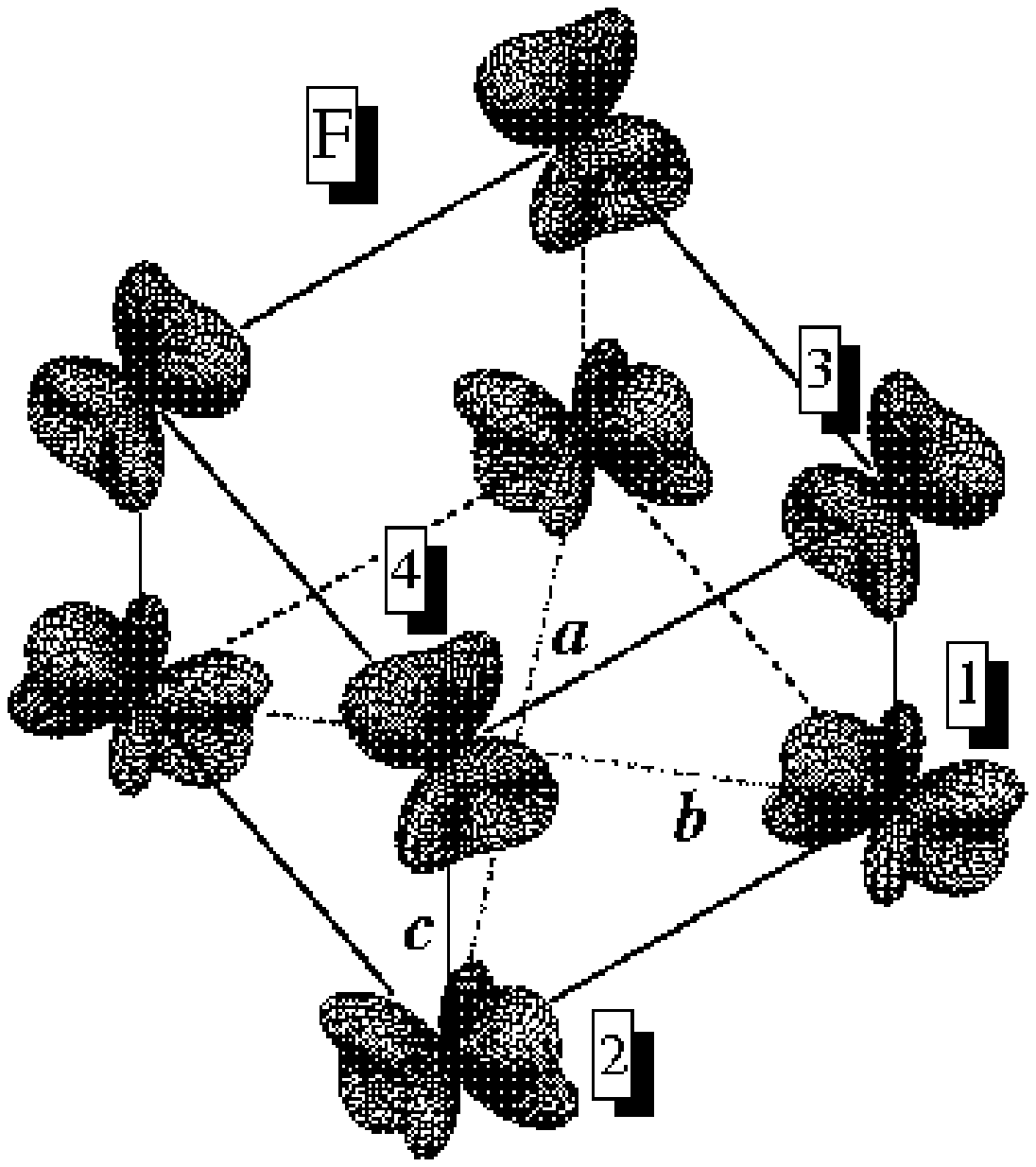}}
\resizebox{5cm}{!}{\includegraphics{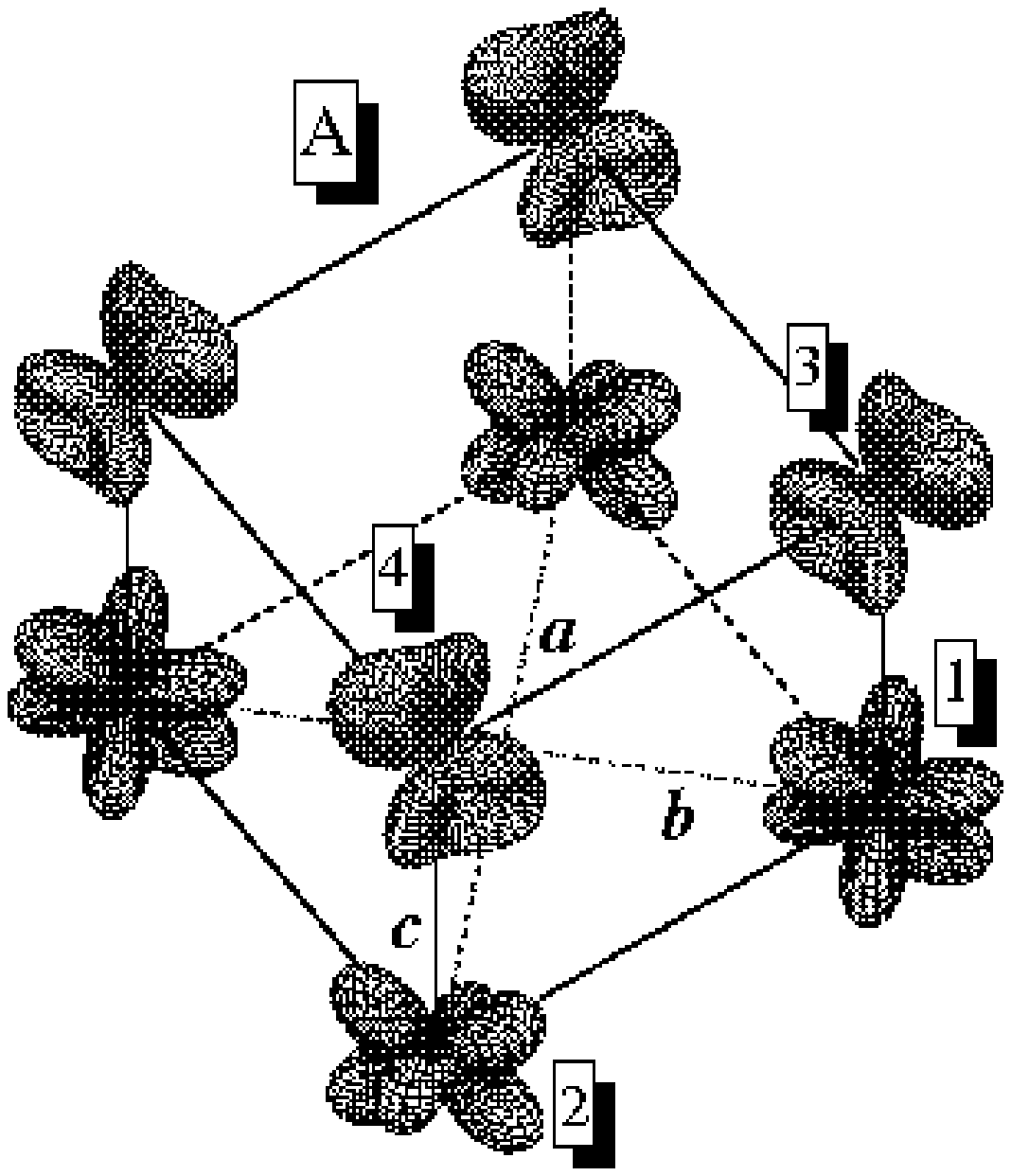}}
\resizebox{5cm}{!}{\includegraphics{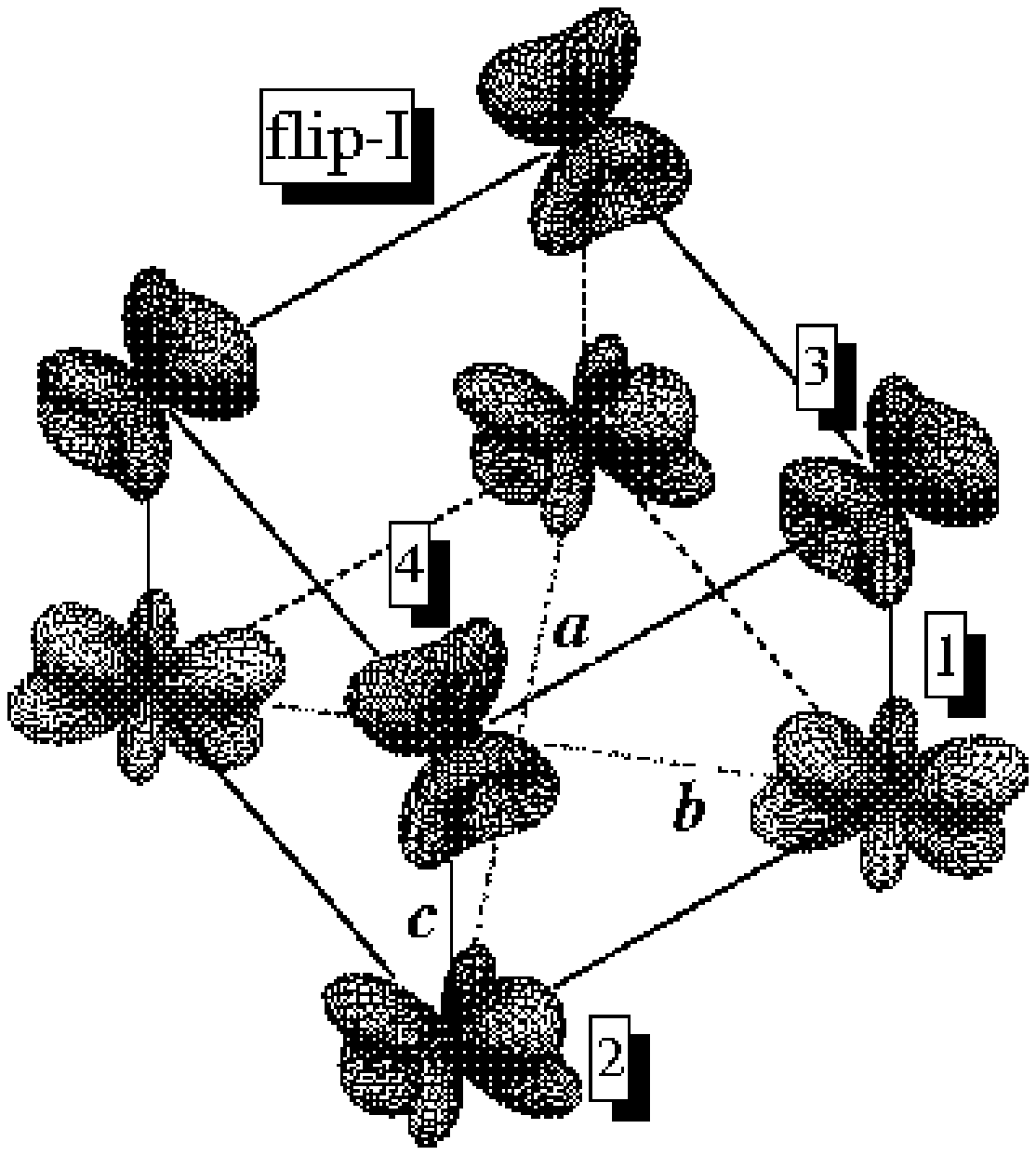}}
\end{center}
\begin{center}
\resizebox{5cm}{!}{\includegraphics{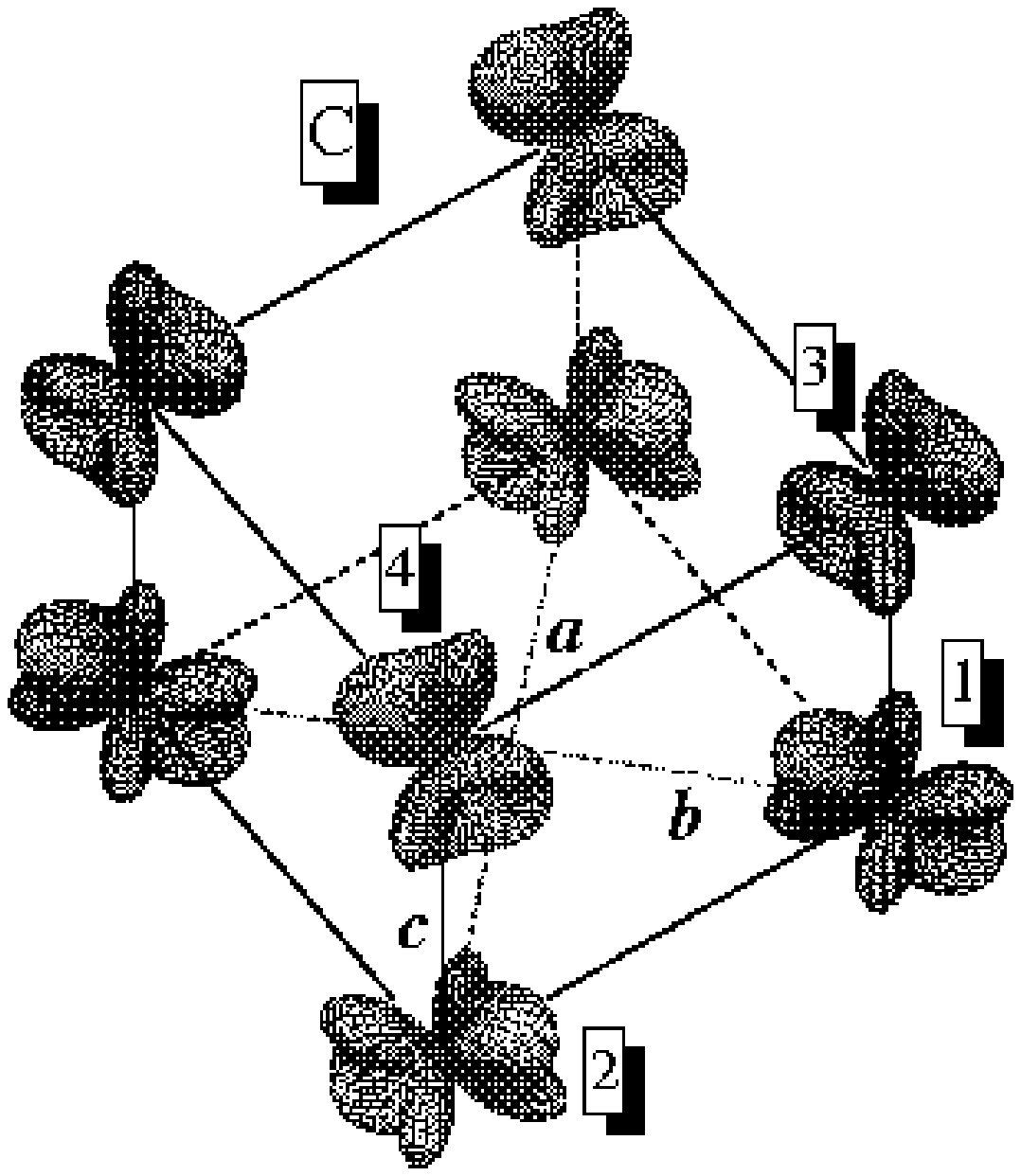}}
\resizebox{5cm}{!}{\includegraphics{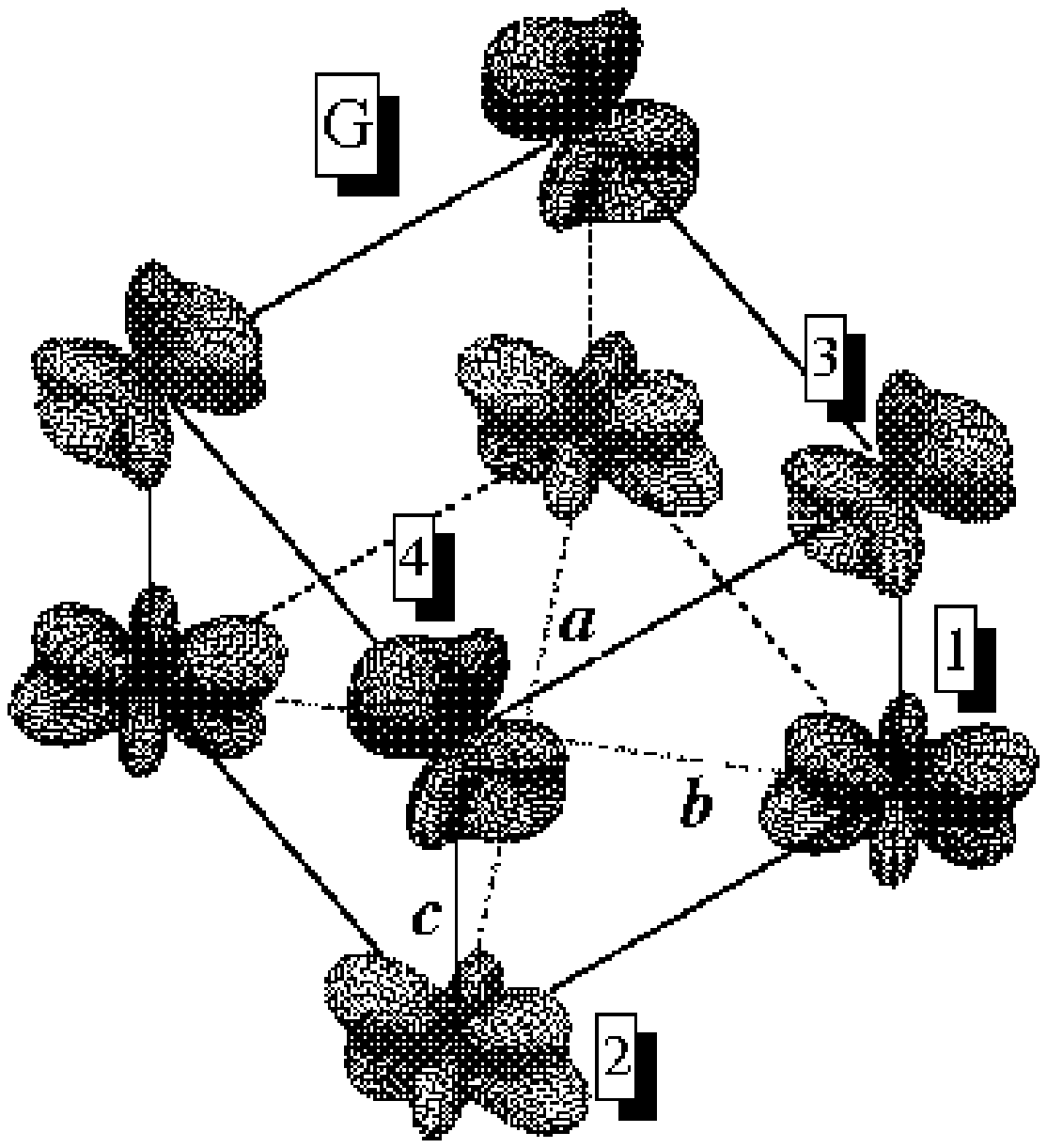}}
\resizebox{5cm}{!}{\includegraphics{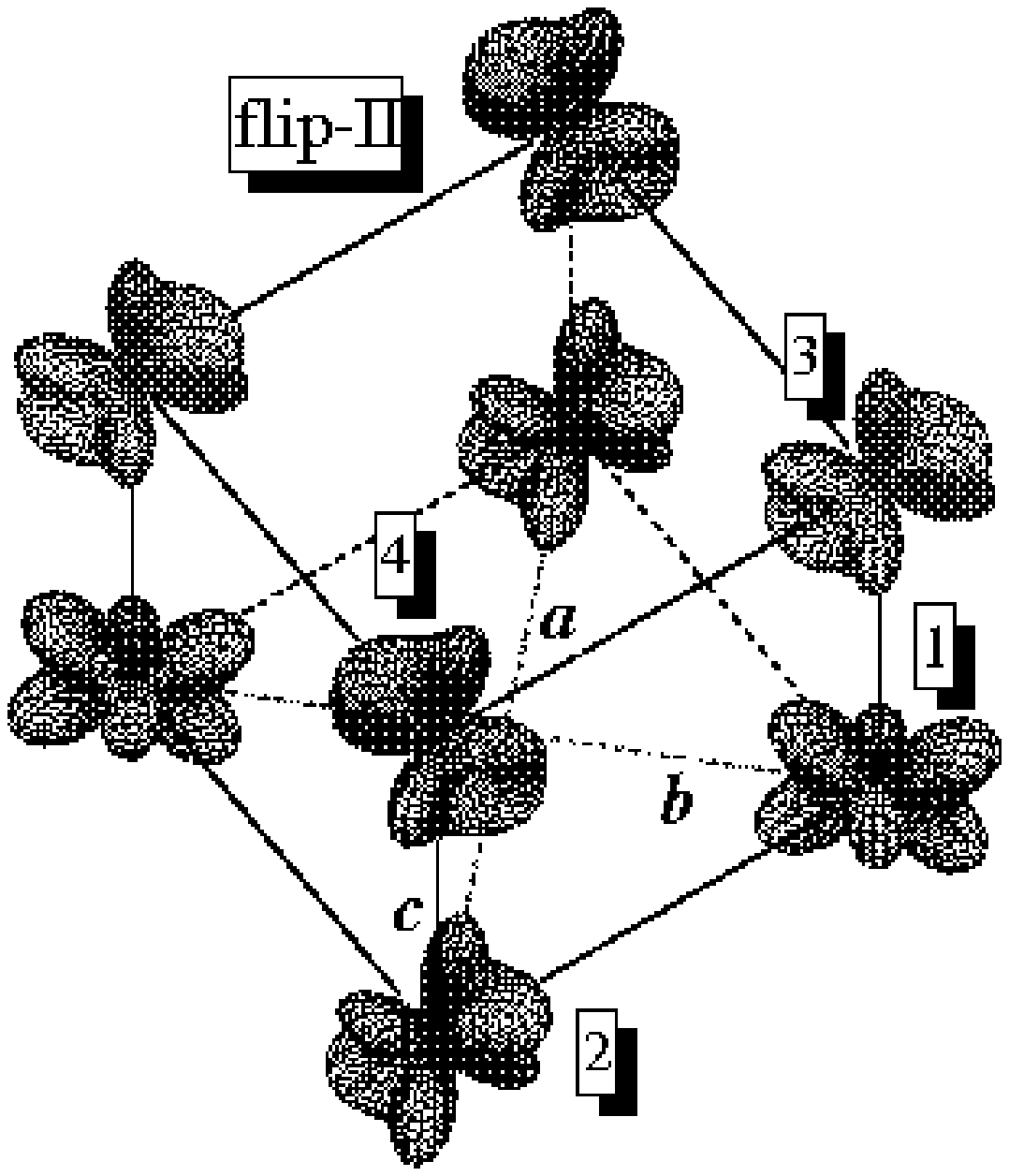}}
\end{center}
\caption{\label{fig.OOLaVOm} (Color online)
Distribution of the charge density around V-sites
in various magnetic phases
of LaVO$_3$, as obtained in Hartree-Fock calculations
.
Different magnetic sublattices
are shown by different colors.}
\end{figure}

  These trends are clearly seen in the HF calculations (Fig.~\ref{fig.OOLaVOm}):
the orbital
ordering in the plane (1,2) strongly depend on the magnetic state
and one can clearly distinguish two types of the orbital-ordering pattern
realized, on the one hand, in the states F and C, and, on the other hand,
in the states A and G.
Therefore, it is perhaps right to say that in LaVO$_3$, the
experimental orbital ordering is partly driven by the magnetic
ordering via the Kugel-Khomskii mechanism.\cite{KugelKhomskii}
This seems to agree with the experimental data,
which show that in
La-based compounds,
the orbital ordering develops few degrees below the
magnetic N\'{e}el temperature (T$_{\rm N}$).\cite{Miyasaka}
This is again different from YVO$_3$, for which the orbital-ordering
temperature is substantially higher than T$_{\rm N}$.

  The change of the orbital ordering is
reflected in the behavior of interatomic magnetic interactions, which not only
depend on the magnetic states, but can even change the signs (Table~\ref{tab:LaVO}).
\begin{table}[h!]
\caption{Magnetic interactions ($J_{{\bf RR}'}$),
Hartree-Fock energies ($E_{\rm HF}$), and total energies ($E_{\rm tot}$)
in LaVO$_3$. See Table~\protect\ref{tab:YVOo} for other notations.}
\label{tab:LaVO}
\begin{ruledtabular}
\begin{tabular}{ccccccc}
 phase  & $J_{12}$           & $J_{13}$             & $J_{24}$           & $J_{34}$  & $E_{\rm HF}$     &  $E_{\rm tot}$ \\
\hline
   F    & $-$$5.1$           & $\phantom{-}$$6.6$   & $\phantom{-}$$6.6$ & $-$$1.7$  & $21.0$           & $30.8$         \\
   A    & $\phantom{-}$$3.8$ & $\phantom{-}$$2.1$   & $-$$4.5$           & $-$$2.4$  & $20.6$           & $24.1$         \\
   C    & $-$$4.8$           & $\phantom{-}$$6.0$   & $\phantom{-}$$6.0$ & $-$$1.8$  & $0$              & $0$            \\
   G    & $-$$6.3$           & $-$$4.4$             & $-$$4.4$           & $-$$2.4$  & $\phantom{1}7.6$ & $11.1$         \\
 flip-I & $\phantom{-}$$7.7$ & $\phantom{-}$$0.6$   & $\phantom{-}$$5.8$ & $-$$1.7$  & $\phantom{1}9.8$ & $15.0$         \\
flip-II & $-$$6.2$           & $-$$4.0$             & $\phantom{-}$$5.8$ & $-$$3.1$  & $12.7$           & $16.9$         \\
\end{tabular}
\end{ruledtabular}
\end{table}
In such a situation, the total energy may have several local minima, realized for those
magnetic states where the signs of interatomic magnetic interactions are consistent
with the type of the imposed spin ordering. We have found at least two such minima,
corresponding to the C-type AFM ground state
and the G-type AFM state, which has higher energy. Thus, we do not quite agree with the conclusions
about the complete quenching of the
orbital ordering in these distorted perovskite compounds.\cite{FangNagaosa}
This is not generally true and LaVO$_3$ is clearly an exception.
However, the CF splitting will also prevent the formation of the orbital
singlet states, which were
employed in order to explain the appearance of the C-type AFM antiferromagnetism
in the theory of orbital fluctuations.\cite{Khaliullin01}
Actually, such a singlet state conflicts with the
$C^5_{2h}$ symmetry of the monoclinic phase.

  Since the interatomic magnetic interactions depend on the magnetic state,
the simple Heisenberg model may be used only for the analysis of local
perturbation around each magnetic state. Then, it is reasonable to expect
a gap $\Delta_{\rm SW}$$\approx$$6.8$ meV
between acoustic and optical branches of the spin-wave spectrum, similar to the
one observed in the C-type AFM phase of YVO$_3$.\cite{Ulrich2003}
However, the Heisenberg model is no longer valid for the analysis of the
transition temperature (unlike in YTiO$_3$), which should take into account a
possible change of the the orbital states
in the course of thermodynamic average.

  Similar to YVO$_3$ and YTiO$_3$, the correlation effects play a
very important role also in LaVO$_3$ and additionally stabilize the
C-type AFM ground state.

\subsection{\label{sec:LTO}LaTiO$_3$}

\begin{figure}[t!]
\begin{center}
\resizebox{5cm}{!}{\includegraphics{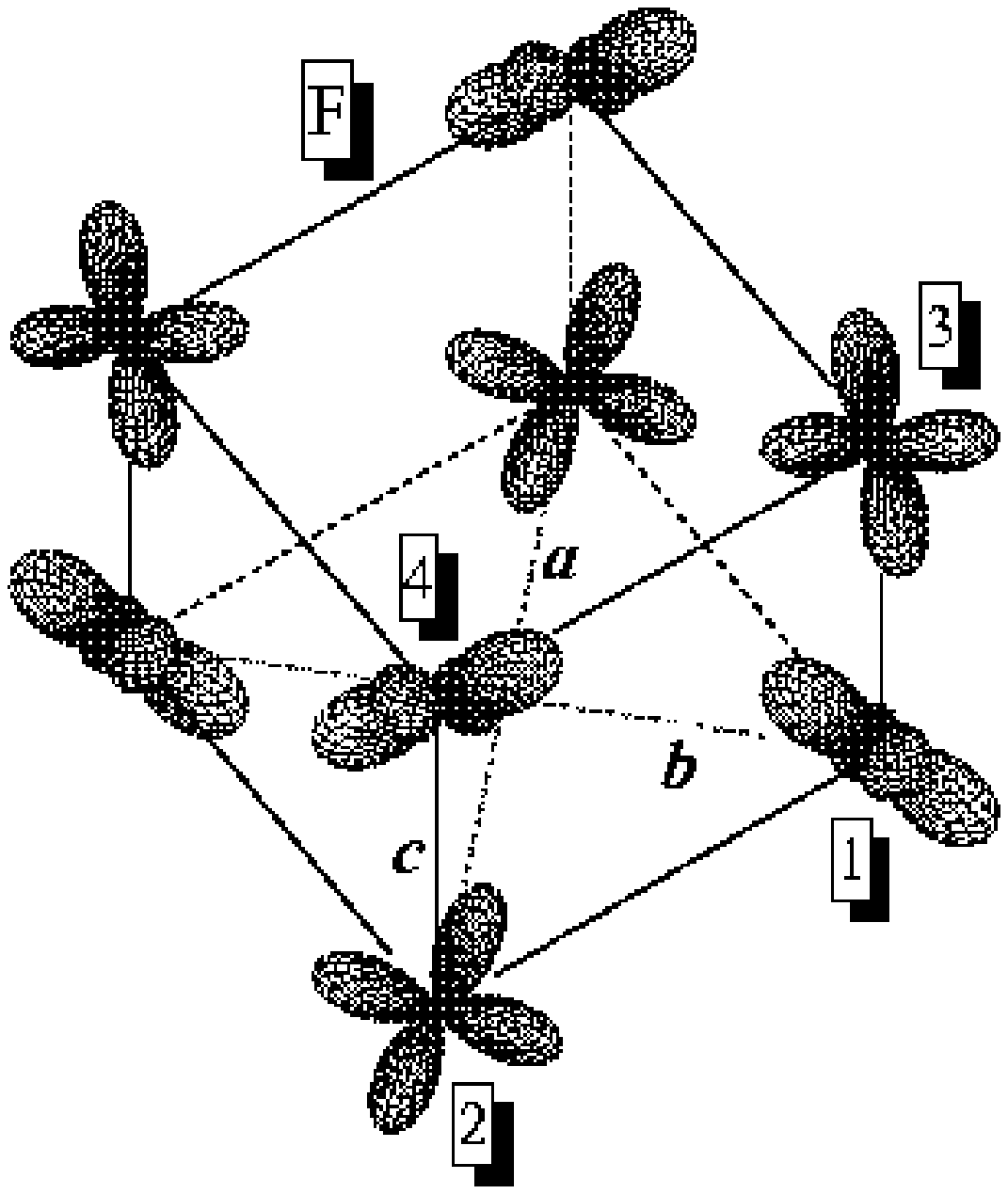}}
\resizebox{5cm}{!}{\includegraphics{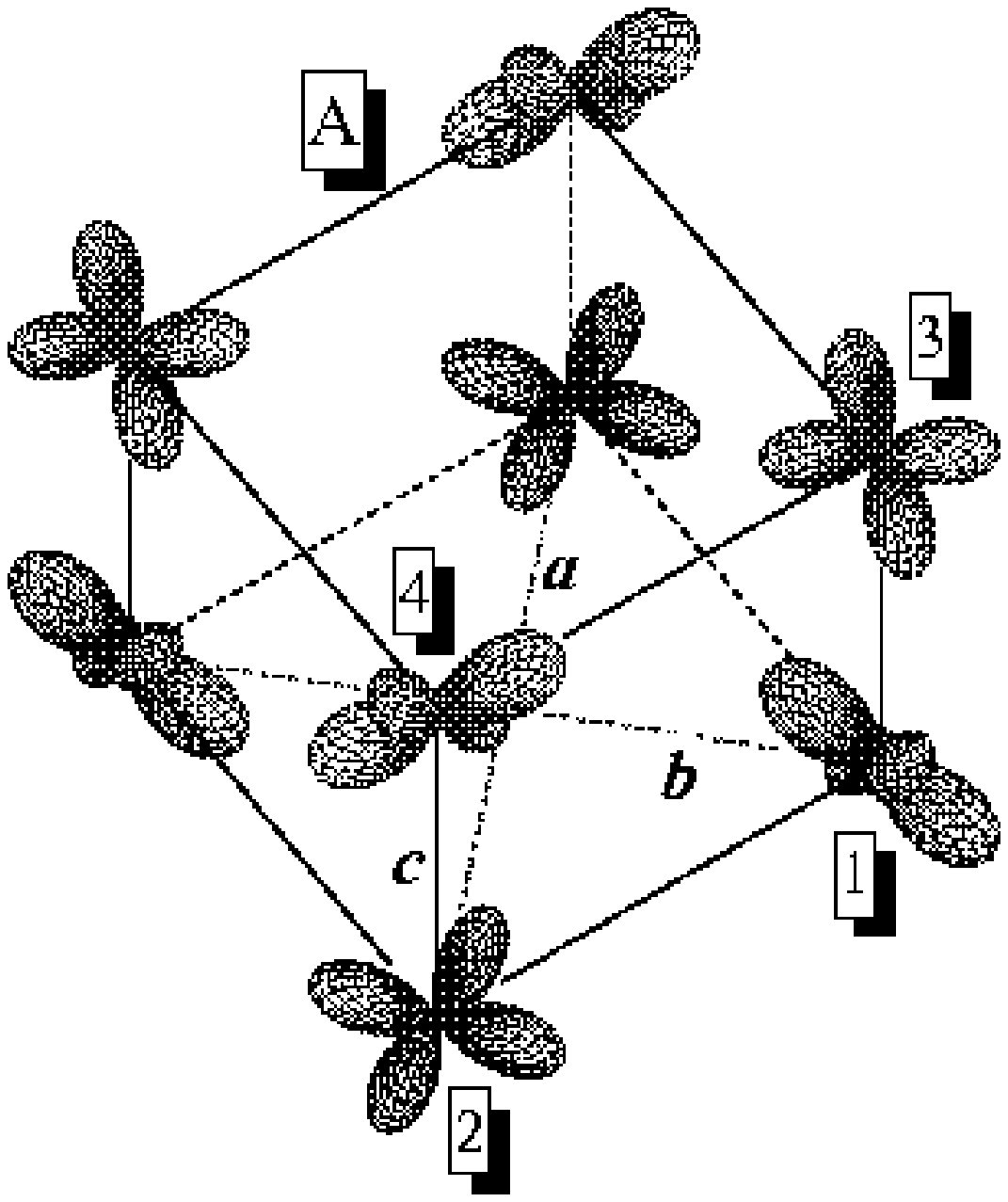}}
\resizebox{5cm}{!}{\makebox{ }}
\end{center}
\begin{center}
\resizebox{5cm}{!}{\includegraphics{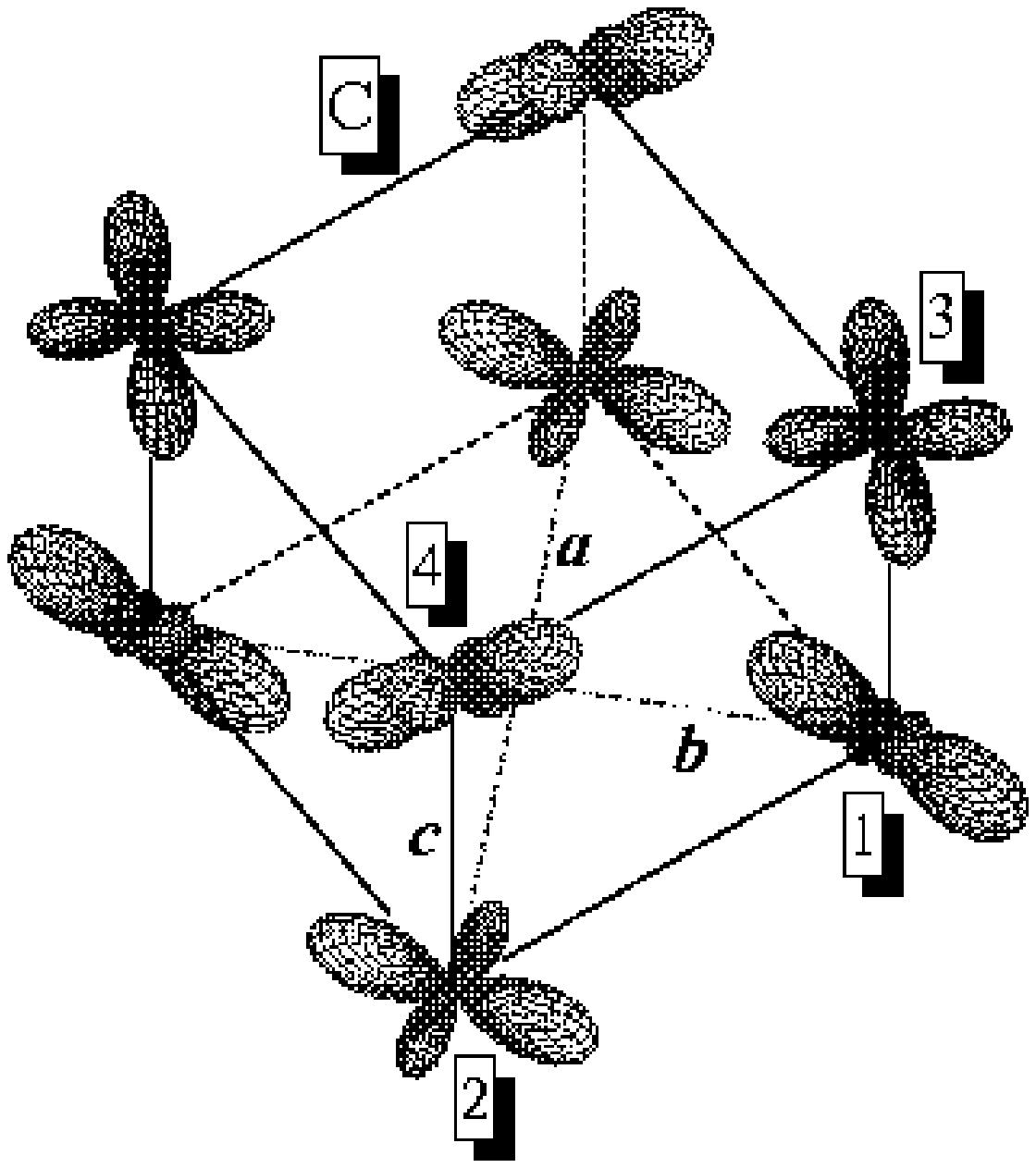}}
\resizebox{5cm}{!}{\includegraphics{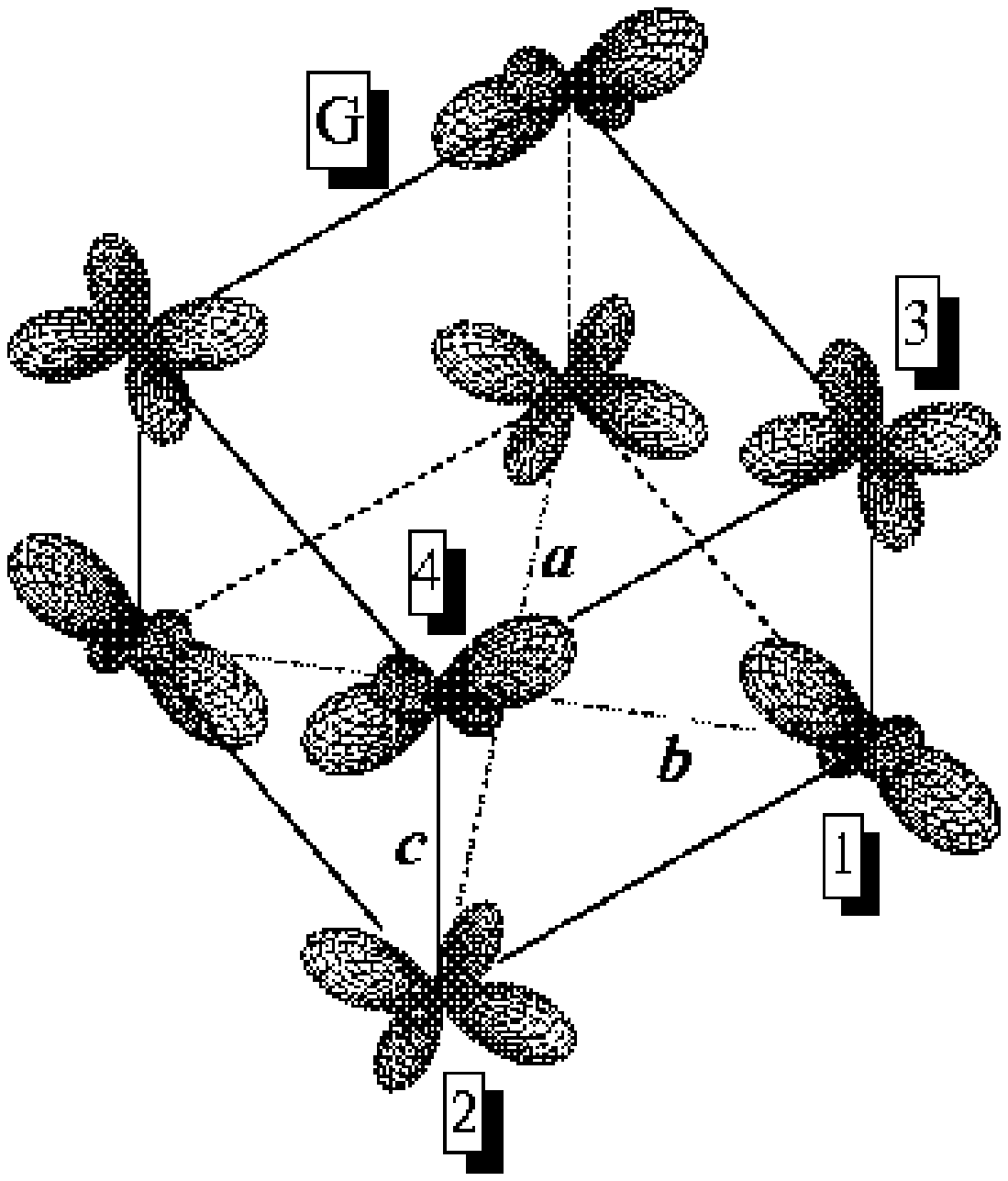}}
\resizebox{5cm}{!}{\includegraphics{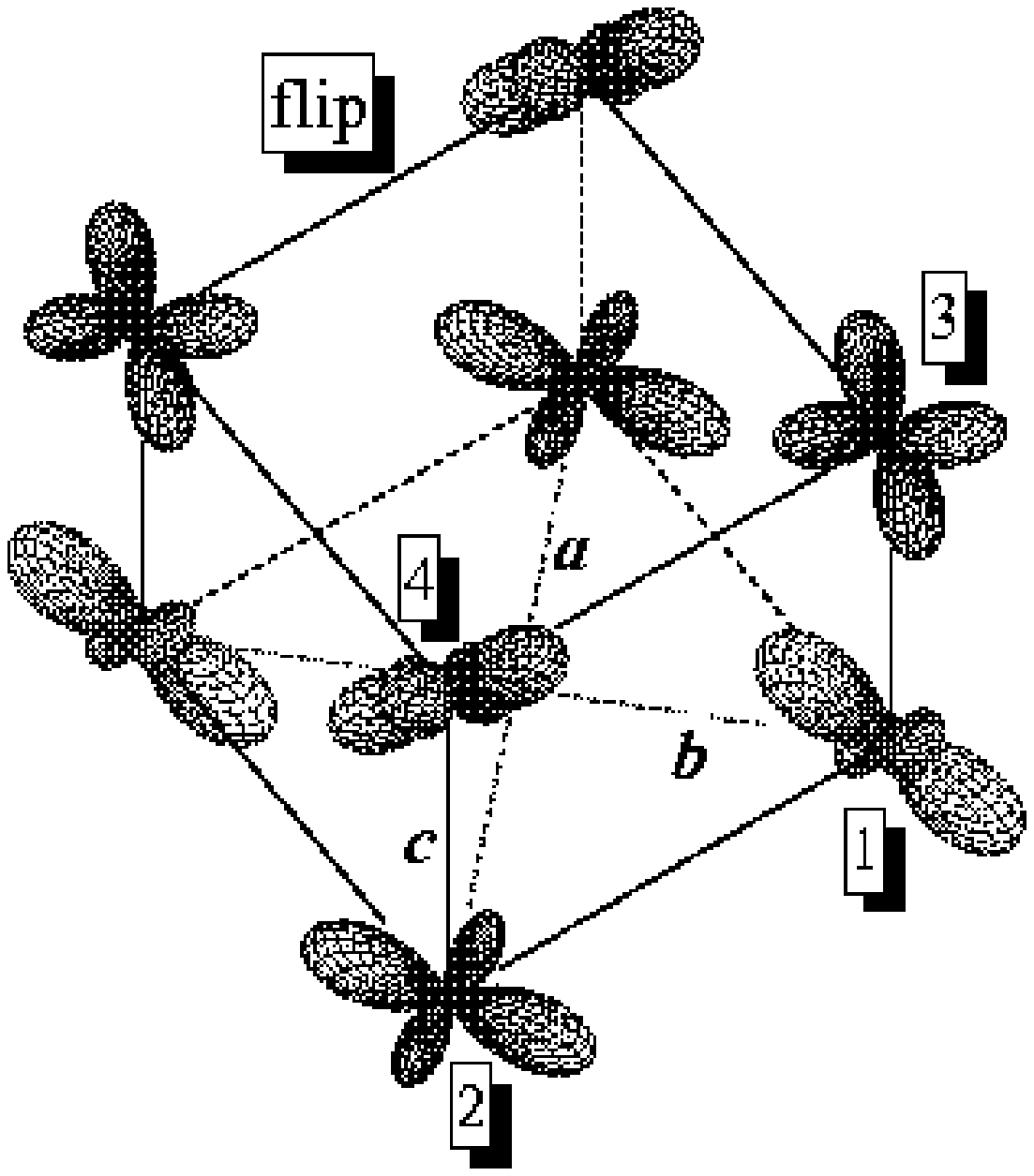}}
\end{center}
\caption{\label{fig.OOLTO} (Color online)
Distribution of the charge density around Ti-sites
in various magnetic phases
of LaTiO$_3$, as obtained in Hartree-Fock calculations.
Different magnetic sublattices
are shown by different colors.}
\end{figure}

  LaTiO$_3$ is a puzzling system. It has the smallest CF splitting
(Fig.~\ref{fig.CFsummary}) among the distorted
perovskite oxides, which formally leaves a room for the orbital
fluctuations. On the other hand, the possible variation of
the orbital order
appears to be bounded by certain
constraint conditions.
For example,
although the orbital ordering depends on the
magnetic state,
this dependence is not particularly large, as it is clearly seen
from the HF calculations,
where the basic shape of the orbital-ordering pattern
remains the same for different magnetic states
(Fig.~\ref{fig.OOLTO}).
There are certainly some variations of the orbital ordering, which can be
seen already on the plot. However, these variations do not seem to change
the qualitative conclusion about the form of interatomic magnetic interactions
and the magnetic ground state of LaTiO$_3$.

  Unfortunately, this conclusion is not consistent with the experimental data.
In this sense, there is a clear difference of LaTiO$_3$ from other perovskite
compounds considered in this work, despite the fact that
we have used absolutely the same procedure for construction and solution of
the model Hamiltonian.

  The magnetic ground state is expected to be of the A-type,
as it is clearly seen from the total-energy calculations as well as from the behavior of
interatomic magnetic interactions (Table~\ref{tab:LTO}),
although experimentally,
it is totally antiferromagnetic G-type.\cite{Keimer}
The magnetic interactions are sensitive to the
magnetic ordering. However, other
magnetic states appear to be unstable and the form of interatomic magnetic interactions in each
magnetic state
systematically leads to the A-type antiferromagnetism.
This conclusion is totally consistent with results of our previous work,\cite{PRB04}
which neglects the Madelung contribution to the CF splitting.
\begin{table}[h!]
\caption{Magnetic interactions ($J_{{\bf RR}'}$),
Hartree-Fock energies ($E_{\rm HF}$), total energies ($E_{\rm tot}$),
and superexchange energies ($E_{\rm SE}$)
in LaTiO$_3$.
See Tables~\protect\ref{tab:YVOo} and \protect\ref{tab:YTO} for other notations.}
\label{tab:LTO}
\begin{ruledtabular}
\begin{tabular}{cccccccc}
 phase  & $J_{12}$ & $J_{13}$ & $J_{24}$ & $J_{34}$ & $E_{\rm HF}$     & $E_{\rm tot}$ & $E_{\rm SE}$     \\
\hline
   F    & $4.5$    & $-$$1.2$ & $-$$1.2$ & $4.5$    & $\phantom{1}5.0$ & $17.7$        & $\phantom{1}1.6$ \\
   A    & $3.6$    & $-$$3.3$ & $-$$3.3$ & $3.6$    & $0$              & $0$           & $0$              \\
   C    & $1.0$    & $-$$2.0$ & $-$$2.0$ & $3.4$    & $19.6$           & $26.3$        & $14.2$           \\
   G    & $2.0$    & $-$$4.9$ & $-$$4.9$ & $2.0$    & $11.5$           & $11.0$        & $\phantom{1}9.0$ \\
 flip   & $1.3$    & $-$$4.5$ & $-$$0.4$ & $4.6$    & $\phantom{1}7.7$ & $11.4$        & $\phantom{1}5.7$ \\
\end{tabular}
\end{ruledtabular}
\end{table}

\begin{figure}[t!]
\begin{center}
\resizebox{8cm}{!}{\includegraphics{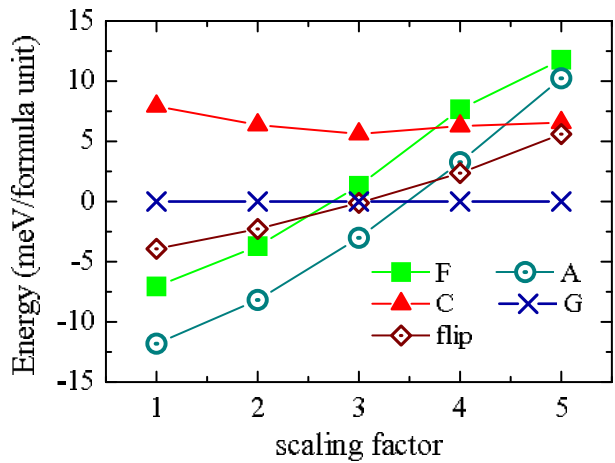}}
\end{center}
\caption{\label{fig.LTOscaling} (Color online)
Total energies measured relative to the G-type
antiferromagnetic state obtained in Hartree-Fock
calculations with exaggerated crystal-field splitting of
the Madelung type. The latter as been multiplied by the
scaling factor shown in the absciss axis.}
\end{figure}

  Then, what is missing in our calculations -- or maybe even more
generally -- in our understanding of the magnetic properties of LaTiO$_3$? Below we discuss
several plausible scenarios.\\
(i) One possibility is that the effect of the crystal distortion maybe still
underestimated. Particularly, we tried to follow the idea of
Refs.~\onlinecite{MochizukiImada} and \onlinecite{Schmitz}, and
additionally
scaled the contribution of the
Madelung
term to the CF splitting by multiplying the right-hand side of
Eq.~(\ref{eqn:CFEI}) by a constant. This corresponds to the change of the
dielectric constant, which was
treated as
an adjustable parameter in Refs.~\onlinecite{MochizukiImada} and \onlinecite{Schmitz}.
We have found that in order to obtain the experimental G-type antiferromagnetic ground
state, the dielectric constant should be reduced by a factor four or five (Fig.~\ref{fig.LTOscaling}).
Then, the exchange interactions become nearly isotropic
($J_{12}$$=$$-$$3.3$ and $J_{13}$$=$$-$$3.5$ meV) and do not depend on the magnetic state.
Hence, LaTiO$_3$ is expected to be a good Heisenberg antiferromagnet,
in agreement
with the experimental inelastic neutron-scattering data. The latter reveals
nearly isotropic behavior of the spin-wave spectrum with
$J_{12} = J_{13} = -$$15.5S^2 \approx -$$3.9$ meV.\cite{Keimer}
Thus, we are totally agree with the authors of
Refs.~\onlinecite{MochizukiImada} and \onlinecite{Schmitz} that the Madelung term alone
could explain many experimental features of LaTiO$_3$.
The only problem is that, according to the electronic
structure calculations, the effect is too small.
One can of course try to blame LDA for this failure.
However, why does this problem occur only for LaTiO$_3$
while for other compounds
our method
works reasonably well?
We believe that if the story about the structural origin of the G-type
antiferromagnetism in LaTiO$_3$ make a sense, it is more likely
that the real magnitude of the crystal distortion in LaTiO$_3$
may be still undisclosed experimentally. This seems to be reasonable, because
the structural data for the distorted perovskite oxides are still in the
process of steady refinement.\cite{Blake,Tsvetkov,Cwik}\\
(ii) Other scenarios are related with the correlation effects, which are
not included in the HF calculations. There is no doubts that they
must play an important role in LaTiO$_3$. However, it seems that
there is no simple scheme which would allow to include these effects
in the electronic structure calculations. The second-order perturbation
theory and the theory of superexchange interactions, which we have tried,
are definitely not enough. They do substantially change the total energies
of the HF method. However, the conclusions strongly depend on the
approximation which we use (Table~\ref{tab:LTO}).
For example, from the second order perturbation theory, it seems to be clear
that the correlation effects tend to stabilize the G-type AFM state:
the correlations change the order of the magnetic states and somewhat lower
the energy of the G-type AFM state relative to the A-state.
The latter trend is also seen the superexchange approach.
However, the superexchange method tends to lower also the energies
of other magnetic states, apparently through the small change of the
orbital ordering.
This is in straight contrast with the more distorted
YTiO$_3$, were two
different methods provided rather
consistent explanation for the
role played by the
correlation effects (Table~\ref{tab:YTO}).\\
(iii) The theory of orbital liquid is just an opposite case to the theory of CF
splitting as these two effects are incompatible with each other.
Although the formation of the orbital liquid in the cubic $t_{2g}$ lattice is
a many-electron effect, the necessary prerequisite, which should exist already at the mean-field level,
is an infinite degeneracy of the magnetic ground state.\cite{KhaliullinMaekawa}
Although our CF splitting for LaTiO$_3$ is small, that is sometimes regarded
as the strong support for the
orbital liquid theory, we do not observe such a degeneracy of the Hartree-Fock ground state:
all HF calculations steadily converge to a
single solution for
the orbital-ordering pattern, which is shown in Fig.~\ref{fig.OOLTO}, irrespectively on the
starting conditions and the size of the supercell. Therefore,
although the correlation effects beyond the HF approximation are
certainly important
in the case of LaTiO$_3$,
it does not necessary mean that the ultimate result
of electron
correlations must be
formation of the orbital liquid state.
We hope that even in LaTiO$_3$, the correlation effects can be
systematically included using the regular perturbation theory expansion
around the \textit{nondegenerate} HF ground state, while the second order of this expansion is
simply not enough.
In such a situation,
it is perhaps more practical
to derive an approximate ground state
from the diagonalization of a many-electron Hamiltonian matrix constructed
in the basis of some limited number of specially selected Slater determinants,
as it is done for example in the path-integral renormalization group method.\cite{Imai,PIRG}

  Thus, one of the challenging problems in the theory of distorted perovskite
oxides remains to be the explanation of the G-type AFM ground state in LaTiO$_3$.
Definitely, we need a more rigorous theory for the
correlation effects. But, will it be enough,
or do we need a more radical refinement of our starting
model, given by Eq.~(\ref{eqn:Hmanybody}), in the case of LaTiO$_3$?

\subsection{\label{sec:spinorbit}Spin-Orbit Interaction and Magnetic Ground State}

  The spin-orbit interaction in the distorted perovskite oxides will
generally leads to the noncollinear spin arangement,
which obeys certain symmetry rules.\cite{PhysicaB1997}
The spin magnetic moments, aligned
along one of the
orthorhombic axes,
will be subjected to certain rotational forces,
coming both from the Dzyaloshinsky-Moriya interactions
and from the single-ion anisotropy energies,\cite{DzyaloshinskyMoriya,PRL96}
which lead to the spin canting and the appearance of nonvanishing
components of the spin-magnetization density along two
other directions.
The type of the magnetic ordering for all three projection of the
spin-magnetization density is generally different.
Thus, each magnetic structure can be generally abbreviated as
$X$-$Y$-$Z$, where $X$, $Y$, and $Z$ is the
type of the magnetic ordering (F, A, C, or G) formed by the projections
of the spin magnetic moments onto the orthorhombic
axes ${\bf a}$, ${\bf b}$, and ${\bf c}$, respectively.
The orbital magnetic structure has the same symmetry, although it has
a different origin of the canting, which comes mainly from the interplay
of the spin-orbit interaction with the CF splitting at each transition
site. The spin and orbital magnetic moments are not generally
collinear to each other.\cite{PRB97}

  Results of HF calculations, which take into account the spin-orbit interaction, are
summarized in Table~\ref{tab:HFGS}.
\begin{table}[h!]
\caption{The type of the magnetic ground state,
the vectors of the spin ($\boldsymbol{\mu}_S$) and orbital
($\boldsymbol{\mu}_L$) magnetic moments (in $\mu_B$), and the values of the band gap
($E_g$, in eV)
obtained in Hartree-Fock calculations. For the magnetic ground state, three capital letters denote
the type of
the magnetic ordering for three projections of the magnetic moments onto the
orthorhombic axes ${\bf a}$,  ${\bf b}$, and ${\bf c}$, respectively.
For the spin and orbital magnetic moments
at the transition-metal sites, these three projections are specified by the vectors
$\boldsymbol{\mu}_S$ and $\boldsymbol{\mu}_L$, respectively.
The positions of the transition-metal sites are shown in Fig.~\protect\ref{fig.structure}.}
\label{tab:HFGS}
\begin{ruledtabular}
\begin{tabular}{ccccccc}
 compound  & phase & site & ground state & $\boldsymbol{\mu}_S$                                                    &
                                           $\boldsymbol{\mu}_L$                                                    & $E_g$   \\
\hline
 YTiO$_3$  &  o    & 1    & G-A-F  & ($-$$0.00$,$-$$0.09$,$\phantom{-}$$0.84$)            &
                                           ($-$$0.05$,$\phantom{-}$$0.01$,$-$$0.03$)                      & $1.1$   \\
LaTiO$_3$  &  o    & 1    & C-F-A  & ($\phantom{-}$$0.02$,$-$$0.17$,$\phantom{-}$$0.76$)  &
                                           ($\phantom{-}$$0.10$,$\phantom{-}$$0.03$,$-$$0.08$)            & $0.6$   \\
  YVO$_3$  &  o    & 1    & F-C-G  & ($-$$0.02$,$\phantom{-}$$0.00$,$\phantom{-}$$1.65$)  &
                                           ($\phantom{-}$$0.00$,$-$$0.00$,$-$$0.17$)                      & $1.2$   \\
  YVO$_3$  &  m    & 1    & C-A-C  & ($-$$0.74$,$\phantom{-}$$0.08$,$\phantom{-}$$1.48$)  &
                                           ($\phantom{-}$$0.07$,$-$$0.04$,$-$$0.16$)                      & $1.0$   \\
           &       & 3    &              & ($-$$0.78$,$-$$0.03$,$\phantom{-}$$1.48$)            &
                                           ($\phantom{-}$$0.05$,$\phantom{-}$$0.04$,$-$$0.07$)            &         \\
 LaVO$_3$  &  m    & 1    & A-C-A  & ($-$$0.05$,$\phantom{-}$$1.64$,$-$$0.05$)            &
                                           ($\phantom{-}$$0.09$,$-$$0.18$,$\phantom{-}$$0.11$)            & $0.9$   \\
           &       & 3    &              & ($\phantom{-}$$0.06$,$\phantom{-}$$1.63$,$\phantom{-}$$0.05$)            &
                                           ($-$$0.05$,$-$$0.09$,$-$$0.06$)                                &         \\
\end{tabular}
\end{ruledtabular}
\end{table}
The magnetic ground state of
YTiO$_3$ is G-A-F, in agreement with the neutron scattering
data.\cite{Ulrich2002}
The ferromagnetic moment is aligned along the ${\bf c}$-axis,
and the canting angle is relatively small.
The absolute values of the spin and orbital magnetic moments are
different from those reported in our previous work.\cite{PRB04}
The difference is related with the additional transformation
to the LMTO basis, which allows to calculate the local magnetic moments
directly
at the transition-metal sites, whereas in Ref.~\onlinecite{PRB04}
they have been calculated in the Wannier basis, which had a
substantial weight at the oxygen sites.
This comparison clearly shows that the covalent effects play a very
important role and allow to explain a substantial reduction
of the local magnetic moments at the transition-metal sites.
As it was already discussed in Sec.~\ref{sec:YTO},
the correlation effects in YTiO$_3$ favor the AFM coupling and
systematically
lower the energies of all
AFM states relative to the ferromagnetic one
(see Table \ref{tab:YTO}).
After including the spin-orbit interaction, this will lead to the additional
spin canting away from the collinear ferromagnetic arrangement.
For example, in the framework of
superexchange approach we have obtained the following
values of the spin and orbital magnetic moments (in $\mu_B$, referred to the site 1):
$\boldsymbol{\mu}_S$$=$$(-0.02,-0.29,\phantom{-}0.77)$ and
$\boldsymbol{\mu}_L$$=$$(-0.05,\phantom{-}0.02,-0.03)$.
It readily explains the experimental values
reported for the F and G components of the magnetic moments,
correspondingly along the ${\bf c}$ and ${\bf a}$-directions,\cite{Ulrich2002}
while the A-type AFM component along the ${\bf b}$-direction
is somewhat larger in our calculations.
The reason is not completely clear.
The orbital magnetic moments in YTiO$_3$ are strongly quenched by the
crystal field.

  The orthorhombic YVO$_3$ has nearly-collinear magnetic structure, where
the G-type AFM moment is aligned along the ${\bf c}$-axis.
The total magnetic moment
$(\boldsymbol{\mu}_S$$+$$\boldsymbol{\mu}_L) \| {\bf c}$$=$$1.48\mu_B$
parallel to the ${\bf c}$-axis
is in excellent agreement with the experimental value of $1.45\mu_B$
reported by Blake~\textit{et~al.} and corresponding to T$=$$65$ K.\cite{Blake}
Ulrich~\textit{et al.} reported somewhat larger value of
$1.72\mu_B$, also oriented along ${\bf c}$.
The weak ferromagnetic component along the ${\bf a}$-direction
has been also observed experimentally.\cite{Ren}
Details of magnetic ordering in the monoclinic phase of YVO$_3$ are
somewhat controversial.
Blake~\textit{et al.} reported the C-type AFM ordering for both
${\bf b}$- and ${\bf c}$-components
of the magnetic moments in the orthorhombic $Pbnm$ notations,
which correspond to the ${\bf a}$- and ${\bf c}$-directions
in the monoclinic $P2_1/a$ notations.\cite{Blake}
This is totally consistent with our finding.
The quantitative difference can be naturally explained by the finite temperature
effects in the intermediate phase.\cite{FangNagaosa}
Furthermore, we
predict the A-type \textit{antiferrimagnetic}
ordering for the remaining ${\bf b}$-component (in the $P2_1/a$ notations),
which implies nearly antiferromagnetic alignment of
the ${\bf b}$-projections of the magnetic moments in the adjacent
${\bf ab}$-planes. However, since the sites $1$ and $3$ are not fully equivalent
in the monoclinic structure,
the ${\bf b}$-projections do not
compensate each other, and the system exposes the net magnetic moment, which
can couple to the magnetic field.\cite{Ren}
The \textit{antiferrimagnetic} ordering along the ${\bf b}$-axis is
also consistent with the temperature-induced magnetization reversal
behavior of YVO$_3$ observed by Ren,\cite{Ren}
and could be a natural explanation for this effect.
More generally, it is right to say that the magnetic coupling
along the ${\bf c}$-direction of the monoclinic phase is always either
ferr\textit{i}magnetic or antiferr\textit{i}magnetic, because the sites $1$ and $3$
are \textit{not equivalent}.

  The C-A-C magnetic ground state appears to be different from the
magnetic structure reported by
Ulrich~\textit{et al.}, who have observed the
C-type AFM ordering for the ${\bf a}$- and ${\bf b}$-components,
and the G-type AFM ordering for the remaining ${\bf c}$-component
(apparently, in the $Pbnm$ notations).\cite{Ulrich2003}
One possible explanation for this difference could be the coexistence
of several magnetic states in a narrow energy range. Such a behavior
has been indeed observed in our HF calculations: in addition to the C-A-C
state we could obtain another self-consistent solution corresponding
to the A-C-F phase with a slightly higher energy
(about $0.04$ meV per one formula unit, which is of the order of the
magnetocrystalline anisotropy energy). The new phase has the following
magnetic moments (in $\mu_B$, where the first and second
lines corresponds to the sites $1$ and $3$, respectively):
$$
\begin{array}{cc}
\boldsymbol{\mu}_S^1=(          -0.07,\phantom{-}1.65,\phantom{-}0.05), &
\boldsymbol{\mu}_L^1=(\phantom{-}0.05,          -0.15,          -0.02), \\
\boldsymbol{\mu}_S^3=(\phantom{-}0.07,\phantom{-}1.67,\phantom{-}0.03), &
\boldsymbol{\mu}_L^3=(          -0.09,          -0.07,          -0.02). \\
\end{array}
$$
Although the exact form of the magnetic state A-C-F is still different from the observation
by Ulrich~\textit{et al.},\cite{Ulrich2003}
we can speculate that they probably used a different experimental
setup which yielded the realization of another magnetic phase, which was
different from the finding by Blake~\textit{et~al.} (Ref.~\onlinecite{Blake})
and Ren~\textit{et~al.} (Ref.~\onlinecite{Ren}).
At least, the C-type AFM component parallel to the
${\bf ab}$-plane is qualitatively consistent with the form of the A-C-F
phase obtain in our HF conclusions, and to certain extent the results of
HF calculations can be further modified by the
correlation effects.

  In the case of LaVO$_3$ we could find three stable solutions: A-C-A, C-A-C, and C-F-G.
The first two are similar to the A-C-F and C-A-C states, emerging in monoclinic YVO$_3$.
The only difference is that the A-C-A state has lower energy
and, therefore, is realized as the ground state in the case of LaVO$_3$.
Like in YVO$_3$, the A-type state corresponds to the \textit{antiferrimagnetic}
ordering. Therefore, LaVO$_3$ is expected to have the net magnetic
moment in the ${\bf ac}$-plane. The third (C-F-G) solution has
nearly collinear spin structure:
$$
\begin{array}{cc}
\boldsymbol{\mu}_S^1=(          -0.01,          -0.06,\phantom{-}1.60), &
\boldsymbol{\mu}_L^1=(\phantom{-}0.01,\phantom{-}0.13,          -0.34), \\
\boldsymbol{\mu}_S^3=(          -0.03,          -0.02,          -1.61), &
\boldsymbol{\mu}_L^3=(\phantom{-}0.01,\phantom{-}0.04,\phantom{-}0.14). \\
\end{array}
$$
It corresponds to the stable G-type AFM phase emerging without the spin-orbit interaction
(see Table~\ref{tab:YVOo}).
Nevertheless, the A-C-A state appears to be well separated from the C-A-C and C-F-G
states, correspondingly by $0.19$ and $6.50$ meV per one formula unit.

  The quenching
of the orbital magnetic moments in two different sublattices of monoclinic
YVO$_3$ and LaVO$_3$ nicely correlates with
the values of the CF splitting (Fig.~\ref{fig.CFsummary}), where
larger CF splitting at the site 3 results in smaller orbital magnetization.

  It is probably meaningless to discuss the relativistic effects in LaTiO$_3$,
where we could not reproduce the correct magnetic ground state. We could
agree with the
criticism risen by Haverkort~\textit{et~al.} (Ref.~\onlinecite{Haverkort})
that our CF splitting alone
does not explain details of their spin-resolved photoemission spectra
(actually, our value of the parameter
$\langle {\bf L} \cdot {\bf S} \rangle$,
obtained in the HF calculations after the transformation to the LMTO basis and radial
integration over the Ti-sphere is about $-$$0.13$,
which
exceeds the experimental value by factor two).
However, it does not make a sense to present as an alternative
the results of calculations
yielding the same A-type AFM ground state,\cite{Streltsov}
which totally
agrees with our finding and (unfortunately) disagrees
with the experiment, even though these calculations yield
somewhat larger value of the CF splitting for LaTiO$_3$.
As it has been already discussed in Sec.~\ref{sec:kinetic},
different values of the CF splitting
obtained by different authors are most probably related with
the nonunique choice
of the Wannier functions for the $t_{2g}$ bands of the
distorted perovskite oxides. One can formally adjust the
theoretical CF splitting
in order to meet certain demands of some particular class of the experimental data.
However, will it solve a more fundamental
problem of the magnetic ground state of LaTiO$_3$?

  In addition, we show in Table~\ref{tab:HFGS} the values of the band gap obtained in the HF calculations
for the magnetic ground state. For YTiO$_3$, YVO$_3$, and LaVO$_3$
there is a good agreement with the experimental
optical data.\cite{Tsvetkov,optics_exp}
However,
for LaTiO$_3$
the experimental gap
is much smaller ($\sim$$0.1$ eV).\cite{optics_exp}
This may indicate again at the particular importance of correlation effects in LaTiO$_3$.

\section{\label{sec:summary}Summary and Concluding Remarks}

  The main purpose of this work was
to make a bridge between first-principles
electronic structure calculations and model approaches for the
strongly-correlated systems, and illustrate how it works for the series
of distorted $t_{2g}$ perovskite oxides.
The whole plan included three major steps:
\textit{first-principles electronic structure calculations} $\rightarrow$
\textit{construction of the model Hamiltonian} for the isolated $t_{2g}$ bands
near the Fermi level $\rightarrow$ \textit{solution of this model Hamiltonian}
using several different approaches.
The choice of the distorted $t_{2g}$ perovskite oxides
was motivated by the fact that
they represent a good example of systems where
it is practically impossible to construct a relevant model
without the impact from the first-principles calculations:
simply, the lattice distortion is too complex and there are too many
model parameters, which cannot be fixed in unbiased way.
In this sense, we strongly believe that any
theoretical model for such complex oxide materials
should be based on the results of first-principles electronic structure calculations.
Otherwise it could be just an abstract mathematical construction deprived of
clear physical grounds.

  The present work clearly demonstrates that nowadays the idea of constructing the parameter-free
model Hamiltonians for the strongly-correlated systems is quite feasible:
all model parameters in our work, including the intraatomic Coulomb interactions,
have been derived from the first-principles calculations using the
method proposed in Ref.~\onlinecite{condmat05}. Apart from the approximations inherent to the
method, the procedure of constructing the model
Hamiltonian was totally parameter-free, and our analysis \textit{did not rely on the use of any adjustable parameters}.
Therefore, it is remarkable that using such a parameter-free approach we could propose
a consistent explanation for a number of
puzzling properties of the distorted $t_{2g}$ perovskite oxides.
These first results are really encouraging and we would like to hope that the method proposed in Ref.~\onlinecite{condmat05}
can be successfully applied in the future for the analysis of electronic and magnetic properties
of other narrow-band compounds.

  At the mean-field Hartree-Fock level,
the results of
all-electron
full-potential
LDA$+$$U$ calculations (Refs.~\onlinecite{SawadaTerakura,FangNagaosa})
can be successfully reproduced in our
model approach for the isolated $t_{2g}$ bands.
We could easily rationalize the main results of these relatively heavy
calculations and elucidate the main microscopic interactions
responsible for the formation of different magnetic structures in
the case of
YTiO$_3$, YVO$_3$, and LaVO$_3$.
We argue that the nonsphericity of the Madelung potential
should be
an indispensable ingredient of electronic structure calculations, and the
results of commonly used atomic-spheres-approximation should be corrected
in order to include these effects.

  The crystal distortion plays an important role in the physics of
$t_{2g}$ perovskite oxides. At least for YTiO$_3$, YVO$_3$, and LaVO$_3$,
the knowledge of the experimental lattice parameters and the atomic positions
greatly helps in explaining the magnetic properties of
these compounds. Of course, some questions still remain. Particularly,
what is the origin of this distortion? Why is it so different for different
compounds, that is finally manifested in the formation of completely different
magnetic structures?
Very similar arguments have been used for LaMnO$_3$, which is a characteristic example
of the Jahn-Teller distorted $e_g$ perovskite oxides.
Particularly,
it was argued that the experimental distortion not
only stabilize the A-type AFM ground state of LaMnO$_3$,\cite{PRL96,LaMnO3_band,Maezono,GorkovKresin}
but is also responsible for the opening of the band gap.\cite{LaMnO3_band,GorkovKresin}
However, the
direction of the Jahn-Teller distortion in LaMnO$_3$
can be naturally understood in terms of
anharmonicity of the electron-lattice interaction.\cite{KugelKhomskii,Kanamori3,Millis}
In this sense, LaMnO$_3$ is an easy example.
Then, is it possible to
rationalize the behavior of
$t_{2g}$ perovskite oxides in a similar way and come up with
some suitable electron-lattice model,
which would explain not only the direction of the lattice distortion
in each particular compound, but also the difference between these compounds?

  Finally, we emphasize the importance of correlation effects in the $t_{2g}$ band of
distorted perovskite oxides. Although the mean-field Hartree-Fock approach provides a
satisfactory description for the magnetic properties of YTiO$_3$, YVO$_3$, and LaVO$_3$, the inclusion of
the correlation effects systematically improves the agreement with the experimental data
for all three compounds.
Definitely,
LaTiO$_3$ is an exceptional example for which we could not obtain the correct
G-type AFM ground state neither at the level of Hartree-Fock approach nor after
including some correlation effects, though in a very approximate form.
However, we expect that the situation may still change by systematically employing
more rigorous theories
for the correlation effects.
The approximations considered in the present work were simply not enough in the case of LaTiO$_3$.

\begin{acknowledgments}
I am grateful to Professor Masatoshi Imada for valuable
discussions.
\end{acknowledgments}


\begin{thebibliography}{10}
\bibitem{comment.7}
There are certain indications that the actual symmetry
of the intermediate phase of YVO$_3$ can be even lower
than $P2_1/a$.\cite{Tsvetkov}
\bibitem{comment.6}
For example,
from the viewpoint of superexchange interactions, it is more natural
to define the ``orthogonal'' orbitals as the ones which are decoupled
via the matrix of transfer integrals between neighboring transition-metal sites:
$\langle \phi_{\bf R} | \hat{h}_{{\bf RR}'} | \phi_{{\bf R}'} \rangle$$=$$0$,
which is not always the same as the regular orthogonality condition.\cite{PRB04}
\bibitem{comment.2}
The Wannir functions have been calculated using the method proposed in Ref.~\onlinecite{condmat05}.
In these calculations, the Wannier functions for the $t_{2g}$ bands have been orthogonalized to
the O($2p$), Ti($e_g$), and Y($4d$) bands, in the case of YTiO$_3$;
O($2p$), Ti($e_g$), and La($5d6sp$) bands, in the case of LaTiO$_3$;
O($2p$), V($e_g$), and Y($4d$) bands, in the case of YVO$_3$; and
O($2p$), V($e_g$), and La($5d$) bands, in the case of LaVO$_3$. We have confirmed that
such a accuracy is sufficient for calculations of
the crystal-field splitting caused by the nonsphericity of the Madelung potential (\ref{eqn:CFEI}).
\bibitem{comment.5}
Note that $2/(E_{\rm S}$$+$$E_{\rm T})$$>$$1/E_{\rm S}$$+$$1/E_{\rm T}$.
Therefore, according to Eq.~(\ref{eqn:Esuperexchange}), two
separate excitations into the singlet (S) and triplet (T) states
are more favorable energetically rather than the excitation into the
Hartree-Fock state
with an averaged energy.
\bibitem{comment.3}
Note, however, that the LDA$+$$U$ calculations
reported in Refs.~\onlinecite{SawadaTerakura}
and \onlinecite{FangNagaosa}
for YVO$_3$
overestimate the total energy difference between
different magnetic states, apparently due to the double-counting of the nonsphericity effects.
By default,
in the full-potential implementation of the LDA$+$$U$ method, the nonsphericity of the
on-site Coulomb and exchange-correlation potential is taken into account twice:
in the LDA-part and in the correction for the on-site Coulomb interactions.
Therefore, the LDA-part should be additionally averaged
inside some atomic sphere
and we belive that it is more
physical to treat this part in ASA.\cite{condmat05}
This problem mainly concerns with the $d^2$ compounds, which
are subjected to the Hund's rule effects
coming from the nonspherical Coulomb interactions between two
$t_{2g}$ electrons.
A more general problem is related with the technical implementations of the
LDA$+$$U$ method (e.g. the choice of the parameter $U$ and the basis orbitals
for the correction), which probably explain a small difference existing
between our results and Ref.~\onlinecite{SawadaTerakura} for
YTiO$_3$.
\bibitem{comment.4}
$8J_{12}$$=$$E({\rm F})$$-$$E({\rm G})$$+$$E({\rm A})$$-$$E({\rm C})$
and
$4J_{13}$$=$$E({\rm F})$$-$$E({\rm G})$$-$$E({\rm A})$$+$$E({\rm C})$,
where $E$ is the total energy either in the pure HF approximation or after including
the correlation effects in the second order of perturbation theory.
\bibitem{comment.1}
Note that our scheme of calculation of the exchange parameters is based on the
infinitesimal rotations of the spin magnetic moments, which implies that the
spins are classical. Therefore, the total energy of the
Heisenberg model has the form (\ref{eqn:EHeisenberg}),
where $J_{{\bf RR}'}$ are
related with the regular Heisenberg parameters $\tilde{J}_{{\bf RR}'} $
as $J_{{\bf RR}'}$$=$$S^2\tilde{J}_{{\bf RR}'}$.
Therefore, to be consistent we our definition of the exchange parameters, we
use the classical
expression also for the transition temperature and replace $S(S$$+$$1)$
by $S^2$.
In order to use the quantum-mechanical definition for the transition temperature,
one should do a proper mapping of the total-energy change onto the Heisenberg model, which
is much more complex, because
in quantum case, the total energy
remains proportional to $S^2$ only for the
ferromagnetically coupled spins.
The antiferromagnetic arangement yields an additional prefactor
$\alpha$, which is specific to the geometry of the system and bounded
by the following conditions: $(1$$+$$1/S)$$\le$$\alpha$$\le$$1$.
The latter
are far from restrictive when the spin is small.
Therefore, the quantum effects, which are especially important for the spin $1/2$,
will modify not only the
expression for the transition temperature, but also the
definition
of the parameters of the Heisenberg model themselves.
\end{thebibliography}
\end{document}